\documentclass[a4paper,10.5pt]{article}
\usepackage{theorem}
\usepackage{bm}
\usepackage{mathdots}
\usepackage{latexsym,amssymb,amsfonts,amsmath,cite}
\usepackage[dvipdfmx]{graphicx}

\newcommand{\ket}[1] {| #1 \rangle}
\newcommand{\bra}[1] {\langle #1 |}
\usepackage{color}

\newtheorem{theorem}{Theorem}
\newtheorem{lemma}{Lemma}
\newtheorem{corollary}[theorem]{Corollary}
\newtheorem{proposition}[theorem]{Proposition}

\begin{document}
\title{{\Large {\bf Relation between two-phase quantum walks 
and the topological invariant 
}
}}

\author{ 
{\small 
Takako Endo,$^{1}$ 
\footnote{endo-takako-sr@ynu.ac.jp
}\quad  
Norio Konno,$^{2}$ 
\footnote{konno@ynu.ac.jp 
}\quad 
Hideaki Obuse,$^{3}$ 
\footnote{hideaki.obuse@eng.hokudai.ac.jp 
}\quad 
}\\ 
{\scriptsize $^{1}$ 
Department of Applied Mathematics, Faculty of Engineering, Yokohama National University
}\\
{\scriptsize Hodogaya, Yokohama 240-8501, Japan 
} \\
{\scriptsize $^{2}$ 
Department of Applied Mathematics, Faculty of Engineering, Yokohama National University
}\\
{\scriptsize Hodogaya, Yokohama 240-8501, Japan
} \\
{\scriptsize $^{3}$ 
Department of Applied Physics, Hokkaido University
}\\
{\scriptsize Kita 13, Nishi 8, Kita-ku, Sapporo Hokkaido 060-8628, Japan
} \\
} 

\vskip 1cm

\date{\empty}
\pagestyle{plain}

\maketitle

\par\noindent
\begin{small}
\baselineskip=24pt
\par\noindent
{\bf Abstract}. 
We study a position-dependent discrete-time quantum walk (QW) in one dimension, whose
time-evolution operator is built up from two coin operators 
which are distinguished by phase factors from $x\geq0$ and $x\leq-1$.  
We call the QW the {\it complete two-phase QW} to discern from the
 two-phase QW with one defect\cite{endosan,maman}.
Because of its localization properties, the two-phase QWs can be considered as an ideal mathematical
 model of topological insulators which are novel
 quantum states of matter characterized by topological invariants.
Employing the complete two-phase QW, we present the stationary measure, and two kinds of limit theorems
 concerning {\it localization} and the {\it ballistic spreading}, which
are the characteristic behaviors in the long-time limit of
 discrete-time QWs in one dimension.
As a consequence, we obtain the mathematical expression of the whole
 picture of the asymptotic behavior of the walker, including
 dependences on initial states, in the long-time limit. 
We also clarify relevant symmetries, which are essential for topological insulators, of the
 complete two-phase QW, and then derive the topological invariant.
Having established both mathematical rigorous results and the
topological invariant of the complete two-phase QW, we 
provide solid arguments to understand localization of QWs in term of topological invariant.
Furthermore, by applying a concept of {\it topological protections}, we
 clarify that localization of the two-phase QW with one defect, studied in the previous work\cite{endosan}, can be related to
 localization of the complete two-phase QW under symmetry preserving perturbations.

\end{small}

\baselineskip=24pt

\setcounter{equation}{0}
\section{Introduction}
Quantum walks (QWs) are considered as quantum counterparts of classical random walks. 
There are mainly two kinds of QWs, that is, the discrete-time QW and continuous-time one\cite{ide,segawa}. 
In this paper, we focus on the discrete one.  
It was mathematically shown that quantum search algorithms constructed
by QWs provide faster computations
than the corresponding classical algorithms. 
As other applications, QWs are ideal platforms to study energy
transportation efficiency of photosynthesis\cite{mo}, and the Anderson
localization in disordered systems\cite{crespi13,schreiber11}, for instance.
Owing to the rich applications, it is of great importance to study the
asymptotic behavior of QWs, however, it would be difficult to implement the states in the long-time limit by experiment.
Moreover, because of its quantumness, it is difficult to understand QWs intuitively.
Therefore, to understand the asymptotic behavior of QWs, it is
exceedingly important both to numerically simulate the time-step evolution of QWs, and obtain limit theorems mathematically.\\
\indent
So far, two kinds of limit theorems have described the characteristic properties of QWs mathematically.
The one is the limit theorem expressing localization. 
Localization is one of the typical properties of discrete-time QWs, which was first studied by Inui {\it et al.} \cite{inui} mathematically and numerically. 
The detailed definition of localization is devoted to \cite{scholz,joye} for example. 
The other is the weak convergence theorem, which expresses ballistic spreadings of the walker by the weak limit measure, and is fully explained in \cite{segawa}, for instance. 
We should note that the weak limit measure is consisted by the Dirac measure part corresponding to localization and absolutely continuous part, corresponding to the ballistic spreading.
The weak convergence theorem for space-homogeneous QWs in one dimension, such as Hadamard walk\cite{konnoweak}, Grover walk\cite{kota}, was derived. \\
\indent
Nowadays, the research on the asymptotic behavior of space-inhomogeneous
QWs is one of the hottest topics in the filed of QWs. Here we review 
previous studies of inhomogeneous QWs.

The study of limit measures for inhomogeneous QWs has been started from the
one-dimensional QW with a single defect since
2009\cite{konno09,konno10}.
Konno introduced one-dimensional QWs of which the coin operator is
position dependent so that the coin operator except at the origin is the
Hadamard matrix, but one at the origin has extra phases $e^{\pm i
\omega}$ to diagonal elements\cite{konno09} or to off-diagonal
elements\cite{konno10} of the Hadamard matrix.
Konno has derived, by using the path-counting method, the weak limit
theorem and the vanishing return
probability at the origin in long time limit for the former
model, while the return probability for the latter model remains finite.
The same conclusion for the return probability has been confirmed by using the CGMV method for the
one-dimensional QW with a single defect whose coin is a general $U(2)$
matrix\cite{cantero}.
The eigenvalues and eigenvectors corresponding localization have been also derived for the
one-dimensional QW with a point defect where the defect coin has an opposite sign
to the Hadamard matrix\cite{ahlbrecht}.

In addition to  the stationary
measure,
Konno {\it
et al.} \cite{segawa}, for the first time, gave the time-averaged limit measure and the weak
convergence theorem showing localization for 
a one-dimensional QWs with one defect coin whose
determinant is minus one, taking advantage of the
generating function 
of the weights of passages. Wojcik {\it et al.} \cite{wojcik} studied
the one-dimensional QWs with one defect which
is realized by multiplying an extra phase $e^{i \omega}$ by the Hadamard
matrix at the origin, called the Wojcik model, and
showed analytically and numerically that the Wojcik model exhibits astonishing
localization effects for changing the phase of the defect. Then, Endo and
Konno\cite{watanabe, endo} proved the prominent localization effect
mathematically, and in subsequent, they gave the weak convergence
theorem for the Wojcik model, which completed the whole description of
the asymptotic behavior. Endo {\it et al.} \cite{madameendo} got a
stationary measure of the QW with one defect whose quantum coin is
defined by the Hadamard matrix at $x\neq0$ and the rotation matrix at
$x=0$. Recently, Endo {\it et al.} \cite{endosan, maman} studied the
two-phase QW with one defect which has two different time-evolution
operators in positive and negative spatial regions, in addition to  another operator at the origin. 
They derived the limit
theorems concerning localization and the ballistic behavior, and they
clarified the effect of the two different quantum coins and initial coin
state of the walker on the asymptotic behavior.\\
\indent
Recently, localization appearing in the space-inhomogeneous QWs has attracted
attentions from a quite different direction, i.e., topological
invariants characterized by eigenvectors of the QW\cite{kitagawa12,kitagawa12exp,kitagawa10}.
This approach has emerged from recent developments on topological
phases of matter in the condensed matter physics, known as topological
insulators and superconductors\cite{bernevig13,chiu15,hasan10,qi11}. 
Since the QW can be considered as the simplified theoretical model of
topological insulators (more preciously, Floquet topological insulators\cite{kitagawafloquet10,lindner11}),
the topological phase of quantum walks has been intensively studied\cite{asboth12,asboth15,asboth13,cedzich15,chandru15,edge15,obuse15,obuse11,rakovszky15,tarasinski14}.  
In the field
of theoretical physics on the topological insulators, 
a fundamental principle, so-called the
bulk-edge correspondence, predicts the existence of localized states
at the interface where two
adjacent spatial regions are characterized by different
topological numbers.

In case of topological insulators where the Fermi energy is in the bulk
energy gap, the edge states predicted by the bulk-edge correspondence dominate electronic properties. In contrast,
in case of quantum walks where the initial state is usually a point like
state and then there is no corresponding Fermi energy, the bulk-edge
correspondence can predict the existence of edge states, but cannot determine how much edge states contribute to the
dynamics, {\it i.e.}, how much the edge states overlap with the
initial state. To this end, we need to know the initial state dependences of
the probability distribution at large time steps,
which can be provided by the time-averaged limit measure.  

A recent work on the two-phase QW with one defect\cite{endosan, maman}
gives the time-averaged limit measure with a point like initial state.  
In the two-phase QW with one defect, however, there are three distinct spatial
regions, i.e., positive, negative spatial regions and the origin. 
On the topological phase of matter, the topological invariant is
generally defined for a system with a finite spatial extent, roughly speaking, 
enough larger than the localization length of localized states.
Thereby, the single defect at the origin may prevent to apply the argument of topological
invariants straightforwardly. Furthermore, such defect coin makes the experimental setup difficult.
\\
\indent
In the present work, we study a complete two-phase QW which
is defined by modifying the two-phase QW with one defect so that the coin operator at the origin
is replaced with that in the positive space part.
One of the important advantages to study the complete two-phase QW is that the model enables us to directly discuss the
mathematically rigorous results on localization of the QW on the line in term of  topological invariants.
In addition, it would be worth to mention that the specific setup of the complete two-phase QW is
already realized in the QW by the bulk optics\cite{kitagawa12exp} and
optical fiber loops\cite{barkhofen16}(though the coin operator in the
latter one is slightly different from that of the present work).
\\
\indent
First, we derive mathematically rigorous results on the
time-averaged limit, stationary,
and weak limit measures of the
complete two-phase QW. 
There have been constructed several kinds of popular techniques to mathematically investigate the asymptotic behaviors of QWs, such as 
the Fourier analysis\cite{schudo}, the CGMV method\cite{cantero}, the stationary phase method\cite{nayak}, the pass counting method\cite{konnopath}, and the generating function methods\cite{watanabe,segawa}. 
These methods are epoch-making, however, there are a lot of strict conditions. 
For example, we can only analyze space-homogeneous QWs by the Fourier analysis\cite{schudo}. 
Motivated by the past studies, we take advantage of two kinds of the generating function methods\cite{watanabe,segawa}.
There is a possibility to analyze various kinds of space-inhomogeneous QWs by the methods, however, it has not been clear the types of QWs that can be analyzed by the generating function methods. 
We can also analyze inhomogeneous QWs via the CGMV method, still the CGMV method allows only for the general discussion of localization properties for the typical QWs with one defect on the line. 
One of the generating function methods offers not only the time-averaged limit theorem which describes localization, but also the weak convergence theorem for QWs. The other generating function method provides the stationary measure which corresponds to the stationary distribution.
It is the first application of the generating function methods for the analysis of inhomogeneous QW without defect.\\
\indent
Next, we study topological invariants of the complete two-phase QW.
So far, coin operators whose matrix elements are real numbers are mostly
employed to study the topological invariant of quantum walks, since
these coin operators apparently satisfy necessary conditions required
from relevant symmetries to establish a non-trivial topological invariant. However, we present
that, by applying a proper unitary transform into the time-evolution
operator, the coin operator with complex
numbers  of the complete two-phase QW can satisfy the condition.
Furthermore, we find that, since this unitary
transform should be applied to the whole space of the complete two-phase QW, the bulk-edge correspondence predicting the
localized states, is applicable when the phases of the two-phase QW
satisfy a specific condition,
nevertheless the time-averaged limit measure shows the presence of
localized states unless the QW is homogeneous.\\
\indent
Taking into account the mathematical rigorous results and the
topological invariant of the complete two-phase QW, we argue the relation between the localization of the complete
two-phase QW and the topological invariant with a help of the bulk-edge correspondence.
We confirm that the bulk-edge correspondence agrees well with 
theorems for the stationary and the time-averaged limit measures.
We also clarify the symmetry preserving perturbative coin operator for
the complete two-phase QW. Accordingly, we clarify the relation of
localization of the two-phase QW with one defect and one of the
complete two-phase QW. 
\\
\indent
The rest of this paper is organized as follows. In Section \ref{modelresult},
we define the complete two-phase QW which is the main target in
this paper, and present our mathematically rigorous results. 
The topological invariant of the corresponding time-evolution operator is studied in Section \ref{topological-invariant}.
Then, we present several examples to argue our results in
Section \ref{examples}. 
Section \ref{conclusion} is devoted to conclusions.

\section{Model and mathematically rigorous results}

\label{modelresult}

In this section, we first give the definition of the complete two-phase QW, and
then present mathematically rigorous results on the time-averaged limit, the stationary, and the weak limit measures.

\subsection{the complete two-phase QW}

In this paper, we treat a two-state discrete-time QW in one dimension
whose one-time step is defined by a unitary time-evolution operator $U^{(s)}$:
\begin{align}
U^{(s)} = \sum_{x\in\mathbb{Z}}|x\rangle\langle x|\otimes U_{x},
\label{eq:time-evolution}
\end{align}
where $U_x$ is called {\it the coin operator} expressed by
\begin{align}
U_{x}=\left\{
\dfrac{1}{\sqrt{2}}\begin{bmatrix}
1 & e^{i\sigma_{+}} \\
e^{-i\sigma_{+}} & -1 \\
\end{bmatrix}_{x\geq 0},\;
\dfrac{1}{\sqrt{2}}\begin{bmatrix}
1 & e^{i\sigma_{-}} \\
e^{-i\sigma_{-}} & -1\\
\end{bmatrix}_{x\leq -1}\right\},\label{2phase_def}
\end{align}
with $\sigma_{\pm}\in[0,2\pi)$.
As a discrete-time QW, the walker has a coin state at position $x\in\mathbb{Z}$ and time $t\in\mathbb{Z}_{\geq 0}$ expressed by a two-dimensional vector:
\begin{align*} \Psi_{t}(x)=
\begin{bmatrix}
\Psi_{t}^{L}(x) \\ 
\Psi_{t}^{R}(x)
\end{bmatrix} \in\mathbb{C}^{2}
\quad(x\in \mathbb{Z}\;\;\;\Psi_{t}^{L}(x),\Psi_{t}^{R}(x)\in \mathbb{C}),\end{align*}
where $\mathbb{C}$ is the set of complex numbers, and $\mathbb{Z}$ is the set of integers. We should note that the walker steps to the left or right according to the recurrence formula
\begin{align*}
\Psi_{t+1} (x) = P_{x+1} \Psi_t (x+1) + Q_{x-1} \Psi_t (x-1) \quad (x \in \mathbb{Z}),
\end{align*}
where
\begin{align*}P_x =\left\{
 \dfrac{1}{\sqrt{2}}
\begin{bmatrix} 
1 & e^{i\sigma_{+}} \\ 
0 & 0 
\end{bmatrix}
_{x\geq 0},\;
 \dfrac{1}{\sqrt{2}}
\begin{bmatrix} 
1 & e^{i\sigma_{-}} \\ 
0 & 0 
\end{bmatrix}_{x\leq -1}
\right\},
\quad Q_x = \left\{
\dfrac{1}{\sqrt{2}}
\begin{bmatrix} 
0 & 0 \\ 
e^{-i\sigma_{+} }& -1
\end{bmatrix}_{x\geq 0},\;
\dfrac{1}{\sqrt{2}}
\begin{bmatrix} 
0 & 0 \\ 
e^{-i\sigma_{-} }& -1
\end{bmatrix}_{x\leq -1}\right\},\end{align*}
with $U_x = P_x + Q_x$. 

We note that $P_x$ and $Q_x$ correspond to the left and right movements, respectively, and 
the walker steps differently in the spatial regions with the phase parameters $\sigma_{+}$ and $\sigma_{-}$, that is, $x\geq0$ and $x\leq-1$ . The QW does not
have defect at the origin, which is in marked contrast to the two-phase QW
with one defect\cite{endosan, maman}. Hereafter, we call the QW $\it the\; complete
\;two$-$\it phase\; QW$.
Putting $\sigma_{+}=\sigma_{-}=0$, the model becomes the Hadamard walk, which has already been intensively studied\cite{segawa,konnoweak}. 
At first, we derive limit theorems concerning localization for our QW,
that is, the time-averaged limit measure and the stationary measure.
Then, we show the weak convergence theorem describing the ballistic
spreading in the distribution of the position in a re-scaled space, which
contributes to mathematically express the whole description of the
behavior of the walker in the long-time limit. 
We also show numerical results for some concrete phase parameters and initial states to see
what our analytical results suggest, especially, to relate the complete two-phase QW to the topological phases in Section \ref{examples}.


\subsection{rigorous result $1$: the time-averaged limit measure}
Let $P(X_{t}=x)$ be the probability that the walker exists at position $x\in\mathbb{Z}$ at time $t\in\mathbb{Z}_{\geq 0}$, where $\{X_{t}\}$ is a set which is defined by $P(X_{t}=x)=\|\Psi_{t}(x)\|^{2}$.
Then, localization of one-dimensional QWs is defined by 
\[\lim_{t\to \infty} \sup P(X_{t}=0)>0.\]
Furthermore, localization can be also mathematically described by
the time-averaged limit measure\cite{endo}. 
More explicitly, the QW starting at the origin shows localization if and only if
 $\overline{\mu}_{\infty}(0)$ is strictly positive:
\[\overline{\mu}_{\infty}(0)=\lim_{T\to \infty}\frac{1}{T}\sum^{T-1}_{t=0}P(X_{t}=0)>0.\]
Now we give the time-averaged limit measure for the complete two-phase QW, the first of the rigorous results in our study.
Let $\Psi_{0}(x)=\delta_{0}(x)\varphi_{0}$ with $\varphi_{0}={}^T\![\alpha,\beta]$ be the initial coin state, where
$\alpha,\beta\in\mathbb{C}$, and  $\overline{\mu}_{\infty}(x)$ be the
time-averaged limit measure at position $x\in\mathbb{Z}$. Throughout
this work, we assume that the walker starts at the origin, and we put
$\alpha=a e^{i\phi_{1}},\;\beta=b e^{i\phi_{2}}$ with $a^{2}+b^{2}=1$,
and $\phi_{j}\in\mathbb{R}\;(j=1,2)$, where $a,b\in\mathbb{R}$. Here
$\mathbb{R}$ is the set of real numbers. The range of $\sigma_+$ is
changed to $\sigma_+ \in [\sigma_-,\sigma_- + 2\pi)$ so that Theorem
\ref{timeavelimit} can be simply expressed, 
though the general expression in the case of $\sigma_{\pm}\in[0,2\pi)$ is available in Appendix A.
\begin{theorem}
\label{timeavelimit}
\noindent 
Put $\sigma=(\sigma_{+}-\sigma_{-})/2 \in [0,\pi)$ and $\tilde{\phi}_{12}=\phi_{1}-\phi_{2}\;$.
Then, we have 
\begin{eqnarray}
\overline{\mu}_{\infty}(x)\!\!\!&=&\!\!\!\dfrac{A(\sigma) \zeta(\sigma_+,\sigma_-,a,b,\tilde{\phi}_{12})}{\left\{3-2\sqrt{2}\eta_-(\sigma)\right\}^{|x+1/2|}},
\label{2-phase.timeaveraged}
\end{eqnarray}
where
\begin{eqnarray*}
\left\{
\begin{array}{l}
\zeta(\sigma_+,\sigma_-,a,b,\tilde{\phi}_{12})=2-\sqrt{2}\left\{a^{2}\eta_{+}(\sigma)+b^{2}\eta_{-}(\sigma)\right\}
-2\sqrt{2}ab\sin\varphi(\sigma) \sin\left(\tilde{\phi}_{12}-\dfrac{\sigma_{+}+\sigma_{-}}{2}\right),\\
A(\sigma)= \dfrac{2\left\{1-\sqrt{2}\eta_{-}(\sigma)\right\}^{2}}
{
\left\{3-2\sqrt{2}\eta_-(\sigma)\right\}^{3/2}
\left[5+\cos 2\sigma-2\sqrt{2}\left\{\eta_{+}(\sigma)+\eta_{-}(\sigma)\right\}\right]
},\\
\eta_\pm(\sigma)=\cos\left\{ \varphi(\sigma)\pm\sigma \right\},\\
\varphi(\sigma) = - \arccos\left(\dfrac{1}{\sqrt{2}}\cos \sigma\right).
\end{array}
\right.
\end{eqnarray*}
\end{theorem}
{\bf Remark 1}: Only the denominator of Eq.\ (\ref{2-phase.timeaveraged}) depends
on the position $x$, while factors in the numerator, $A(\sigma)$ and
$\zeta(\sigma_+,\sigma_-,a,b,\tilde{\phi}_{12})$, do not.\\
{\bf Remark 2}: $\zeta(\sigma_+,\sigma_-,a,b,\tilde{\phi}_{12})$ depends on all parameters including
of the initial state, respectively.\\
{\bf Remark 3}: The time-averaged limit measure has an exponential
decay for the position, whose ratio is given by $1/\{3-2\sqrt{2}\eta_{-}(\sigma)\}$. We note that when $\sigma=0$,
the system is homogeneous and 
the ratio of decay becomes one and $A(0)=0$ (no
localization because of $\overline{\mu}_\infty(x)=0$), while for the other value of $\sigma$, the ratio of decay
is always larger than one. \\
{\bf Remark 4}: The time-averaged limit measure 
shows the symmetric distribution at $x=-1/2$,
which is in marked contrast to that of the
two-phase QW with one defect which has an origin symmetry
\cite{endosan}.

\begin{corollary}
Even in
inhomogeneous case, by choosing appropriate phase parameters, $\sigma_+$, $\sigma_-$ and the initial state, the time-averaged limit measure becomes
zero and localization at the
origin may not happen 
when $\zeta(\sigma_+, \sigma_-, a,b,\tilde{\phi}_{12})$ becomes zero.
\end{corollary}

To show this, at first, we derive a condition of the local minimum and
maximum of $\zeta$ by differentiating $\zeta$ by
$\tilde{\phi}_{12}$ and $\vartheta$ after rewriting $a =
\cos(\vartheta)$ and $b=\sin(\vartheta)$ where $\vartheta \in
[0,\pi/2]$.
The condition of the local
minimum or maximum of $\zeta(\sigma_+,\sigma_-,a,b,\tilde{\phi}_{12})$
for $\tilde{\phi}_{12}$ and $\vartheta$ is given by
\begin{align}
 \tilde{\phi}_{12} &= \frac{\sigma_+ + \sigma_-}{2} +
 \frac{2n+1}{2}\pi,\label{eq:local-minimum1}\\
\vartheta &=
 \frac{(-1)^{n+1}}{2} \arctan\left( \frac{1 }{\sin \sigma}\right) + \frac{\pi}{2}\,\frac{1+(-1)^n}{2},
\label{eq:local-minimum2}
\end{align}
where $n \in \mathbb{Z}$.
Substituting Eqs.\ (\ref{eq:local-minimum1}) and
(\ref{eq:local-minimum2}) into $\zeta$ in Eq.\
(\ref{2-phase.timeaveraged}), we obtain
\begin{equation}
 \zeta(a,b,\phi,\sigma_+,\sigma_-) = (2-\cos^2 \sigma)\left\{1+(-1)^n\right\}.
\end{equation}
Accordingly,  the value of $\zeta$ at a local minimum point determined from
Eqs. (\ref{eq:local-minimum1}) and (\ref{eq:local-minimum2})
when $n$ is odd is zero
for arbitrary $\sigma_+$ and $\sigma_-$, while
$\zeta$ has a local maximum at a point determined when $n$ is even. Therefore, there exists
a special initial coin state for arbitrary $\sigma_+$ and $\sigma_-$,
which satisfies $\overline{\mu}_\infty(x)=0$. Also, we can find an
initial coin state so that $\overline{\mu}_\infty(x)$ at $x=-1,0$
becomes largest for given $\sigma_+$ and $\sigma_-$.
Since the argument by topological
invariance, as explained in Sec.\ 3, is not useful to understand the initial coin state dependence of
localization, Eq.\ (\ref{2-phase.timeaveraged}) is important to clearly
observe localization of quantum walks in numerical calculations and experiments. \\
{\bf Remark 5:} $\zeta(\sigma_+, \sigma_-,a,b,\tilde{\phi}_{12})$ depends on the
relative phase difference $\sigma=(\sigma_+ - \sigma_-)/2$, however, is independent of
each phase parameter $\sigma_{+}$ or $\sigma_{-}$ if $ab=0$. \\
{\bf Remark 6:} We cannot determine the probability distribution only from the time-averaged limit measure, since $\sum_{x\in\mathbb{Z}}\overline{\mu}_{\infty}(x)<1$ holds. 
Appendix A is devoted to the derivation of  Theorem \ref{timeavelimit}.


\subsection{rigorous result $2$: the stationary measure}
In this subsection, we present the second of our rigorous results, the stationary
measure of the complete two-phase QW.\\
\indent
By employing $P_x$ and $Q_x$, the time-evolution
operator $U^{(s)}$ is written in the matrix form:
\[U^{(s)}=\begin{bmatrix}
\ddots&\vdots&\vdots&\vdots&\vdots&\vdots&\iddots\\
\cdots&O&P_{-1}&O&O&O&\cdots\\
\cdots&Q_{-2}&O&P_{0}&O&O&\cdots\\
\cdots&O&Q_{-1}&O&P_{1}&O&\cdots\\
\cdots&O&O&Q_{0}&O&P_{2}&\cdots\\
\cdots&O&O&O&Q_{1}&O&\cdots\\
\iddots&\vdots&\vdots&\vdots&\vdots&\vdots&\ddots
\end{bmatrix}\;\;\;
with\;\;\;O=\begin{bmatrix}0&0\\0&0\end{bmatrix}.\]
Now let us consider the generalized eigenequation
\begin{align}U^{(s)}\Psi=\lambda\Psi,\label{2phase-eigenvalue.prob.}\end{align}
where $\lambda\;(\in\mathbb{C})$ with a restriction $|\lambda|=1$ is the eigenvalue of $U^{(s)}$ and $\Psi$ is the eigenvector defined by
\[\Psi= {}^T\!\left[\cdots,\begin{bmatrix}
\Psi^{L}(-1) \\ 
\Psi^{R}(-1)
\end{bmatrix} ,\begin{bmatrix}
\Psi^{L}(0) \\ 
\Psi^{R}(0)
\end{bmatrix},\begin{bmatrix}
\Psi^{L}(1) \\ 
\Psi^{R}(1)
\end{bmatrix} ,\cdots\right]\in(\mathbb{C}^{2})^{\mathbb{Z}},\]
where $T$ means the transposed operation.
First of all, we give the stationary measure 
of our complete two-phase QW.
The stationary measure at position $x \in {\mathbb Z}$ is defined by
$\mu(x)=|\Psi^{L}(x)|^2 + |\Psi^{R}(x)|^2$\;\cite{watanabe}.
The derivation of Theorem \ref{statmeasure} is based on the {\it
splitted generating function method} (the SGF method)\cite{watanabe},
which is provided in Appendix B.
The solutions of the generalized eigenequation \eqref{2phase-eigenvalue.prob.} are given in Proposition \ref{prob-meas} in Appendix B.
\begin{theorem}
\label{statmeasure}
Let
\begin{align*}
p=1-e^{-4i\sigma}-4e^{-2i\sigma},\quad
q=1+e^{-4i\sigma}-6e^{-2i\sigma},\quad
r^{(\pm)}=1\pm e^{2i\sigma},\end{align*}
and
$c\in\mathbb{R}_{+}$ with $\mathbb{R}_{+}=(0, \infty)$. Now put
\[\theta^{(+)}=\dfrac{r^{(+)}-e^{2i\sigma}\sqrt{q}}{\sqrt{r^{(-)}+e^{2i\sigma}\sqrt{q}} \sqrt{e^{2i\sigma}p+r^{(-)}\sqrt{q}}},
\]
and
\[\theta^{(-)}=\dfrac{r^{(+)}+e^{2i\sigma}\sqrt{q}}{\sqrt{r^{(-)}-e^{2i\sigma}\sqrt{q}} \sqrt{e^{2i\sigma}p-r^{(-)}\sqrt{q}}}.\]
Then, by letting $\lambda^{(j)}\;(j=1,2,3,4)$ be the eigenvalues and $\Psi^{(j)}\;(j=1,2,3,4)$ be the eigenvectors, we obtain the stationary measure depending on the eigenvalues and eigenvectors as follows:
\begin{enumerate}
\item For $\lambda^{(1)}=\sqrt{\dfrac{e^{2i\sigma}\{p+e^{-2i\sigma}r^{(-)}\sqrt{q}\}}{2(-e^{-2i\sigma}r^{(-)}-\sqrt{q})}}$ and $\Psi^{(1)}(0)={}^T\![\alpha,\;\beta]={}^T\!\dfrac{c}{\sqrt{2}}
\left[1,\;\dfrac{e^{-i\sigma_{-}}}{2}(r^{(-)} e^{-2i\sigma}+\sqrt{q})\right]$, and $\lambda^{(2)}=-\lambda^{(1)}$ and $\Psi^{(2)}(0)=\Psi^{(1)}(0)$, we have
\begin{eqnarray*}\mu(x)=\left\{ \begin{array}{ll}
\dfrac{c^{2}}{4}\left\{4(1+\sin^{2}\sigma)+\Re\{(e^{2 i\sigma}-1)\sqrt{1+e^{-4 i\sigma}-6e^{-2 i\sigma}}\}\right\}|\theta^{(+)}|^{2x}& (x\geq0),\\
\\
\dfrac{c^{2}}{4}\left\{4(1+5\sin^{2}\sigma+4\sin^{4}\sigma)+(3+4\sin^{2}\sigma)\Re\{(e^{2 i\sigma}-1)\sqrt{1+e^{-4 i\sigma}-6e^{-2 i\sigma}}\}\right\}|\theta^{(+)}|^{2|x|}& (x\leq-1).
\end{array} \right.\end{eqnarray*}
\item For $\lambda^{(3)}=\sqrt{\dfrac{e^{2i\sigma}\{p-e^{-2i\sigma}r^{(-)}\sqrt{q}\}}{2(-e^{-2i\sigma}r^{(-)}+\sqrt{q})}}$ and $\Psi^{(3)}(0)={}^T\![\alpha,\;\beta]={}^T\!\dfrac{c}{\sqrt{2}}\left[1,\;\dfrac{e^{-i\sigma_{-}}}{2}(r^{(-)} e^{-2i\sigma}-\sqrt{q})\right] $, and $\lambda^{(4)}=-\lambda^{(3)}$ and $\Psi^{(3)}(0)=\Psi^{(4)}(0)$, we have
\begin{eqnarray*}\mu(x)=\left\{ \begin{array}{ll}
\dfrac{c^{2}}{4}\left\{4(1+\sin^{2}\sigma)-\Re\{(e^{2 i\sigma}-1)\sqrt{1+e^{-4 i\sigma}-6e^{-2 i\sigma}}\}\right\}|\theta^{(-)}|^{2x}& (x\geq0),\\
\\
\dfrac{c^{2}}{4}\left\{4(1+5\sin^{2}\sigma+4\sin^{4}\sigma)-(3+4\sin^{2}\sigma)\Re\{(e^{2 i\sigma}-1)\sqrt{1+e^{-4 i\sigma}-6e^{-2 i\sigma}}\}\right\}|\theta^{(-)}|^{2|x|}& (x\leq-1).
\end{array} \right.\end{eqnarray*}
\end{enumerate}
\end{theorem}
We should note that there always exist four eigenvalues $\lambda^{(j)}\, (j=1,2,3,4)$, if we irrespectively take into account the eigenvalues whose eigenvectors diverge or do not diverge when $|x|$ goes to infinity.  
Also, the stationary measure does not have an origin
symmetry in general, however, is attenuated or diverged
exponentially for the position with the same decay or divergence rate
both in $x\geq1$ and $x\leq-1$.  
 We should note that $\sum_{x\in\mathbb{Z}}\mu(x)$ strongly depends on $c(\in\mathbb{R}_+)$, and by choosing appropriate $c$ and $\lambda^{(j)}\;(j=1,2,3,4)$, \[\sum_{x\in\mathbb{Z}}\mu(x)<1\] holds, 
which indicates that there is a possibility that we can investigate localization also by the stationary measure. 

Here we have a conjecture which implies that the stationary measure is symmetric around $x=-1/2$ independent of each phase parameter $\sigma_{+}$ and $\sigma_{-}$. We will mention the {\it $x=-1/2$ symmetry} in Subsections $4.1$ and $4.2$ with specific examples. 

\noindent
{\bf Conjecture}
\begin{enumerate}
\item For $\lambda^{(1)}=\sqrt{\dfrac{e^{2i\sigma}\{p+e^{-2i\sigma}r^{(-)}\sqrt{q}\}}{2(-e^{-2i\sigma}r^{(-)}-\sqrt{q})}}$ and $\Psi^{(1)}(0)={}^T\![\alpha,\;\beta]={}^T\!\dfrac{c}{\sqrt{2}}
\left[1,\;\dfrac{e^{-i\sigma_{-}}}{2}(r^{(-)} e^{-2i\sigma}+\sqrt{q})\right]$, and $\lambda^{(2)}=-\lambda^{(1)}$ and $\Psi^{(2)}(0)=\Psi^{(1)}(0)$, we have
\begin{eqnarray*}&4(1+\sin^{2}\sigma)+\Re\{(e^{2 i\sigma}-1)\sqrt{1+e^{-4i\sigma}-6e^{-2 i\sigma}}\}\\
&=
\{4(1+5\sin^{2}\sigma+4\sin^{4}\sigma)+(3+4\sin^{2}\sigma)\Re\{(e^{2 i\sigma}-1)\sqrt{1+e^{-4 i\sigma}-6e^{-2i\sigma}}\}\}|\theta^{(+)}|^{2}.\end{eqnarray*}
\item For $\lambda^{(3)}=\sqrt{\dfrac{e^{2i\sigma}\{p-e^{-2i\sigma}r^{(-)}\sqrt{q}\}}{2(-e^{-2i\sigma}r^{(-)}+\sqrt{q})}}$ and $\Psi^{(3)}(0)={}^T\![\alpha,\;\beta]={}^T\!\dfrac{c}{\sqrt{2}}\left[1,\;\dfrac{e^{-i\sigma_{-}}}{2}(r^{(-)} e^{-2i\sigma}-\sqrt{q})\right] $, and $\lambda^{(4)}=-\lambda^{(3)}$ and $\Psi^{(3)}(0)=\Psi^{(4)}(0)$, we have
\begin{eqnarray*}&4(1+\sin^{2}\sigma)-\Re\{(e^{2 i\sigma}-1)\sqrt{1+e^{-4 i\sigma}-6e^{-2 i\sigma}}\}\\
&=
\{4(1+5\sin^{2}\sigma+4\sin^{4}\sigma)-(3+4\sin^{2}\sigma)\Re\{(e^{2i\sigma}-1)\sqrt{1+e^{-4 i\sigma}-6e^{-2 i\sigma}}\}\}|\theta^{(-)}|^{2}.\end{eqnarray*}
\end{enumerate}


\subsection{rigorous result $3$: the weak convergence theorem}
Put $C=\sum_{x}\overline{\mu}_{\infty}(x),$
where $\overline{\mu}_{\infty}(x)$ is the time-averaged limit measure at position $x\in\mathbb{Z}$ obtained by Theorem \ref{timeavelimit}.
From now on, we present the weak convergence theorem for the missing part $1-C$ with $0\leq C\leq 1.$ 
In general, the weak convergence theorem describes the ballistic behavior of the QW\cite{konnoweak}. Throughout this subsection, we assume that the walker starts from the origin with the initial coin state $\varphi_{0}={}^T\![\alpha,\beta]$, where $\alpha,\beta\in\mathbb{C}$. 
Put $\alpha=ae^{\phi_{1}},\;\beta=be^{\phi_{2}}$ with $a,b\geq0,\;a^{2}+b^{2}=1$ and $\phi_{1}, \phi_{2}\in\mathbb{R}$. 

\begin{theorem}
\label{weaklimit}
Let $\tilde{\phi}_{12}=\phi_{1}-\phi_{2}\;$.
For the complete two-phase QW, $X_{t}/t$ converges weakly to the random variable $Z$ which has the following density function:
\begin{align}\mu(x)=C\delta_{0}(x)+w(x)f_{K}(x;1/\sqrt{2}),\label{weakmeasure}\end{align}
where 
\[f_{K}(x;1/\sqrt{2})=\dfrac{1}{\pi(1-x^{2})\sqrt{1-2x^{2}}},\]
and
\begin{eqnarray}
w(x)=\left\{ \begin{array}{ll}
\dfrac{t^{(+)}_{2}x^{3}+t^{(+)}_{1}x^{2}}{s_{1}x^{2}+s_{0}}& (x\geq0), \\
\dfrac{t^{(-)}_{2}x^{3}+t^{(-)}_{1}x^{2}}{s_{1}x^{2}+s_{0}} & (x<0), \\
\end{array} \right.
\end{eqnarray}
with
\begin{eqnarray*}
\left\{
\begin{array}{l}
s_{1}=\cos^{2}\sigma,\quad s_{0}=\sin^{2}\sigma,\\
t^{(+)}_{2}=1-2a^2-2ab\cos(\tilde{\phi}_{12}-\sigma_{+}),\quad t^{(-)}_{2}=1-2a^{2}-2ab\cos(\tilde{\phi}_{12}-\sigma_{+}),\\
t^{(+)}_{1}=1+4a^{2}\sin^{2}\sigma-2ab\left\{\cos(\tilde{\phi}_{12}-\sigma_{+})-\cos(\tilde{\phi}_{12}-\sigma_{-})\right\},\quad
t^{(-)}_{1}=1.
\end{array}
\right.
\end{eqnarray*}
\end{theorem}
\noindent
Here $w(x)f_{K}(x;1/\sqrt{2})$ is an absolutely continuous part of $\mu(x)$.
We emphasize that the weak limit measure is generally asymmetric for the
origin, and heavily depends on the phase parameters and initial
state. Furthermore, like the time-averaged limit measure, if $ab=0$, then, the weak limit measure becomes independent of each phase parameter $\sigma_{+}$ or $\sigma_{-}$, in other words, we can write down the weight function $w(x)$ by the relative phase difference $\sigma$ and initial state. For instance, if we start the walk with the mixed state $\varphi_{0}={}^T\![1,0]$ or $\varphi_{0}={}^T\![0,1]$ with probability $1/2$, which is same as the initial state in Grimmett et al.\cite{schudo}, then the same argument holds.
We provide with the proof of Theorem \ref{weaklimit} in Appendix C by using the {\it time-space generating function method}\cite{segawa}.

\section{Topological invariants of the complete two-phase QW}
\label{topological-invariant}
In this section, we investigate the topological invariant 
of the complete two-phase QW.

As we derived in the previous section,
the time-averaged limit measure of the complete two-phase QW
exhibits localization around the origin. 
Recently, there has been a development to understand localization of QWs
as a localized surface state originating to the topological invariant
which is inherited from the time-evolution operator\cite{kitagawa10,kitagawa12,kitagawa12exp}.
In order to relate the topological invariant to the localized states, we
use a fundamental principle, called  the {\it bulk-edge correspondence}. This principle
states that  
the absolute value of difference of topological
numbers in two adjacent spatial regions gives a lower limit of
the number of eigenvalues/eigenvectors exhibiting localization in the vicinity of the interface.
It would be very interesting to directly compare the mathematically derived
stationary and time-averaged limit measures with the prediction of the bulk-edge
correspondence, which motivates us to derive the topological invariants
of the complete two-phase QW.

\subsection{relevant symmetries of the time-evolution operator to
  establish topological phases}

Since each spatial region of the complete two-phase QW should be characterized by own
topological number, we  separately calculate the topological number
on the regions with the phase $\sigma_+$ or $\sigma_-$ in the
coin operator.
First, we calculate the topological number for the region with the phase $\sigma_+$.
Put the coin operator for $x\ge 0$ as
\begin{equation*}
 U_{+} =
\frac{1}{\sqrt{2}}
\begin{bmatrix}
1 & e^{i \sigma_+} \\
e^{-i \sigma_+} & -1
\end{bmatrix}.
\end{equation*}
We assume that the whole of line has the same coin $U_{+}$,
because the calculation of the topological number is simplified for a
system with translation symmetry.
Topological invariant for the region $x
\le -1$ is easily obtained from the result for $x \ge 0$ by replacing $\sigma_+$ with $\sigma_-$.

In the argument of the ordinary topological phase of matter,
it is important to identify symmetries of the system, that is,
time-reversal, particle-hole, and chiral
symmetries\cite{bernevig13,chiu15,hasan10,qi11}.
We call these three symmetries {\it relevant symmetries for
topological phases}.
While these symmetries give constraints on the Hamiltonian, 
now we derive the corresponding constraints on
the time-evolution operator of QWs. 
The detail of the derivation is explained in Appendix D.
In order to make clear the relation to physics, 
we introduce quasi-energy $\varepsilon \in \mathbb{R}$, by the terminology in physics,  which is defined from
the eigenvalue $\lambda$ of the time-evolution operator in Eq.\
(\ref{2phase-eigenvalue.prob.}) as
\begin{equation*}
 \lambda = e^{-i \varepsilon},
\end{equation*}
Note that quasi-energy $\varepsilon$ has $2\pi$ periodicity.
The conditions required from the relevant symmetries for topological
 phases on the time-evolution operator are summarized as
\begin{eqnarray}
 T\, U^{(s)}\, T^{-1} =&
  \left(U^{(s)}\right)^{-1}  &\text{  (Time-reversal symmetry)},
\label{eq:TRS}\\ 
 P\, e^{i \varepsilon_P} U^{(s)}\, P^{-1} =&
 e^{i\varepsilon_P}\cdot  U^{(s)} & \text{  (Particle-hole symmetry)},
\label{eq:PHS}\\ 
\Gamma\,e^{i \varepsilon_\Gamma}  U^{(s)}\, \Gamma^{-1} =&
  \left(e^{i \varepsilon_\Gamma} \cdot U^{(s)}\right)^{-1} &\text{  (Chiral symmetry) }.
\label{eq:chiral}
\end{eqnarray}
Here the symmetry operators $T$ and $P$ are anti-unitary operators
(i.e., they should contain a complex conjugate operator $K$), while
$\Gamma$ is a unitary one.
In Eqs.\ (\ref{eq:PHS}) and (\ref{eq:chiral}), we assume that the
time-evolution operator $U^{(s)}$ satisfies the eigenvalue equation
$e^{i \varepsilon_X} U^{(s)} {\Psi} = e^{i \varepsilon} \Psi$, 
where $X$ stands for $P$ or $\Gamma$, 
with an eigenvector $\Psi$ and $\varepsilon_P, \varepsilon_\Gamma \in \mathbb{R}$. 
If so, Eqs.\ (\ref{eq:PHS}) and (\ref{eq:chiral})
guarantee that 
$e^{i \varepsilon_X} U^{(s)} (X \Psi) = e^{-i \varepsilon} (X \Psi)$.
Therefore, a pair of quasi-energies  with
opposite signs $\pm \varepsilon$ around the symmetric point
$\varepsilon_P$ and/or $\varepsilon_\Gamma$ appears.
Generally, $\varepsilon_P$ and $\varepsilon_\Gamma$ are set to be zero, however, let us
consider arbitrary values,
because the recent work on the topological phase of Hadamard walks \cite{obuse15}
pointed out its importance.

As explained in Ref.\ \cite{asboth13}, in order to clarify symmetries of the time-evolution operator, we should
redefine the time-evolution operator by shifting the origin of time to fit in the {\it
symmetry time frame}. 
For our QW, this corresponds to the situation that a half of the first
coin operation is absorbed into the initial state. 
Then, the redefined time-evolution operator for the one-time step is expressed as

\begin{eqnarray*}
 U^{(s)}_{+} = (\mathbb{I}_\text{p}\otimes(U_{+})^{1/2})\, S\, (\mathbb{I}_\text{p}\otimes(U_{+})^{1/2}),
\end{eqnarray*}
where $\mathbb{I}_\text{p} = \sum_{x\in \mathbb{Z}}|x\rangle \langle x|$.
As explained in Appendix E, by employing $U^{(s)}_{+}$ and
setting proper symmetric point of quasi-energies, 
we identify the symmetry operators satisfying 
particle-hole and chiral symmetries in Eqs.\ (\ref{eq:PHS}) and
(\ref{eq:chiral}), respectively:
\begin{align}
 P&= \sum_x \ket{x}\bra{x}\otimes\left(V_{\sigma^\prime} \cdot
				     \tau_0\, K
				     \cdot V_{\sigma^\prime}^{-1}\right)\quad \text{with }\varepsilon_P=0,\label{eq:PHS op.}\\
\Gamma &= \sum_x  \ket{x}\bra{x}\otimes \left( V_{\sigma^\prime} \cdot
					  \tau_1 \cdot
					  V_{\sigma^\prime}^{-1}\right)\quad \text{with }\varepsilon_\Gamma=-\pi/2
\label{eq:chiral op.},\end{align}
where
\begin{eqnarray}
 V_{\sigma^\prime}=
\begin{bmatrix}
e^{i\sigma^\prime/2}   & 0\\
0 & e^{-i\sigma^\prime/2}
 \end{bmatrix}, \quad \sigma^\prime=\sigma_+.
\label{eq:V}
\end{eqnarray}
A two-dimensional identity matrix $\tau_0$ and Pauli matrices
\begin{eqnarray*}
 \tau_1 = \begin{bmatrix}
0	   & 1\\
1 & 0
\end{bmatrix},\quad
 \tau_2 = \begin{bmatrix}
0	   & -i\\
i & 0
\end{bmatrix},\quad
 \tau_3 = \begin{bmatrix}
1	   & 0\\
0 & -1
\end{bmatrix},
\end{eqnarray*}
act on the coin space.
We note that, while $\sigma^\prime=\sigma_+ + m\pi$ $(m \in
\mathbb{Z})$ is a more general expression of $\sigma^\prime$ in Eq.\ (\ref{eq:V}), we fix $m=0$ for simplicity.

While both chiral and particle-hole symmetries of the
 time-evolution operator can be identified, 
the values of $\varepsilon_P$ and $\varepsilon_\Gamma$ in
 Eqs.\ (\ref{eq:PHS op.}) and (\ref{eq:chiral op.}) are
 different. 
This means that the time-evolution operator does not possess both chiral and particle-hole
 symmetries at the same symmetric point of quasi-energy.
If two of the three relevant symmetries for topological phases are established, the other symmetry is
 automatically confirmed by combining the two identified symmetry operators. 
In our case, either chiral or particle-hole symmetries is established
 at a specific quasi-energy. Thereby, time-reversal
 symmetry cannot be retained.

For further arguments, we need to choose the symmetric point of
quasi-energy, that is, $\varepsilon_P=0$ or $\varepsilon_\Gamma=-\pi/2$.
The proper symmetric point of quasi-energy to study
topological invariant can be chosen from the distribution of the
eigenvalue of $U_+^{(s)}$,
because the topological numbers can be well defined unless
absolutely continuous spectra exist at the symmetric points of
quasi-energy, i.e., the quasi-energy gap should open by the terminology in physics.
In Appendix F, the eigenvalue problem of homogeneous $U_+^{(s)}$ is
solved
by the Fourier analysis.
We obtain the eigenvalue the time evolution operator in the Fourier space
\begin{align}
\lambda &=e^{i[\pm \varepsilon(k)+\pi/2]}=i[\cos\{\varepsilon(k)\} \pm
 i\sin\{\varepsilon(k)\}], \nonumber \\
&\cos[\varepsilon(k)] = \sqrt{1-\dfrac{1}{2}\cos^2(k)}, \label{eq:bulk spectrum}
\end{align}
where $k \in [0:2\pi)$ is the wave number.
We see that 
the absolutely continuous spectra include quasi-energy
$\varepsilon=0,\pi$ ($\lambda=\pm 1$), however, do not exist at
$\varepsilon=\pm\pi/2$ ($\lambda=\pm i$).
Therefore, we focus on the time-evolution operator possessing chiral
symmetry, where the
topological numbers are well defined due to the presence of the
quasi-energy gap around $\varepsilon=\pm \pi/2$.
Then, the time-evolution
 operator $U_+^{(s)}$, we focus on hereafter in this section, possesses only chiral symmetry, and then
 belongs to the chiral unitary class (class AIII in the Cartan classification).
In the table of the classification of topological phases\cite{schnyder08}, it is known
 that class AIII can possess ${\mathbb Z}$ topological phases in the one
 dimensional system.

\subsection{topological invariants and the bulk-edge correspondence}

Having identified that the system possesses chiral symmetry,
topological invariants for the above time-evolution operator are given
by calculating winding numbers\cite{asboth13}.  Because of $2 \pi$
periodicity of  quasi-energy, we remark that the QW with chiral symmetry or particle-hole symmetries possesses
two symmetric points  of the quasi-energy at $\varepsilon_\Gamma=-\pi/2$ as well as at $\varepsilon_\Gamma+\pi=\pi/2$.
Thus, we need to introduce two topological numbers.
For the spatial region $x\ge0$, topological numbers
$\nu_{\pm \pi/2}^{+}$ at quasi-energies $\varepsilon=\pm \pi/2$
$(\lambda=\mp i)$
are derived as
\begin{eqnarray}
 (\nu_{-\pi/2}^{+}, \nu_{+\pi/2}^+) = 
(1,0).
\label{eq:topological invariant +}
\end{eqnarray}
The detail of the calculation is presented in Appendix F.

Topological invariants for the spatial region $x\le -1$ of the
complete two-phase QW are obtained by applying Eq.\
(\ref{eq:topological invariant +}) with a replacement $\sigma_+$ with
$\sigma_-$.
However, 
we should keep using the same chiral symmetry operator $\Gamma$ 
for the whole of the system, otherwise chiral
symmetry is broken. In other words, the same $\sigma^\prime$ in
$V_{\sigma^\prime}$ has to be applied for 
$x\ge0$ and $x\le -1$. This restricts 
the phase $\sigma_-$ so as to  satisfy
\begin{equation}
 \sigma_-=\sigma_+ + n\pi \quad(n \in {\mathbb Z}),
\label{eq:constraint sigma_-}
\end{equation}
to maintain chiral symmetry.
According to Appendix F, topological numbers $\nu_{\pm \pi/2}^-$ at quasi-energies
$\varepsilon=\pm \pi/2$ for the spatial region with $x\le -1$ are
summarized as
\begin{eqnarray}
 (\nu_{-\pi/2}^-, \nu_{+\pi/2}^-) = \left\{
\begin{array}{ll}
(1,0) & \quad n : \text{ even},\\
(0,1) & \quad n : \text{ odd}.
\end{array}
\right.
\label{eq:topological invariant -}
\end{eqnarray}

As we mentioned at the beginning of this section, the bulk-edge
correspondence predicts a minimum number of localized states
from the absolute value of the difference of topological numbers of
adjacent two regions. Thereby,
on one hand, when $n$ is even, the two regions have the same topological numbers
$(\nu_{-\pi/2}^{\pm}, \nu_{+\pi/2}^{\pm})=(1,0)$, and then 
localized eigenstates are not expected from
the bulk-edge correspondence. 
On the other hand, when $n$ is odd, the two regions have
different topological numbers whose difference is $|\nu_{\pm \pi/2}^+ - \nu_{\pm
\pi/2}^-|=1$, and then, two localized eigenstates, one at 
$\varepsilon=-\pi/2$ ($\lambda=i$) and the other at $\varepsilon=\pi/2$
($\lambda=-i$),  in the vicinity of the
origin are predicted by the bulk-edge correspondence.

If $n$ in Eq.\ (\ref{eq:constraint sigma_-}) is not an
integer, chiral symmetry of the two-phase
QW is broken. 
In this case, localized states at $\varepsilon=\pm \pi/2$ originating to
the topological invariant are not expected from the bulk-edge
correspondence.
While this does not exclude a possibility having
localized states at $\varepsilon=\pm \pi/2$ since the bulk-edge
correspondence gives a lower bound of the number of edge states.
However, most generally, quasi-energy of such localized states deviates from $\varepsilon = \pm \pi/2$.

\subsection{robustness of localized states by topological protections}

One of the important properties of localized states originating to
the topological invariant is robustness against perturbations satisfying
the following conditions; 
\begin{enumerate}\item the perturbation should 
preserve the same relevant symmetries for topological phases with the non-perturbative time-evolution
operator.
\item the perturbation remains quasi-energy gaps open.
\end{enumerate}\noindent
In case of one dimension, this robustness can be expressed more clearly
such that the eigenvalue of
the localized state remains unchanged and the eigenvector
remains localized under the perturbations. Therefore, this property can be
utilized to verify whether localized states at $\varepsilon=\pm \pi/2$
originate to the topological invariant  or not.

Now we consider the perturbations for the complete two-phase QW. 
We assume that only the coin operator introduces the perturbation. 
Since the complete two-phase QW possesses chiral symmetry,
we need to show that the perturbative coin operator should also satisfy
the condition demanded from chiral symmetry.
Taking account of Eqs. (\ref{eq:chiral})
and (\ref{eq:chiral op.}),  the coin operator $U_p$, as a
source of perturbation, should satisfy the following relation:
\begin{equation}
\Gamma (\mathbb{I}_\text{p} \otimes U_p \tau_3) \Gamma^{-1} = (\mathbb{I}_\text{p}\otimes U_p \tau_3)^{-1}.
\label{eq:condition perturbation}
\end{equation}
We identify that the coin operator $U_p$ with the position
dependent parameters $\theta_x, \omega_x \in [0,2\pi)$ 
satisfies Eq.\ (\ref{eq:condition perturbation}):
\begin{eqnarray}
 U_p(\theta_x,\omega_x) = 
\begin{bmatrix}
e^{i\omega_x}\cos (\theta_x) & e^{i \sigma_p} \sin (\theta_x)  \\
e^{-i \sigma_p} \sin (\theta_x)  & -e^{-i\omega_x}\cos (\theta_x)
\end{bmatrix},\quad \sigma_p=\sigma_+ \text{ or }\sigma_-. 
\label{eq:perturbative coin}
\end{eqnarray}
We note that, when $\theta_x=\pi/4$, $\omega_x=0$, and $\sigma_p=\sigma_+$ or
$\sigma_-$, $U_p$ is identical with the coin
operator of
the complete two-phase QW.
We examine robustness of localized states by employing the 
perturbative coin operator $U_p$ with random $\theta_x$
in the next section.


\section{Relation between localization and topological invariants}
\label{examples}
In this section, we show several examples of the complete two-phase QW
with specific phases to see what our 
analytical results suggest.
Especially, we argue the relation between localization
and topological invariants of the complete two-phase QW. 
In addition, we investigate the relation to the two-phase QW with one defect.
\par\indent
\par\noindent
\subsection{in case of the complete two-phase QW with chiral symmetry}
Next we treat the complete two-phase QW whose quantum coin is given by 

\begin{align}
U_{x}=\left\{
U_{+}=\dfrac{1}{\sqrt{2}}\begin{bmatrix}1 & -i\\ i & -1 \end{bmatrix}_{x=0,1,2,\cdots},\;
U_{-}=\dfrac{1}{\sqrt{2}}\begin{bmatrix}1 & i\\ -i& -1 \end{bmatrix}_{x=-1,-2,\cdots}
\right\},
\label{eq:U Ex2}
\end{align}
\par\noindent
We obtain the QW by putting $\sigma_{+}=3\pi/2$ and $\sigma_{-}=\pi/2$ in Eq.\ \eqref{2phase_def}.
Because of $\sigma_+ - \sigma_- = \pi$, the two-phase QW retains
chiral symmetry associated with the symmetric point $\varepsilon_\Gamma=-\pi/2$.
Taking account of Eqs.\
(\ref{eq:topological invariant +}) and (\ref{eq:topological invariant -}) and  applying the bulk-edge
correspondence,  non-degenerate localized eigenstates with eigenvalues
$\lambda = \pm i$ are predicted. \\
\indent
Here we consider the stationary measure. By inputting $\sigma_{+}=3\pi/2$ and $\sigma_{-}=\pi/2$ into Theorem \ref{statmeasure}, we obtain the doubly degenerated eigenvalues $\lambda=\pm i$. Also from Theorem \ref{statmeasure}, we have the stationary measure for $\lambda=\pm i$,
\begin{align}
\mu(x) = 
\left\{
\begin{array}{ll}
c^2 (2+\sqrt{2}) \left(\dfrac{1}{3+2\sqrt{2}}\right)^x &\quad (x \ge
0), \\ \\
c^2 (2+\sqrt{2}) \left(\dfrac{1}{3+2\sqrt{2}}\right)^{|x|-1}&\quad (x \le -1), \\
\end{array}
\right.
\label{eq:stationary measure with symmetry}
\end{align}
and 
\begin{align}
\mu(x) = 
\left\{
\begin{array}{ll}
c^2 (2-\sqrt{2}) \left(\dfrac{1}{3-2\sqrt{2}}\right)^x &\quad (x \ge
0), \\ \\
c^2 (2-\sqrt{2})\left(\dfrac{1}{3-2\sqrt{2}}\right)^{|x|-1} &\quad (x \le -1). \\
\end{array}
\right.\label{eq:mu(x)3,4 chiral case}
\end{align}

Thereby, values of the eigenvalues by
Theorem \ref{statmeasure} are consistent with the prediction by the
bulk-edge correspondence, while they are doubly degenerated. 
We clearly see 
that both the stationary distributions with chiral symmetry are symmetric at
$x=-1/2$.
However, the decay rate of the stationary measure corresponding to the
degenerated eigenvalues are different.   
From Eqs.\ (\ref{eq:stationary measure with symmetry}) and
(\ref{eq:mu(x)3,4 chiral case}), 
we find that, 
the stationary measure for one of the degenerated pair $\lambda=\pm i$
shows the
exponential decay 
with the rate $1/(3+2\sqrt{2})$ 
from the origin where the topological number varies, while the other
one grows exponentially
with the rate $1/(3-2\sqrt{2})$ which is larger than one. The latter one cannot be regarded as a
localized state in a view point of the bulk-edge correspondence since
the measure near the origin is rather small.
Furthermore, from a physical viewpoint, a stationary measure
exhibiting divergences of
the measure for the point at infinity is generally considered
as an unphysical solution because of the contradiction to the
normalization condition.
Hence, we can discard one of the degenerated states, which exhibits the
divergent solution.
Accordingly, we obtain the
non-degenerated two eigenstates for $\lambda=\pm i$ and confirm the
complete agreement with the prediction of bulk-edge correspondence.\\
\indent
Next, we focus on the time-averaged limit measure.
Let the initial coin state be $\varphi_{0}={}^T\![1,0]$.
According to Theorem \ref{timeavelimit}, we obtain the time-averaged limit measure by
\begin{eqnarray}\overline{\mu}_{\infty}(x)
&=\left\{ \begin{array}{ll}
\dfrac{2+\sqrt{2}}{2(3+2\sqrt{2})^{x+2}} & (x\geq0), \\
\\
\dfrac{2+\sqrt{2}}{2(3+2\sqrt{2})^{|x|+1}} & (x\leq-1), \\
\end{array} \right.\label{eq:time-ave measure with symmetry}
\end{eqnarray}
and as a result, the coefficient of the delta function is given by
 \begin{align}C=\sum_{x}\overline{\mu}_{\infty}(x)=\sum_{j=0}^{\infty}\dfrac{2+\sqrt{2}}{2(3+2\sqrt{2})^{j+2}}
  +\sum_{k=1}^{\infty}\dfrac{2+\sqrt{2}}{2(3+2\sqrt{2})^{k+1}}=0.12132...>0.
\end{align}
We remark that the time-averaged limit measure in Eq.\ (\ref{eq:time-ave
measure with symmetry}) is identical
with the stationary measure in Eq.\ (\ref{eq:stationary measure with
symmetry}) by inputting $c^2=1/[2(3+2\sqrt{2})^2]$.
Even from this point of view, the stationary
measure exhibiting the divergence at infinity is excluded. 

Owing to Theorem \ref{weaklimit}, we have the weight function of the probability distribution by
\[w(x)=\left\{ \begin{array}{ll}
x^{2}(5-x) & (x\geq0), \\
\\
x^{2}(1-x) & (x<0). \\
\end{array} \right.\]
\noindent
Hence we see
\begin{align}\int_{-\frac{1}{\sqrt{2}}}^{\frac{1}{\sqrt{2}}}w(x)f_{K}(x;1/\sqrt{2})
dx =0.87868...,\end{align}
which leads to 
\[C+\int_{-\frac{1}{\sqrt{2}}}^{\frac{1}{\sqrt{2}}}w(x)f_{K}(x;1/\sqrt{2}) dx=1.\]
\indent
Here we present the numerical results of the time-averaged limit measure,
and we see that the numerical results gradually near to our analytical
result at a very low speed (Fig. 1). Next we show the numerical results
of the probability distribution at time $10000$ in re-scaled space
$(x/10000, 10000P_{10000}(x))$, as well as $w(x)f_{K}(x;1/\sqrt{2})$,
which is an absolutely continuous part of the weak limit measure
$\mu(dx)$, in Fig.\ 2. We see that the curve representing
$w(x)f_{K}(x;1/\sqrt{2})$ seems to be on the middle of the probability distribution on each position, which suggests that our analytical result is mathematically correct. 
We should note that the weak limit measure represents the asymmetry of
the probability distribution (Fig. 2). We emphasize that the
walker is trapped near the origin in a short time, and the distribution
form does not change after many steps.\\
\indent
We also numerically calculate the eigenvalues of the two-phase QW with the
coin operator in Eq.\ (\ref{eq:U Ex2}) on the path
in the range of  $-N \le x \le N-1$ with $N=100$ (see Appendix G for details).
As shown in Fig.\ \ref{fig:spectrum2},
 we confirm that most eigenvalues are densely
distributed on the continuous spectrum in Eq.\ (\ref{eq:bulk spectrum}).
Most importantly, there exist non-degenerated two eigenvalues
at $\lambda=\pm i$.
We also confirm that the numerically calculated stationary measure
with the eigenvalue $\lambda=i$
almost overlaps with 
the stationary measure in Eq.\ (\ref{eq:stationary measure with symmetry}),
as shown in Fig.\ 4.
As we mentioned, the localized state originating to the topological
invariant should be robust against perturbations of the system. 
We check this by using the perturbative coin operator $U_p$ in Eq.\
(\ref{eq:perturbative coin}). We consider that $\theta_x$ of $U_p$
consists of two parts:
\[
 \theta_x=\theta_0+\delta \theta_x,
\]
which are the constant $\theta_0=\pi/4$ and independent and identically
distributed random variable $\delta \theta_x$ in the range of
$[-\pi/4 : \pi/4]$. Note that we choose the rather narrow range for
$\delta \theta_x$ so that the quasi-energy gaps remain open.
We choose $\sigma_p = \sigma_-=\pi/2$ for $x \le -1$ and $\sigma_p=\sigma_+=3\pi/2$
for $x \ge 0$, and $\omega_x=0$ for all $x$.
Figure 5 shows eigenvalues of the two-phase QW with the above
perturbative coin on the path. We clearly see that two eigenvalues
corresponding to localization at $\lambda=\pm i$ remain unchanged under the
perturbation, while the others do not. 
Taking into account these results, we confirm the validity of the
bulk-edge correspond for the complete two-phase QW with chiral symmetry.\\
\indent Finally, we show the initial state dependence of localization by
focusing on $\zeta(\sigma_+, \sigma_-, a, b,\tilde{\phi}_{12})$ in Eq.\
(\ref{2-phase.timeaveraged}). 
According to Eqs.\ (\ref{eq:local-minimum1}) and (\ref{eq:local-minimum2}), we find
that
$\zeta$ becomes a local maximum whose value is $\zeta=4$  at $a=\cos(3\pi/8)$ and
$\tilde{\phi}_{12}=3\pi/2$, while
the value of a local minimum of $\zeta$ becomes zero at
$a=\cos(\pi/8)$ and $\tilde{\phi}_{12}=\pi/2$.
As marked by crosses in Fig.\ \ref{fig:zeta_w-chiral} 
showing
$a$ and $\tilde{\phi}_{12}$ dependences of $\zeta$ by putting
$\sigma_+=3\pi/2, \sigma_-=\pi/2$, the local maximum (minimum)
corresponds to the global maximum (minimum).
In the former case, similar to Fig.\ 1, numerical result of the time-averaged limit
probability upto finite times as shown in Fig.\ 7 agrees well with the time-averaged limit
measure which exhibits localization.   
However, in the latter case,  the time-averaged limit
measure  becomes zero because of $\zeta=0$.
Thereby, numerical results of the time-averaged limit
probability in Fig.\ 8 do not shows a peak around $x=-1/2$. Moreover,
the time-averaged limit probability even near $x=-1/2$ decreases as
increasing time and is inversely proportional to time.  Therefore, we consider that the time-averaged limit
probability becomes zero at $t\rightarrow 0$, which agrees with $\overline{\mu}_\infty(x)=0$.
We remind that even in this case the stationary measure shows
localization as shown in Fig.\ 4, which is consistent with the prediction by the
bulk-edge correspondence.
Therefore, $\zeta(\sigma_+,\sigma_-,a,b,\tilde{\phi}_{12})$ provides an
important information to find the optimal condition of the initial coin
state to observe localization in numerical calculations and even in experiments.

\begin{figure}[t]
\begin{minipage}{0.47\hsize}
\centerline{\includegraphics[width=7.1cm]{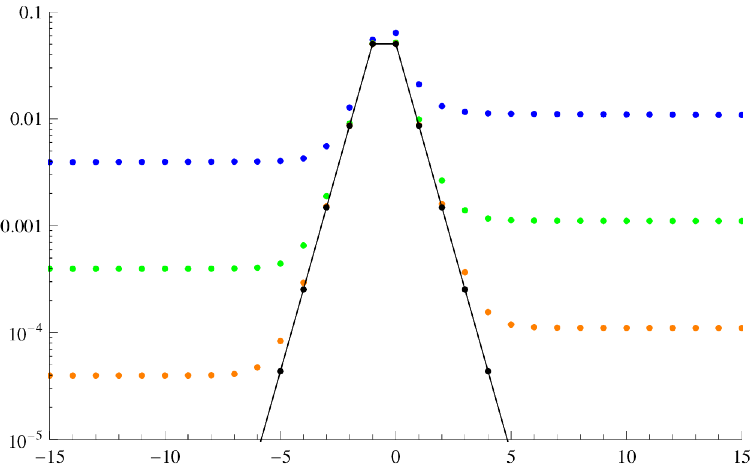}}
\vspace*{13pt}
\caption{\label{fig3}$(\sigma_{+}=3\pi/2,\sigma_{-}=\pi/2,\; {}^T\![\alpha, \beta]= {}^T\![1, 0])$
 The time averaged limit measure (black line) and time-averaged
  probabilities upto time 100 (blue dots), 1000 (green dots), and 10000
  (orange dots) with the initial coin state ${}^T\![\alpha, \beta]= {}^T\![1, 0]$.}
\end{minipage}
 \hfill
 \begin{minipage}{0.47\hsize}
 \centerline{\includegraphics[width=7.1cm]{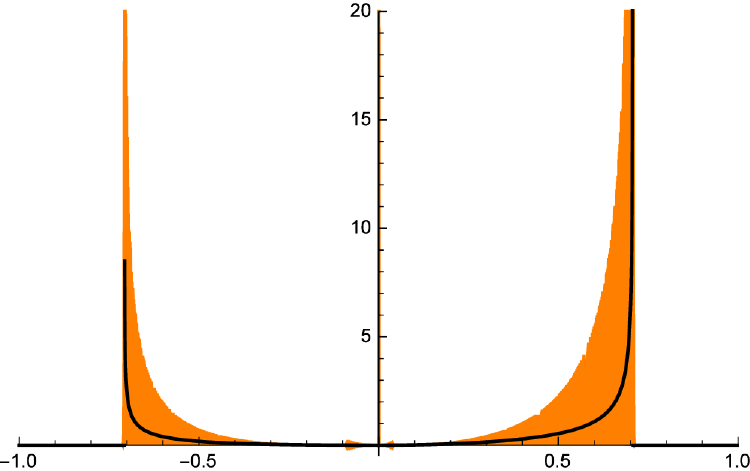}} 
 \vspace*{13pt}
 \caption{\label{fig4}$(\sigma_{+}=3\pi/2, \sigma_{-}=\pi/2,\;
  {}^T\![\alpha, \beta]= {}^T\![1, 0])$ Orange curve: Probability distribution in a re-scaled space $(x/10000, 10000P_{10000}(x))$ at time $10000$, Black curve: $w(x)f_{K}(x; 1/\sqrt{2})$.}
 \end{minipage}
 \end{figure}
 \begin{figure}[h]
  \begin{minipage}{0.47\hsize}
    \centerline{\includegraphics[width=6.1cm]{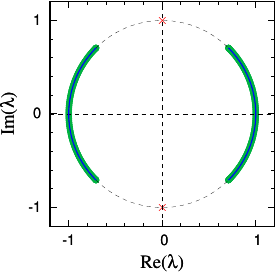}}
 \vspace*{13pt}
 \caption{\label{fig:spectrum2}$(\sigma_{+}=3\pi/2, \sigma_{-}=\pi/2)$
 Green dense dots and red crosses : Numerically calculated
 eigenvalues $\lambda$ of the time-evolution operator of the complete
 two-phase QW with chiral symmetry on the path.
 Eigenvalues corresponding to the exponentially localized
  eigenvector are  highlighted by red crosses.
 Blue curves : The eigenvalues in Eq.\ (\ref{eq:bulk spectrum}).
 The dashed curve represents a unit circle on the
 complex plane.}
  \end{minipage}
  \hfill
 \begin{minipage}{0.47\hsize}
    \centerline{\includegraphics[width=7.6cm]{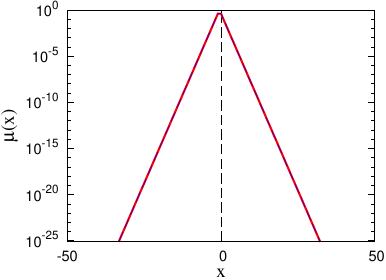}} 
 \vspace*{13pt}
 \caption{\label{fig:edestate2}$(\sigma_{+}=3\pi/2, \sigma_{-}=\pi/2)$
 Red solid line : Probability distribution of the numerically calculated eigenvector
  with the eigenvalue $\lambda=i$.
 Blue dashed line : The stationary measure in Eq.\
  (\ref{eq:stationary measure with symmetry}) with
 the normalization constant $c^2=1/(4+3\sqrt{2})$.} 
 \end{minipage}
 \end{figure}

 \begin{figure}[h]
  \begin{minipage}{0.47\hsize}
    \centerline{\includegraphics[width=6.1cm]{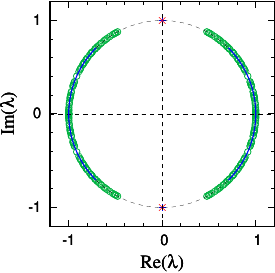}}
\vspace*{13pt}
\caption{\label{fig:spectrum random}$(\sigma_{+}=3\pi/2,
 \sigma_{-}=\pi/2)$
Green open dots and red crosses: Numerically calculated
eigenvalue $\lambda$ of the  time-evolution operator of the complete
two-phase QW with the perturbative coin on the path. 
Blue curves : The eigenvalues in Eq.\ (\ref{eq:bulk spectrum}).}
\end{minipage}
\hfill
\begin{minipage}{0.47\hsize}
\centerline{\includegraphics[width=6.1cm]{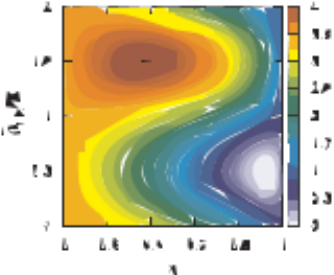}}
\vspace*{13pt}
\caption{\label{fig:zeta_w-chiral}
\label{fig:spectrum random}$(\sigma_{+}=3\pi/2,
 \sigma_{-}=\pi/2)$
$a$ and $\tilde{\phi}_{12}$ dependences of
 $\zeta(\sigma_+, \sigma_-,a,b,\tilde{\phi}_{12})$ in Eq.\
 (\ref{2-phase.timeaveraged}). The red and black crosses represent the local
 minimum and maximum of $\zeta$, respectively.}
\end{minipage}
\end{figure} 
\begin{figure}[h]
\begin{minipage}{0.47\hsize}
\centerline{\includegraphics[width=6.1cm]{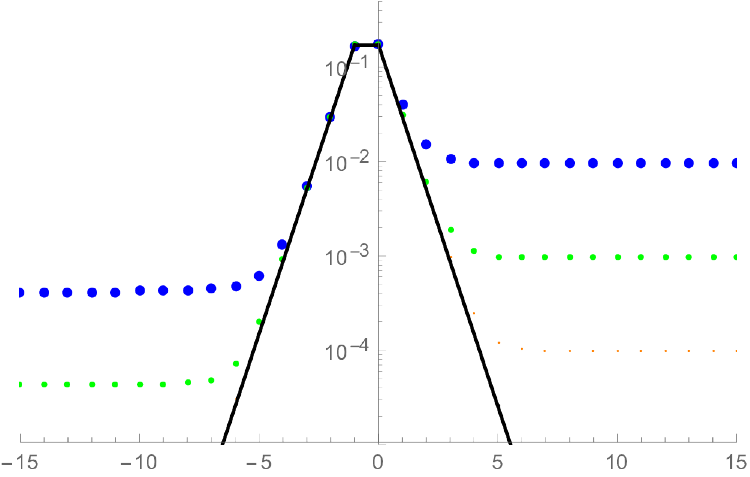}}
\vspace*{13pt}
\caption{
\label{fig:spectrum random}$(\sigma_{+}=3\pi/2,
 \sigma_{-}=\pi/2,\; {}^T\![\alpha, \beta]= {}^T\![-i\cos(3\pi/8), \sin(3\pi/8)])$
 The time averaged limit measure (black line) and time-averaged
 probabilities upto time 100 (blue dots), 1000 (green dots), and 10000
 (orange dots) with the initial coin state ${}^T\![\alpha, \beta]= {}^T\![-i\cos(3\pi/8), \sin(3\pi/8)]$.}
\end{minipage}
 \hfill
 \begin{minipage}{0.47\hsize}
\centerline{\includegraphics[width=6.1cm]{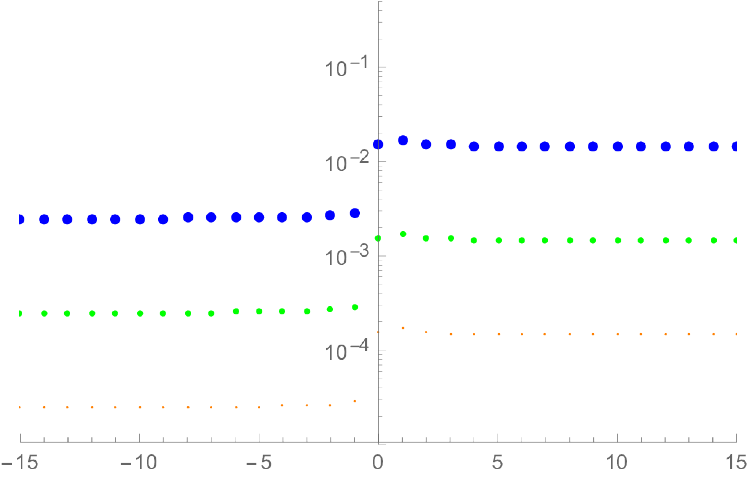}}
\vspace*{13pt}
\caption{$(\sigma_{+}=3\pi/2, \sigma_{-}=\pi/2,\;
 {}^T\![\alpha, \beta]= {}^T\![i\cos(\pi/8), \sin(\pi/8)])$
The time-averaged 
 probabilities upto time 100 (blue dots), 1000 (green dots), and 10000
 (orange dots) with the initial coin state ${}^T\![\alpha, \beta]=
 {}^T\![i\cos(\pi/8), \sin(\pi/8)]$.
 Note that the time averaged limit measure is not shown because it is zero in this case.}
 \end{minipage}
\end{figure}

\par\indent
\par\noindent
\subsection{in case of the complete two-phase QW without the relevant
  symmetries for topological phases}

Here we focus on the complete two-phase QW whose quantum coin is expressed by
 \begin{align}
U_{x}=\left\{ \begin{array}{ll}
U_{+}=\dfrac{1}{\sqrt{2}}\begin{bmatrix}1 & -1\\ -1& -1 \end{bmatrix}\quad (x=0,1,2,\cdots) \\
\\
U_{-}=\dfrac{1}{\sqrt{2}}\begin{bmatrix}1 & i\\ -i & -1 \end{bmatrix} \quad (x=-1,-2,\cdots)
\end{array} \right..
\label{eq:U Ex3}
\end{align}
\par\noindent
We obtain the QW by putting $\sigma_{+}=\pi$ and $\sigma_{-}=\pi/2$ in Eq. \eqref{2phase_def}.
Because of $\sigma_+ - \sigma_- = \pi/2 \ne n \pi$ $(n \in \mathbb{Z})$, the two-phase QW  does not
retain any relevant symmetries for topological phases, and 
localized states with eigenvalues $\lambda=\pm i$ are not expected from the
bulk-edge correspondence.\\
\indent
At first, we consider the stationary measure. 
By inputting $\sigma_{+}=\pi$ and $\sigma_{-}=\pi/2$ into Theorem \ref{statmeasure}, we obtain the eigenvalues
\[
\lambda=\pm\sqrt{-\frac{1}{2}-\frac{\sqrt{3}}{2}i},\quad\pm\sqrt{-\frac{1}{2}+\frac{\sqrt{3}}{2}i}.
\]
We also obtain the stationary measure for $\lambda=\pm\sqrt{-\frac{1}{2}-\frac{\sqrt{3}}{2}i}$ 
\begin{align}
\mu(x)
& =
\left\{
\begin{array}{ll}
\dfrac{c^2}{2} (3-\sqrt{3}) \left(\dfrac{1}{2-\sqrt{3}}\right)^x &\quad (x \ge
 0),\\ \\
\dfrac{c^2}{2} (3-\sqrt{3}) \left(\dfrac{1}{2-\sqrt{3}}\right)^{|x|-1} &\quad (x \le -1), \\
\end{array}
\right.
\label{eq:mu(x)1,2 non-chiral case1}
\end{align}
and that for $\lambda=\pm\sqrt{-\frac{1}{2}+\frac{\sqrt{3}}{2}i}$ by
\begin{eqnarray}
\mu(x) 
&=
\left\{
\begin{array}{ll}
\dfrac{c^2}{2} (3+\sqrt{3}) \left(\dfrac{1}{2+\sqrt{3}}\right)^x &\quad (x \ge
 0), \\ \\
\dfrac{c^2}{2} (3+\sqrt{3}) \left(\dfrac{1}{2+\sqrt{3}}\right)^{|x|-1} &\quad (x \le -1). \\
\end{array}
\right.\label{eq:mu(x)3,4 non-chiral case1}
\end{eqnarray}

The fact that the all eigenvalues derived from Theorem \ref{statmeasure} deviate from $\lambda=\pm i$ is
consistent with the result by the bulk-edge correspondence.
The stationary measure in Eq.\ (\ref{eq:mu(x)1,2 non-chiral case1})
exponentially diverges as $|x|$ approaches to infinity, while 
that in Eq.\ (\ref{eq:mu(x)3,4 non-chiral case1}) shows the exponential
decay from the origin, thus, localization. However, this localization
cannot be related to the topological invariant, and then the novel topological protection under
perturbations is never expected. 
By the way, we see from Eq.\ (\ref{eq:mu(x)3,4 non-chiral case1}) that the stationary distribution without the relevant symmetry has also symmetry around $x=-1/2$, which is identical with that of the stationary measure with chiral symmetry.\\
\indent
Then, we focus on the time-averaged limit measure.
Let the initial coin state be $\varphi_{0}={}^T\![1,0]$.
Theorem \ref{timeavelimit} gives the time-averaged limit measure of the complete two-phase QW without the relevant symmetries by

\begin{align}\overline{\mu}_{\infty}(x)
&=\left\{ \begin{array}{ll}
\dfrac{(3+\sqrt{3})}{6 (2+\sqrt{3})^{x+2}} & (x\geq0),\\
\\
\dfrac{(3+\sqrt{3})}{6 (2+\sqrt{3})^{|x|+1}} & (x\leq-1).
\end{array} \right.\label{eq:time-ave measure without symmetry}
\end{align}
We remark that the above time-averaged limit measure agrees
with the stationary measure in Eq.\ (\ref{eq:mu(x)3,4 non-chiral case1})
by putting 
\[c^{2}=\dfrac{1}{3(2+\sqrt{3})^2}.
\]
From this point of view, the stationary
measure exhibiting the divergence at infinity is excluded.
\noindent
Here we get the coefficient of the delta function by
\begin{align*}C=\sum_{x}\overline{\mu}_{\infty}(x)=0.154701...>0.
\end{align*}

According to Theorem \ref{weaklimit}, the weight function is given by
\[w(x)=\left\{ \begin{array}{ll}
\dfrac{2x^{2}(3-x)}{x^{2}+1} & (x\geq0), \\
\\
\dfrac{2x^{2}(1-x)}{x^{2}+1} & (x<0). \\
\end{array} \right.\]
\noindent
Hence we see
\begin{align}\int_{-\frac{1}{\sqrt{2}}}^{\frac{1}{\sqrt{2}}}w(x)f_{K}(x;1/\sqrt{2})
dx=0.845299.... 
\end{align} 
Therefore, we have
\[C+\int_{-\frac{1}{\sqrt{2}}}^{\frac{1}{\sqrt{2}}}w(x)f_{K}(x;1/\sqrt{2}) dx=1.\]

Here we present the numerical results of the time-averaged limit
measure in Fig. 9 and  the probability distribution at time $10000$ in
re-scaled space $(x/10000, 10000P_{10000}(x))$ in Fig.\ 10
when the initial state is $[1,0]$.
The observed behaviors are consistent with those in the previous
subsection for the complete two-phase QW with chiral symmetry.
Similar to the previous case, $\zeta$ in Eq.\ (\ref{2-phase.timeaveraged}) strongly depends on the initial
states as shown in Fig. \ref{fig:zeta_wo-chiral}.

Again, we numerically calculate the eigenvalue of the two-phase QW with the
coin operator in Eq.\ (\ref{eq:U Ex3}) on the path in the range of  $-N \le x \le N-1$ with $N=100$.
As shown in Fig.\ \ref{fig:spectrum3},%
 we confirm that densely distributed eigenvalues are 
consistent with Eq.\ (\ref{eq:bulk spectrum}) and there exist two isolated eigenvalues
at $\lambda \ne \pm i$. 
These isolated eigenvalues seem to be consistent with two of four
eigenvalues derived by Theorem \ref{statmeasure}. 
The other two eigenvalues from Theorem \ref{statmeasure} correspond to the unphysical solutions due to
the divergence of the stationary measure.
We also check that the probability distribution
calculated from the isolated eigenvector on the upper half plane is
exponentially localized and its exponential decay rate is consistent
with that of the stationary measure in Eq.\
(\ref{eq:mu(x)3,4 non-chiral case1}).

Finally, we check robustness of these localized states against the
perturbative coin operator $U_p$.
We employ the same condition with the previous subsection, except for
the value of $\sigma_+$.  
Figure 14 shows the eigenvalue of the two-phase QW with the above
perturbative coin on the path. We clearly see that 
all eigenvalues, 
including those corresponding to localization, changes their values under the
perturbation, in contrast to the previous case.
Taking account of the above observations, we concluded that
the two-phase QW with $\sigma_+ = \pi$ and $\sigma_-=\pi/2$
does have the localized eigenvectors, while
they do not originate to the topological invariant.

\begin{figure}[t]
\begin{minipage}{0.47\hsize}
\centerline{\includegraphics[width=7.1cm]{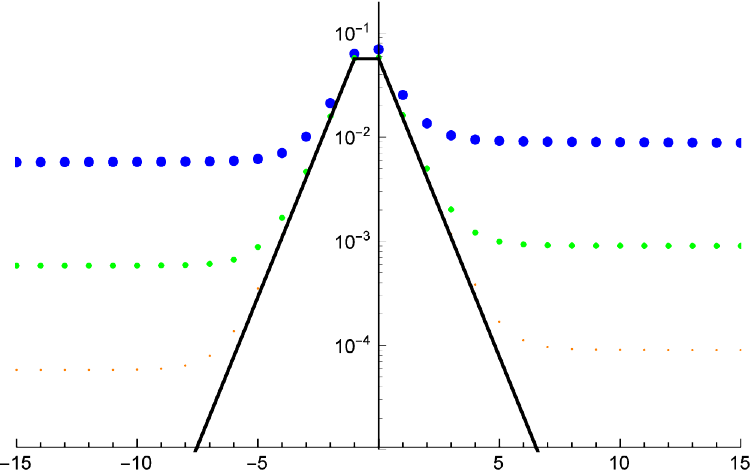}} 
\vspace*{13pt}
\caption{\label{fig7}$(\sigma_{+}=\pi, \sigma_{-}=\pi/2,\;
 {}^T\![\alpha, \beta]= {}^T\![1, 0])$
 The time averaged limit measure (black line) and time-averaged
 probabilities upto time 100 (blue dots), 1000 (green dots), and 10000
 (orange dots) with the initial coin state ${}^T\![\alpha, \beta]= {}^T\![-i\cos(3\pi/8), \sin(3\pi/8)]$.}
\end{minipage}
 \hfill
\begin{minipage}{0.47\hsize}
\centerline{\includegraphics[width=7.1cm]{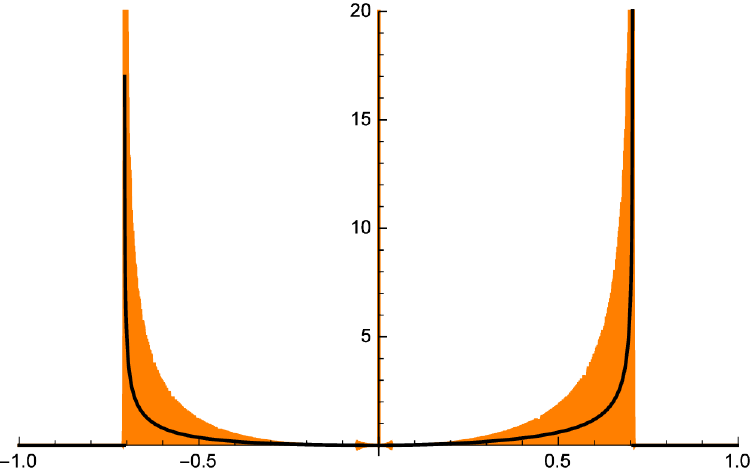}} 
\vspace*{13pt}
\caption{\label{fig8} $(\sigma_{+}=\pi, \sigma_{-}=\pi/2,\;  {}^T\![\alpha, \beta]= {}^T\![1, 0])$
Orange curve: Probability distribution in a re-scaled space
 $(x/10000, 10000P_{10000}(x))$ at time $10000$, Black curve:
 $w(x)f_{K}(x; 1/\sqrt{2})$.}
\end{minipage}
 \end{figure}
\hfill
\begin{figure}[h]
\begin{minipage}{0.47\hsize}
\centerline{\includegraphics[width=6.1cm]{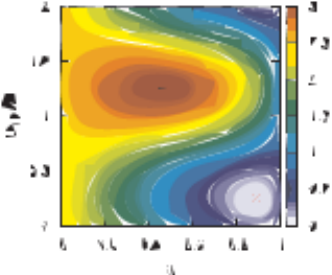}} 
\vspace*{13pt}
\caption{\label{fig:zeta_wo-chiral}
\label{fig:spectrum random}$(\sigma_{+}=\pi,
 \sigma_{-}=\pi/2)$
$a$ and $\tilde{\phi}_{12}$ dependences of
 $\zeta(\sigma_+, \sigma_-,a,b,\tilde{\phi}_{12})$ in Eq.\
 (\ref{2-phase.timeaveraged}). The red and black crosses represent the
 local minimum and maximum of $\zeta$, respectively.}
\end{minipage}
\end{figure} 

\begin{figure}[t]
\begin{minipage}{0.47\hsize}
\centerline{\includegraphics[width=6.1cm]{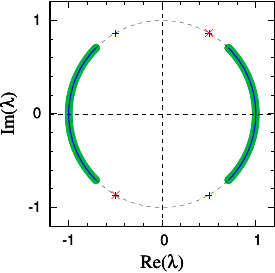}} 
\vspace*{13pt}
\caption{\label{fig:spectrum3}$(\sigma_{+}=\pi, \sigma_{-}=\pi/2)$
Green dense dots and red crosses : Numerically calculated
eigenvalues $\lambda$ of the time-evolution operator of the complete
two-phase QW without the relevant symmetries for
topological phases on the path. Eigenvalues corresponding
to the exponentially localized eigenvector are  highlighted
by red crosses.
Blue curves : The eigenvalues in Eq.\ (\ref{eq:bulk spectrum}).
Blue crosses : The eigenvalues derived by Theorem \ref{statmeasure}.
The dashed curve represents a unit circle on the complex plane.}
\end{minipage}
 \hfill
\begin{minipage}{0.47\hsize}
\centerline{\includegraphics[width=7.1cm]{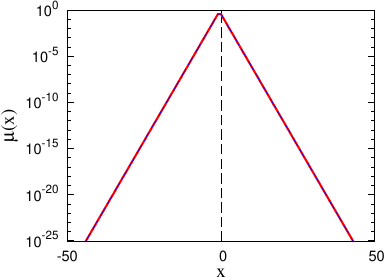}} 
\vspace*{13pt}
\caption{\label{fig:edgestate3}$(\sigma_{+}=\pi, \sigma_{-}=\pi/2)$
Red solid line : Probability distribution of the numerically calculated eigenvector
showing localization on the upper half plane.
Blue dashed line : The stationary measure in Eq.\
(\ref{eq:mu(x)3,4 non-chiral case1}) with
the normalization constant
$c^2=(1+\sqrt{3})/(9+5\sqrt{3})$.}
\end{minipage}
\end{figure}

\begin{figure}[h]
\begin{minipage}{0.5\hsize}
\centerline{\includegraphics[width=6.1cm]{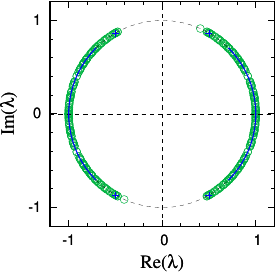}} 
\vspace*{13pt}
\caption{\label{fig:spectrum random3}$(\sigma_{+}=\pi,
\sigma_{-}=\pi/2)$
Green open dots: Numerically calculated
eigenvalue $\lambda$ of the  time-evolution operator of the complete
two-phase QW with the perturbative coin on the path.
Blue curves : eigenvalues in Eq.\ (\ref{eq:bulk spectrum}).
Blue crosses : eigenvalues derived by Theorem \ref{statmeasure}.}
\end{minipage}
\end{figure}

\subsection{localization length}

\begin{figure}[t]
  \begin{minipage}[t]{0.5\textwidth}
\centerline{\includegraphics[width=7.1cm]{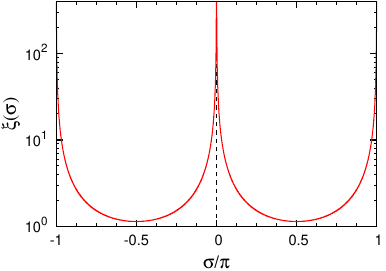}} 
\vspace*{13pt}
\caption{
The phase relativistic value $\sigma$ dependence on the localization length
 $\xi(\sigma)$. Note the relation $\xi(-\sigma)=\xi(\sigma)$.}
\label{fig:xi}
  \end{minipage}
 \end{figure}

In the previous subsections, we showed two examples with different phase
parameter sets of $\sigma_+$ and $\sigma_-$ of the complete two-phase QW. Here we show how
localization depends on the phase relative difference $\sigma= (\sigma_+
- \sigma_-)/2$. 
To this end, we simplify the time-averaged limit measure of the complete two-phase
QW in Theorem \ref{timeavelimit}.
Taking into account the fact that the time-averaged limit measure depends only on the
relativistic value of the phase parameters $\sigma=(\sigma_+ -
\sigma_-)/2$ with the initial coin state $\varphi_0=^T[1,0]$, 
we obtain the simplified exponential form up to the prefactors (which
are slightly different in positive and negative regions and at the origin):
\begin{eqnarray}
 \overline{\mu}_\infty(x) &\propto &
\exp\left(
-\frac{2|x|}{\xi(\sigma)}
\right)
\label{eq:mu_exp}
,\\
\xi(\sigma)^{-1} &=& \ln\left\{
3-2\sqrt{2}a(\sigma)\right\}/2
,\\
a(\sigma) &=& \cos\{|\varphi(\sigma)|+|\sigma|\},\quad \cos\varphi(\sigma)=\frac{1}{\sqrt{2}}\cos(\sigma),
\nonumber
\end{eqnarray} 
where $\xi(\sigma)$ is called $\it the\; localization\; length$ as the
terminology in physics.
Figure \ref{fig:xi}
shows the relative phase difference $\sigma$ dependence on
the localization length $\xi(\sigma)$. We remark that the localization
length becomes minimum at $\sigma=\pm\pi/2$, and  grows rapidly as
 $\sigma$ goes to zero or $\pi$.
At $\sigma=0$ and $\pi$ corresponding to the homogeneous QW, the localization
length $\xi(\sigma)$ diverges, which indicates that localization does
not occur.

\subsection{relation to the two-phase QW with one defect}

Finally, we mention the relation between the complete two-phase QW and
the two-phase QW with one defect studied in \cite{endosan,maman}. 
The two-phase QW with one defect is characterized by 
the following coin operator
\par\indent
\par\noindent
\begin{align}
U_{x}^{\prime}=\left\{ \begin{array}{ll}
\dfrac{1}{\sqrt{2}}\begin{bmatrix}
1 & e^{i\sigma_{+}} \\
e^{-i\sigma_{+}} & -1 \\
\end{bmatrix}&(x\geq 1),\\
\\
\begin{bmatrix}
1 & 0 \\
0 & -1\\
\end{bmatrix}&(x=0),\\
\\
\dfrac{1}{\sqrt{2}}\begin{bmatrix}
1 & e^{i\sigma_{-}} \\
e^{-i\sigma_{-}} & -1\\
\end{bmatrix}&(x\leq -1),\\
\end{array} \right.\label{2phase_def app}
\end{align}
\par\indent
\par\noindent
with $\sigma_{\pm}\in[0,2\pi)$.
Only the difference from the complete two-phase QW is the coin at the
origin $(x=0)$.
The eigenvalue problem of the unitary matrix  $U^{(s)}=S U_x^{\prime}$
is studied in Ref.\ \cite{endosan}, and four eigenvalues
are derived
\begin{equation*}
\lambda^{(1)} =  -\lambda^{(2)} =
\frac{\cos \sigma +(\sin \sigma + \sqrt{2})i}
{\sqrt{3+2\sqrt{2}\sin \sigma }}, \quad
\lambda^{(3)} =  -\lambda^{(4)} =
\frac{\cos \sigma +(\sin \sigma - \sqrt{2})i}
{\sqrt{3-2\sqrt{2}\sin \sigma }},
\end{equation*}
with the relative phase difference $\sigma=(\sigma_+-\sigma_-)/2$.
The corresponding stationary measure is also derived in Ref.\
\cite{endosan} and summarized as follows:
The stationary measure obtained from the eigenvectors of $\lambda^{(1)}$
and $\lambda^{(2)}$ is derived as
\begin{align}
\mu(x) =
\left\{
\begin{array}{ll}
c^2(2+\sqrt{2}\sin \sigma)
\left(\dfrac{1}{3+2\sqrt{2}\sin \sigma}\right)^x & (x \ge 1),\\\\
c^2 & (x = 1),\\\\
c^2\left\{2+\sqrt{2}\sin \left( \dfrac{\sigma_+ + 3\sigma_-}{2}\right)\right\}
\left(\dfrac{1}{3+2\sqrt{2}\sin \sigma}\right)^{|x|} & (x \le -1),\\
\end{array}
\right.
\end{align}
and that from the eigenvectors of $\lambda^{(3)}$ and $\lambda^{(4)}$
becomes
\begin{align}
\mu(x) =
\left\{
\begin{array}{ll}
c^2(2-\sqrt{2}\sin \sigma)
\left(\dfrac{1}{3-2\sqrt{2}\sin \sigma}\right)^x & (x \ge 1),\\\\
c^2 & (x = 1),\\\\
c^2\left\{2-\sqrt{2}\sin \left( \dfrac{\sigma_+ + 3\sigma_-}{2}\right)\right\}
\left(\dfrac{1}{3-2\sqrt{2}\sin \sigma}\right)^{|x|} & (x \le -1).\\
\end{array}
\right.
\end{align}

When $\sigma=\pm\pi/2$, we find double-degenerated eigenvalues $\lambda^{(1)} = \lambda^{(3)}= \pm i$
and $\lambda^{(2)}=\lambda^{(4)}= \mp i$. By looking the stationary
measures, in this case, we distinguish them by the decay rate, that is, 
the exponentially decay or divergence from the origin.
If $\sigma \ne \pm \pi/2$, we have non-degenerated eigenvalues at
$\lambda \ne \pm i$.
These observations are
the same with that of the complete two-phase QW as we see in the
previous subsections.
Furthermore, when $\sigma=\pm\pi/2$, which corresponds to the chiral symmetry, we emphasize that the decay rate of the stationary measure for the complete two-phase QW agrees with that of the two-phase QW with one defect.
Thereby, this gives a hint to find the
relation between the two inhomogeneous QWs.  

Then, we need to understand how the coin operator at $x=0$ of the
two-phase QW with one defect, say the defect coin operator, affects on the localized
states. 
We can confirm that the defect coin operator satisfies Eq.\
(\ref{eq:condition perturbation}), indicating that the defect coin operator is identical with the perturbative coin $U_p$ with
$\theta_x=0$. Assuming that the defect coin operator replaces the coin operator
at a certain position $x$ of the complete two-phase QW with the relevant
symmetries for topological phases $(\sigma=\pm \pi/2)$, this is nothing but the symmetry preserving perturbation.
Therefore, the localized states of
the two-phase QW with one defect can be regarded as the localized states
of the complete two-phase QW survived from the symmetry preserving
perturbation at the origin.

\section{Summary}
\label{conclusion}
In this paper, we treated the complete two-phase QW, which can be considered as an ideal mathematical model of topological insulator.
We obtained a measure and two kinds of limit theorems describing localization and the ballistic behavior.
Indeed, we got the time-averaged limit and stationary measures for our
QW at first.
Since the dependence on the initial states of the time-averaged
limit measure was also explicitly derived,
we gave a condition that localization at the origin is not
occurred even when the two
regions have different topological numbers by setting appropriate
phases of the time-evolution operators and initial state of the
walker. Conversely stating, localization is most enhanced with a proper initial coin state 
Therefore, we conclude that the time-averaged limit measure  derived in
this paper can be used to find an optimal condition of the initial coin state to
clearly observe localization even in actual experiments.\\
\indent In addition, owing to the asymmetric unitary matrices, both the stationary
and time-averaged limit measures are generally asymmetric for the
origin, however, the stationary and time-averaged measures become
symmetric at $x=-1/2$ at least, for the two cases studied in Section 4. This indicates that the defect at the origin of the two-phase
QW with one defect  makes the time-averaged distribution symmetric at the origin\cite{endosan}.
Moreover, we presented the weak convergence theorem which describes the ballistic behavior in the probability distributions for the position of the walker in re-scaled spaces.  \\
\indent We also studied the topological invariant of the complete two-phase QW. We
clarified that the time-evolution operator of the complete two-phase QW
possesses chiral symmetry with $\varepsilon_\Gamma=-\pi/2$,
when the two phases satisfy $\sigma_+-\sigma_-=n\pi$ $(n\in
\mathbb{Z})$.
Therefore, the complete two-phase QW belongs to class AIII in the Cartan
classification. 
Then, we derived the topological numbers for two-specific eigenvalues $\lambda=\pm i$.
Taking into account these results, we compared the mathematical rigorous
results with prediction by the bulk-edge correspondence, and confirmed 
the perfect agreements.
Furthermore, we succeeded to
find the relation between localization of the two-phase QW with one defect
and one of the complete two-phase QW, by considering that the defect
coin operator at the origin as the symmetry preserving perturbation on
the complete two-phase QW. 
In addition, we succeeded to relate the stationary and time-averaged limit measures by using both mathematical and physical consideration, which indicates that we can analyze localization of QWs by the stationary measure.
The approach used in the present work would provide solid arguments to understand
localization of various QWs in term of topological invariant.\\
{\bf Acknowledgments.}
TE is supported by financial support of Postdoctoral Fellowship from Japan Society for the
Promotion of Science.
HO is supported by the ``Topological Materials
Science'' (No.\ JP18H04210) and Grants-in-Aid (No.\ JP18H01140 and No.\ JP18K18733) from
the Japan Society for Promotion of Science.

\begin{small}
\bibliographystyle{jplain}

\end{small}

\flushleft
\noindent {\large{\bf Appendix A}}  \\ \noindent
\label{prooftimeave}
Hereafter, we present the general expression of Theorem \ref{timeavelimit} along with the {\it time-space generating function method}\cite{segawa}. The protocol is similar to that of Section $4$ in Ref. \cite{endosan}.
To begin with, we give some notations.
The coin operator  $U_{x}$ can be divided into two parts by
\[U_{x}=P_{x}+Q_{x},\]
where
\[ 
P_{x}=\begin{bmatrix} 
a_{x} & b_{x} \\
0 & 0 
\end{bmatrix}, \;\;
Q_{x}=\begin{bmatrix}
0 & 0 \\
c_{x} & d_{x}
\end{bmatrix}.\]
Here we introduce a notation of the weight of all the passages of the walker which moves to the left $l$ times and to the right $m$ times till time $t$ as follows\cite{segawa}:
\[\Xi_{t}(l,m)=\sum_{l_{j},m_{j}}P_{x_{l1}}^{l_{1}}Q_{x_{m1}}^{m_{1}}P_{x_{l2}}^{l_{2}}Q_{x_{m_{2}}}^{m_{2}}\cdots P_{x_{lt}}^{l_{t}}Q_{x_{mt}}^{m_{t}},\]\\
with $l+m=t,\;-l+m=x,\;\;\sum_{i}l_{i}=l,\;\sum_{j}m_{j}=m$, and $\;\sum_{\gamma=l_{i},m_{j}}|x_{\gamma}|=x$. 
We remark that the time-averaged limit measure can be written by the
square norm of the residue of the generating function $\tilde{\Xi}_{x}(z)\equiv\sum_{t\geq0}\Xi_{t}(x)z^{t}$, which leads us to complete the proof:
\begin{proposition}\label{monsan}\cite{endo}
We have
\[\overline{\mu}_{\infty}(x)=\sum_{\theta^{(\pm)}_{j}}\left\|Res(\tilde{\Xi}_{x}(z):z=e^{i\theta^{(\pm)}_{j}})\varphi_{0}\right\|^{2},\]
where $\{e^{i\theta^{(\pm)}_{j}}\}$ is the set of the singular points of $\tilde{\Xi}_{x}(z)$.
\end{proposition}
\indent
Now we give useful concrete formula of $\tilde{\Xi}_{x}(z)$, which plays an important role for the proof. 
The derivation of Lemma \ref{kls-2phase} comes from Lemma $3.1$ 
in Ref. \cite{segawa}. In Appendix A, we assume that the walker starts at the origin with the initial coin state $\varphi_{0}={}^T\![\alpha,\beta]$, where $\alpha,\beta\in\mathbb{C}$,
and $|\alpha|^{2}+|\beta|^{2}=1$.
\begin{lemma} \label{kls-2phase}
\begin{enumerate}
\item If $x=0$, we have 
\[\tilde{\Xi}_{0}(z)=\dfrac{1}{1-\dfrac{e^{-i\sigma_{+}}}{\sqrt{2}}\tilde{f}_{0}^{(+)}(z)-\dfrac{e^{i\sigma_{+}}}{\sqrt{2}}\tilde{f}_{0}^{(-)}(z)+\tilde{f}_{0}^{(+)}(z)\tilde{f}_{0}^{(-)}(z)}
\begin{bmatrix} 
1-\dfrac{e^{i\sigma_{+}}}{\sqrt{2}} & -\dfrac{1}{\sqrt{2}}\tilde{f}_{0}^{(+)}(z)\\
\dfrac{1}{\sqrt{2}}\tilde{f}_{0}^{(-)}(z) & 1-\dfrac{e^{-i\sigma_{+}}}{\sqrt{2}}\tilde{f}_{0}^{(+)}(z)\\
\end{bmatrix}.\]
\item If $|x|\geq 1$, we have
\[
\tilde{\Xi}_{x}(z)=\left\{\begin{array}{ll}
(\tilde{\lambda}^{(+)}(z))^{x-1}
\left[
    \begin{array}{c}
      \tilde{\lambda}^{(+)}(z)\tilde{f}_{0}^{(+)}(z)\\
      z \\
    \end{array}
  \right]\left[\dfrac{e^{-i\sigma_{+}}}{\sqrt{2}},-\dfrac{1}{\sqrt{2}}\right]\tilde{\Xi}_{0}(z) & (x\geq 1), \\
  &\\
(\tilde{\lambda}^{(-)}(z))^{|x|-1}
\left[
    \begin{array}{c}
      z \\
      \tilde{\lambda}^{(-)}(z)\tilde{f}_{0}^{(-)}(z) \\
    \end{array}
  \right]\left[\dfrac{1}{\sqrt{2}},\dfrac{e^{i\sigma_{+}}}{\sqrt{2}}\right]\tilde{\Xi}_{0}(z) & (x\leq -1), \\
\end{array} \right.\]
\end{enumerate}
\noindent
with $\tilde{\lambda}^{(+)}(z)=\dfrac{z}{e^{-i\sigma_{+}}\tilde{f}_{0}^{(+)}(z)-\sqrt{2}}$ and $\tilde{\lambda}^{(-)}(z)=\dfrac{z}{\sqrt{2}-e^{i\sigma_{-}}\tilde{f}_{0}^{(-)}(z)}.$
Note that $\tilde{f}_{0}^{(+)}(z)$ and $\tilde{f}_{0}^{(-)}(z)$ satisfy
\[
\left\{
\begin{array}{l}
(\tilde{f}^{(+)}_{0}(z))^{2}-\sqrt{2}e^{i\sigma_{+}}(1+z^{2})\tilde{f}^{(+)}_{0}(z)+e^{2i\sigma_{+}}z^{2}=0,\\
\\
(\tilde{f}^{(-)}_{0}(z))^{2}-\sqrt{2}e^{-i\sigma_{-}}(1+z^{2})\tilde{f}^{(-)}_{0}(z)+e^{-2i\sigma_{-}}z^{2}=0.
\end{array}
\right.
\]
\end{lemma}
Thereby, we obtain
\begin{lemma}
\label{monpal}
$\tilde{f}_{0}^{(+)}(z)$ and $\tilde{f}_{0}^{(-)}(z)$ are written down with respect to $\theta$ as 
\begin{align}\tilde{f}_{0}^{(\pm)}(z)=e^{i(\theta\pm\sigma_{\pm})}\times e^{i\tilde{\phi}(\theta)},\end{align}
with
\begin{align}\sin\tilde{\phi}(\theta)=\operatorname{sgn}(\sin\theta)\sqrt{2\sin\theta^{2}-1},\quad
\cos\tilde{\phi}(\theta)=\sqrt{2}\cos\theta.\label{tildephi}\end{align} 
\end{lemma}
The derivation of Lemma \ref{monpal} is similar to Lemma 3 in Ref. \cite{maman}, and we omit it here.
By taking advantage of Lemmas \ref{kls-2phase} and \ref{monpal}, we obtain the set of the singular points of $\tilde{\Xi}_{x}(z)$:
\begin{lemma}
\label{kai}
Let
\begin{eqnarray*}e^{i\theta^{(\pm)}_{1}}=\pm\left(\dfrac{\sin t^{(+)}(\sigma)}{\sqrt{3-2\sqrt{2}\cos t^{(+)}(\sigma)}}+\dfrac{\sqrt{2}-\cos t^{(+)}(\sigma)}{\sqrt{3-2\sqrt{2}\cos t^{(+)}(\sigma)}}i\right),\\
e^{i\theta^{(\pm)}_{2}}=\pm\left(\dfrac{\sin t^{(-)}(\sigma)}{\sqrt{3-2\sqrt{2}\cos t^{(-)}(\sigma)}}+\dfrac{\sqrt{2}-\cos t^{(-)}(\sigma)}{\sqrt{3-2\sqrt{2}\cos t^{(-)}(\sigma)}}i\right),\end{eqnarray*}
where $t^{(\pm)}(\sigma)=\varphi^{(\pm)}(\sigma)-\sigma$ with 
\[\left\{
\begin{array}{l}
\cos\varphi^{(\pm)}(\sigma)=\dfrac{1}{\sqrt{2}}\cos\sigma,\\
\sin\varphi^{(\pm)}(\sigma)=\pm\sqrt{1-\dfrac{1}{2}\cos^{2}\sigma}.
\end{array}
\right.\]
Then, we have the set of all the singular points of $\;\tilde{\Xi}_{x}(z)$ with $|z|=1$ by
\begin{eqnarray*}
B=\left\{ \begin{array}{ll}
B_{1}=\{e^{i\theta^{(+)}_{1}},e^{i\theta^{(-)}_{1}}\}& \rm{if} \cos t^{(+)}(\sigma)\leq1/\sqrt{2},\\
B_{2}=\{e^{i\theta^{(+)}_{1}},e^{i\theta^{(-)}_{1}},e^{i\theta^{(+)}_{2}},e^{i\theta^{(-)}_{2}}\}& \rm{if} \cos t^{(\pm)}(\sigma)\leq1/\sqrt{2},\\
B_{3}=\{e^{i\theta^{(+)}_{2}},e^{i\theta^{(-)}_{2}}\}&\rm{if} \cos t^{(-)}(\sigma)\leq 1/\sqrt{2}.\\
\end{array} \right.
\end{eqnarray*} 
\end{lemma}
The derivation of Lemma \ref{kai} is similar to that of Lemma 4 in Ref. \cite{endosan}. Here, we omit it.

Next, we derive the residues of $\;\tilde{\Xi}_{x}(z)$ at the singular points. 
For the simplicity, we put the denominator of $\tilde{\Xi}_{0}(z)$ by $\tilde{\Lambda}_{0}(z)\equiv1-e^{-i\sigma_{+}}\tilde{f}_{0}^{(+)}(z)/\sqrt{2}-e^{i\sigma_{+}}\tilde{f}_{0}^{(-)}(z)/\sqrt{2}+\tilde{f}_{0}^{(+)}(z)\tilde{f}_{0}^{(-)}(z)$. 
Note that all the singular points for localization come from the solution of $\tilde{\Lambda}_{0}(z)=0.$

Then, explicit expressions of the square of the absolute value of the residues of $1/\tilde{\Lambda}_{0}(z)$ are given in a similar way as that of Lemma $5$ in Ref. \cite{endosan}, and we obtain Lemma \ref{pari}:
\begin{lemma}
\label{pari}
\begin{enumerate}
\item For $e^{i\theta^{(\pm)}_{1}}$, we have
\begin{align}&\left|Res\left(\frac{1}{\tilde{\Lambda}_{0}(z)}: z=e^{i\theta^{(\pm)}_{1}}\right)\right|^{2}\nonumber\\
&=\frac{\left(1-\sqrt{2}\cos t^{(+)}(\sigma)\right)^{2}}{\left(3-2\sqrt{2}\cos t^{(+)}(\sigma)\right)^{2}\left\{5+\cos 2\sigma-2\sqrt{2}\cos(2\sigma+t^{(+)}(\sigma))-2\sqrt{2}\cos t^{(+)}(\sigma)\right\}}.\end{align}
\item For $e^{i\theta^{(\pm)}_{2}}$, we have
\begin{align}&\left|Res\left(\frac{1}{\tilde{\Lambda}_{0}(z)}: z=e^{i\theta_{2}^{(\pm)}}\right)\right|^{2}\nonumber\\
&=\frac{\left(1-\sqrt{2}\cos t^{(-)}(\sigma)\right)^{2}}{\left(3-2\sqrt{2}\cos t^{(-)}(\sigma)\right)^{2}\left\{5+\cos 2\sigma-2\sqrt{2}\cos(2\sigma+t^{(-)}(\sigma))-2\sqrt{2}\cos t^{(-)}(\sigma)\right\}}.\end{align}
\end{enumerate}
\end{lemma}
Noting Lemma \ref{kls-2phase}-Lemma \ref{pari}, we first show the case of $x=0$ in Theorem $2$ in the following way. By Lemma \ref{kls-2phase}, we have
\[\tilde{\Xi}_{0}(z)\varphi_{0}=\frac{1}{\tilde{\Lambda}_{0}(z)}\begin{bmatrix}\alpha\left(1-\dfrac{e^{i\sigma_{+}}}{\sqrt{2}}\tilde{f}_{0}^{(-)}(z)\right)-\dfrac{\beta}{\sqrt{2}}\tilde{f}_{0}^{(+)}(z)\\
\dfrac{\alpha}{\sqrt{2}}\tilde{f}_{0}^{(-)}(z)+\beta\left(1-\dfrac{e^{-i\sigma_{+}}}{\sqrt{2}}\tilde{f}_{0}^{(+)}(z)\right)\end{bmatrix}.\]
Thus, we get the square norm of the residues by
\begin{eqnarray}\left\|Res(\tilde{\Xi}_{0}(z)\varphi_{0}:z=e^{i\theta^{(\pm)}_{j}})\right\|_{j=1,2}^{2}=\left|Res\left(\dfrac{\alpha\left(1-e^{i\sigma_{+}}\tilde{f}_{0}^{(-)}(z)/\sqrt{2}\right)-\beta\tilde{f}_{0}^{(+)}(z)/\sqrt{2}}{\tilde{\Lambda}_{0}(z)}: z=e^{i\theta^{(\pm)}_{j}}\right)\right|^{2}+\nonumber\\
\left|Res\left(\dfrac{\alpha\tilde{f}_{0}^{(-)}(z)/\sqrt{2}+\beta\left(1-e^{-i\sigma_{+}}\tilde{f}_{0}^{(+)}(z)/\sqrt{2}\right)}{\tilde{\Lambda}_{0}(z)}: z=e^{i\theta^{(\pm)}_{j}}\right)\right|^{2}.\label{squarenorm-residue}\end{eqnarray}
Taking into account \[Res(1/\tilde{\Lambda}_{0}(z): z=e^{i\theta^{(\pm)}_{j}})=\lim_{z\to e^{i\theta^{(\pm)}_{j}}}(z-e^{i\theta^{(\pm)}_{j}})/\tilde{\Lambda}_{0}(z)\quad(j=1,2)\] holds, where $\{e^{i\theta^{(\pm)}_{j}}\}_{j=1,2}$ is the set of the singular points of $\tilde{\Xi}_{x}(z)$, we obtain 
\begin{align}\left|Res\left(\dfrac{1}{\tilde{\Lambda}_{0}(z)}: z=e^{i\theta^{(\pm)}_{j}}\right)\right|_{j=1,2}^{2}=\dfrac{2}{\left|1+\dfrac{\partial\tilde{\phi}(\theta)}{\partial\theta}\right|_{\theta^{(\pm)}_{j}}^{2}\left|1+e^{2i\sigma}-2\sqrt{2}e^{i(2\sigma+\theta^{(\pm)}_{j}+\tilde{\phi}(\theta^{(\pm)}_{j}))}\right|^{2}}.\label{expanding-res}\end{align}
\noindent
Thereby, we see from Eq. \eqref{expanding-res},
\begin{align*}
&\left|Res\left(\dfrac{\alpha\left(1-e^{i\sigma_{+}}\tilde{f}_{0}^{(-)}(z)/\sqrt{2}\right)-\beta\tilde{f}_{0}^{(+)}(z)/\sqrt{2}}{\tilde{\Lambda}_{0}(z)}: z=e^{i\theta^{(\pm)}_{j}}\right)\right|_{j=1,2}^{2}\\
&\hspace{70mm}=\dfrac{2\left|\alpha\left(1-e^{i\sigma_{+}}\tilde{f}_{0}^{(-)}(e^{i\theta^{(\pm)}_{j}})/\sqrt{2}\right)-\beta\tilde{f}_{0}^{(+)}(e^{i\theta^{(\pm)}_{j}})/\sqrt{2}\right|^{2}}{\left|1+\dfrac{\partial\tilde{\phi}(\theta)}{\partial\theta}\right|_{\theta^{(\pm)}_{j}}^{2}\left|1+e^{2i\sigma}-2\sqrt{2}e^{i(2\sigma+\theta^{(\pm)}_{j}+\tilde{\phi}(\theta^{(\pm)}_{j}))}\right|^{2}},\\
\\
&\left|Res\left(\dfrac{\alpha\tilde{f}_{0}^{(-)}(z)/\sqrt{2}+\beta\left(1-e^{-i\sigma_{+}}\tilde{f}_{0}^{(+)}(z)/\sqrt{2}\right)}{\tilde{\Lambda}_{0}(z)}: z=e^{i\theta^{(\pm)}_{j}}\right)\right|_{j=1,2}^{2}\\
&\hspace{70mm}=\dfrac{2\left|\alpha\tilde{f}_{0}^{(-)}(e^{i\theta^{(\pm)}_{j}})/\sqrt{2}+\beta\left(1-e^{-i\sigma_{+}}\tilde{f}_{0}^{(+)}(e^{i\theta^{(\pm)}_{j}})/\sqrt{2}\right)\right|^{2}}{\left|1+\dfrac{\partial\tilde{\phi}(\theta)}{\partial\theta}\right|_{\theta^{(\pm)}_{j}}^{2}\left|1+e^{2i\sigma}-2\sqrt{2}e^{i(2\sigma+\theta^{(\pm)}_{j}+\tilde{\phi}(\theta^{(\pm)}_{j}))}\right|^{2}}.\end{align*}
Here we have

\begin{align*}
&\left|\alpha\left(1-\dfrac{e^{i\sigma_{+}}\tilde{f}_{0}^{(-)}(e^{i\theta^{(\pm)}_{j}})}{\sqrt{2}}\right)-\dfrac{\beta\tilde{f}_{0}^{(+)}(e^{i\theta^{(\pm)}_{j}})}{\sqrt{2}}\right|^{2}\hspace{40mm}\\
&\hspace{15mm}=\left\{ a^{2}\left(\dfrac{3}{2}-\sqrt{2}\cos(2\sigma+\theta^{(\pm)}_{j}+\tilde{\phi}(\theta^{(\pm)}_{j}))\right)+\dfrac{b^{2}}{2}-\sqrt{2}ab\cos(\tilde{\phi}_{12}-\theta^{(\pm)}_{j}-\sigma_{+}-\tilde{\phi}(\theta^{(\pm)}_{j}))+ab\cos(\tilde{\phi}_{12}-\sigma_{-})\right\},\\
&\left|\dfrac{\alpha\tilde{f}_{0}^{(-)}(e^{i\theta^{(\pm)}_{j}})}{\sqrt{2}}+\beta\left(1-\dfrac{e^{-i\sigma_{+}}\tilde{f}_{0}^{(+)}(e^{i\theta^{(\pm)}_{j}})}{\sqrt{2}}\right)\right|^{2}\hspace{10mm}\\
&\hspace{15mm}=\left\{\dfrac{a^{2}}{2}+b^{2}\left(\dfrac{3}{2}-\sqrt{2}\cos(\theta^{(\pm)}_{j}+\tilde{\phi}(\theta^{(\pm)}_{j}))\right)+\sqrt{2}ab\cos(\tilde{\phi}_{12}+\theta^{(\pm)}_{j}-\sigma_{-}+\tilde{\phi}(\theta^{(\pm)}_{j}))-ab\cos(\tilde{\phi}_{12}-\sigma_{-})\right\}.
\end{align*}
Therefore, we get
\begin{align*}&\left\|Res(\tilde{\Xi}_{0}(z)\varphi_{0}: z=e^{i\theta^{(\pm)}_{j}})\right\|_{j=1,2}^{2}=\dfrac{2}{\left|1+\dfrac{\partial\tilde{\phi}(\theta)}{\partial\theta}\right|_{\theta^{(\pm)}_{j}}^{2}\left|1+e^{2i\sigma}-2\sqrt{2}e^{i(2\sigma+\theta^{(\pm)}_{j}+\tilde{\phi}(\theta^{(\pm)}_{j}))}\right|^{2}}\\
&\times\left\{2-\sqrt{2}(a^{2}\cos(2\sigma+\theta^{(\pm)}_{j}+\tilde{\phi}(\theta^{(\pm)}_{j}))+b^{2}\cos(\theta^{(\pm)}_{j}+\tilde{\phi}(\theta^{(\pm)}_{j})))-\sqrt{2}ab(\cos(\tilde{\phi}_{12}-\sigma_{+}-\theta^{(\pm)}_{j}
-\tilde{\phi}(\theta^{(\pm)}_{j}))\right.\\
&\hspace{110mm}\left.-\cos(\tilde{\phi}_{12}-\sigma_{-}+\theta^{(\pm)}_{j}+\tilde{\phi}(\theta^{(\pm)}_{j})))\right\}.\end{align*}
Noting Proposition \ref{monsan},
we obtain the general expression of the case of $x=0$ in Theorem \ref{timeavelimit}.

Here the range of the summation is 
\[\{\theta^{(\pm)}_{j}\in[0,2\pi); e^{i\theta^{(\pm)}_{j}}\in B\}.\]
\\
In the next stage, we give the detail of the derivation for the case of $x\geq1$ in Theorem \ref{timeavelimit}.
By Lemma \ref{kls-2phase}, we see
\[\tilde{\Xi}_{x}(z)\varphi_{0}=\dfrac{(\tilde{\lambda}^{(+)}(z))^{x-1}}{\sqrt{2}\tilde{\Lambda}_{0}(z)}\begin{bmatrix}(\tilde{\lambda}^{(+)}(z))\left\{\alpha e^{i(\theta+\tilde{\phi}(\theta))}-\sqrt{2}\alpha\tilde{f}_{0}^{(+)}(z)\tilde{f}_{0}^{(-)}(z)-\beta\tilde{f}_{0}^{(+)}(z)\right\}\\ z\left\{\alpha e^{-i\sigma_{+}}-\sqrt{2}\alpha\tilde{f}_{0}^{(-)}(z)-\beta\right\}\end{bmatrix}\qquad (x\geq 1).\]
Thereby,  we get the square norm of the residues as
\begin{align}\left\|Res(\tilde{\Xi}_{x}(z)\varphi_{0}: z=e^{i\theta^{(\pm)}_{j}})\right\|^{2}
&=\left|Res\left(\dfrac{(\tilde{\lambda}^{(+)}(z))^{x}\left\{\alpha e^{i(\theta+\tilde{\phi}(\theta))}-\sqrt{2}\alpha\tilde{f}_{0}^{(+)}(z)\tilde{f}_{0}^{(-)}(z)-\beta\tilde{f}_{0}^{(+)}(z)\right\}}{\sqrt{2}\tilde{\Lambda}_{0}(z)}: z=e^{i\theta^{(\pm)}_{j}}\right)\right|^{2}\nonumber\\&+
\left|Res\left(\dfrac{(\tilde{\lambda}^{(+)}(z))^{x-1}z\left\{\alpha e^{-i\sigma_{+}}-\sqrt{2}\alpha\tilde{f}_{0}^{(-)}(z)-\beta\right\}}{\sqrt{2}\tilde{\Lambda}_{0}(z)}: z=e^{i\theta^{(\pm)}_{j}}\right)\right|^{2}.\label{norm}\end{align}

Here we get
\begin{align}\left|Res\left(\dfrac{(\tilde{\lambda}^{(+)}(z))^{x}\left\{\alpha e^{i(\theta+\tilde{\phi}(\theta))}-\sqrt{2}\alpha\tilde{f}_{0}^{(+)}(z)\tilde{f}_{0}^{(-)}(z)-\beta\tilde{f}_{0}^{(+)}(z)\right\}}{\sqrt{2}\tilde{\Lambda}_{0}(z)}: z=e^{i\theta^{(\pm)}_{j}}\right)\right|^{2}\nonumber\hspace{55mm}\\
=\dfrac{\left|\tilde{\lambda}^{(+)}(e^{i\theta^{(\pm)}_{j}})\right|^{2x}\left|\alpha e^{i(\theta^{(\pm)}_{j}+\tilde{\phi}(\theta^{(\pm)}_{j}))}-\sqrt{2}\alpha\tilde{f}_{0}^{(+)}(e^{i\theta^{(\pm)}_{j}})\tilde{f}_{0}^{(-)}(e^{i\theta^{(\pm)}_{j}})-\beta\tilde{f}_{0}^{(+)}(e^{i\theta^{(\pm)}_{j}})\right|^{2}}{\left|1+\dfrac{\partial\tilde{\phi}(\theta)}{\partial\theta}\right|_{\theta^{(\pm)}_{j}}^{2}\left|1+e^{2i\sigma}-2\sqrt{2}e^{i(2\sigma+\theta^{(\pm)}_{j}+\tilde{\phi}(\theta^{(\pm)}_{j}))}\right|^{2}},\label{ookisa1}\end{align}
and
\begin{align}\left|
Res\left(\dfrac{(\tilde{\lambda}^{(+)}(z))^{x-1}z\left\{\alpha e^{-i\sigma_{+}}-\sqrt{2}\alpha\tilde{f}_{0}^{(-)}(z)-\beta\right\}}{\sqrt{2}\tilde{\Lambda}_{0}(z)}: z=e^{i\theta^{(\pm)}_{j}}\right)
\right|^{2}\nonumber\hspace{60mm}\\
=\dfrac{\left|\tilde{\lambda}^{(+)}(e^{i\theta^{(\pm)}_{j}})\right|^{2(x-1)}\left|\alpha e^{-i\sigma_{+}}-\sqrt{2}\alpha\tilde{f}_{0}^{(-)}(e^{i\theta^{(\pm)}_{j}})-\beta\right|^{2}}{\left|1+\dfrac{\partial\tilde{\phi}(\theta)}{\partial\theta}\right|_{\theta^{(\pm)}_{j}}^{2}\left|1+e^{2i\sigma}-2\sqrt{2}e^{i(2\sigma+\theta^{(\pm)}_{j}+\tilde{\phi}(\theta^{(\pm)}_{j}))}\right|^{2}}.\label{ookisa2}\end{align}
Thereby, Eqs. \eqref {norm}, \eqref{ookisa1} and \eqref{ookisa2} provide 
\begin{align*}
\left\|Res(\tilde{\Xi}_{x}(z)\varphi_{0}: z=e^{i\theta^{(\pm)}_{j}})\right\|_{j=1,2}^{2}=\dfrac{\left|\tilde{\lambda}^{(+)}(e^{i\theta^{(\pm)}_{j}})\right|^{2(x-1)}}{\left|1+\dfrac{\partial\tilde{\phi}(\theta)}{\partial\theta}\right|_{\theta^{(\pm)}_{j}}^{2}\left|1+e^{2i\sigma}-2\sqrt{2}e^{i(2\sigma+\theta^{(\pm)}_{j}+\tilde{\phi}(\theta^{(\pm)}_{j}))}\right|^{2}}\left(1+\left|\tilde{\lambda}^{(+)}(e^{i\theta^{(\pm)}_{j}})\right|^{2}\right)\nonumber\hspace{20mm}\\
\times\left\{1+2a^{2}-2\sqrt{2}a^{2}\cos(2\sigma+\theta^{(\pm)}_{j}+\tilde{\phi}(\theta^{(\pm)}_{j}))-2ab\left(\cos(\tilde{\phi}_{12}-\sigma_{+})-\sqrt{2}\cos(\tilde{\phi}_{12}-\sigma_{-}+\theta^{(\pm)}_{j}+\tilde{\phi}(\theta^{(\pm)}_{j}))\right)\right\}.\end{align*}
Hence, we obtain
\begin{align}\overline{\mu}_{\infty}(x)=\sum_{\theta^{(\pm)}_{j}}\dfrac{\left|\tilde{\lambda}^{(+)}(e^{i\theta^{(\pm)}_{j}})\right|^{2(x-1)}}{\left|1+\dfrac{\partial\tilde{\phi}(\theta)}{\partial\theta}\right|_{\theta^{(\pm)}_{j}}^{2}\left|1+e^{2i\sigma}-2\sqrt{2}e^{i(2\sigma+\theta^{(\pm)}_{j}+\tilde{\phi}(\theta^{(\pm)}_{j}))}\right|^{2}}\left(1+\left|\tilde{\lambda}^{(+)}(e^{i\theta^{(\pm)}_{j}})\right|^{2}\right)\nonumber\hspace{40mm}\\
\times\left\{1+2a^{2}-2\sqrt{2}a^{2}\cos(2\sigma+\theta^{(\pm)}_{j}+\tilde{\phi}(\theta^{(\pm)}_{j}))-2ab\left(\cos(\tilde{\phi}_{12}-\sigma_{+})-\sqrt{2}\cos(\tilde{\phi}_{12}-\sigma_{-}+\theta^{(\pm)}_{j}+\tilde{\phi}(\theta^{(\pm)}_{j}))\right)\right\},\nonumber\\\label{jikanheikin}\end{align}
where $\{e^{i\theta^{(\pm)}_{j}}\}_{j=1,2}$ is the set of the singular points of $\tilde{\Xi}_{x}(z).$
Now we compute $\left|\tilde{\lambda}^{(+)}(e^{i\theta^{(\pm)}_{j}})\right|^{2}$.
By the definition of $\tilde{\lambda}^{(+)}(z)$ in Lemma \ref{kls-2phase}, we see
\[\tilde{\lambda}^{(+)}(e^{i\theta})=\dfrac{1}{e^{i\tilde{\phi}(\theta)}-\sqrt{2}e^{-i\theta}},\]
which leads to
\[\left|\tilde{\lambda}^{(+)}(e^{i\theta})\right|^{2}=\dfrac{1}{3-2\sqrt{2}\cos(\theta+\tilde{\phi}(\theta))}.\]
Equation \eqref{tildephi} gives

\begin{eqnarray}
\left|\tilde{\lambda}^{(+)}(e^{i\theta^{(\pm)}_{1}})\right|^{2}=\dfrac{1}{3-2\sqrt{2}\cos t^{(+)}(\sigma)},\quad
\left|\tilde{\lambda}^{(+)}(e^{i\theta^{(\pm)}_{2}})\right|^{2}=\dfrac{1}{3+2\sqrt{2}\cos t^{(-)}(\sigma)}.
\label{lambdaookisa}\end{eqnarray} 
Taking into account Proposition \ref{monsan}, and substituting Lemma \ref{pari} and Eq.  \eqref{lambdaookisa} into Eq. \eqref{jikanheikin}, we obtain the general expression of the case of $x\geq1$ in Theorem \ref{timeavelimit}.
In a similar fashion, we get the case of $x\leq-1$ in Theorem \ref{timeavelimit}, and we complete the proof.
\\
\flushleft
\noindent {\large{\bf Appendix B}}  \\ \noindent
In Appendix B, we give the detail of the derivation of Theorem \ref{statmeasure}.
Throughout Appendix B, let us focus on the generalized eigenequation
\begin{align}U^{(s)}\Psi=\lambda\Psi,\label{eigenvalueproblem}\end{align}
where $\lambda\in\mathbb{C}$ with $|\lambda|=1$, and $\Psi\in(\mathbb{C}^{2})^{\infty}$.
Taking advantage of the SGF method\cite{watanabe}, we solve the generalized eigenequation \eqref{eigenvalueproblem}.
At first, we introduce the generating functions of $\Psi^{j}(x)\;(j=L,R)$:
\begin{eqnarray}
f^{j}_{+}(z)=\sum^{\infty}_{x=1} \Psi^{j}(x)z^{x},\quad
f^{j}_{-}(z)=\sum^{-\infty}_{x=-1} \Psi^{j}(x)z^{x},
\label{2phase-bokansuu}
\end{eqnarray}
\noindent
which provide
\begin{lemma}
\label{2phase-hodai1}
Put
\begin{eqnarray*}A\!\!\!&=&\!\!\!\begin{bmatrix}\lambda-\dfrac{1}{\sqrt{2}z}&-\dfrac{e^{i\sigma_{\pm}}}{\sqrt{2}z}\nonumber\\\
-\dfrac{e^{-i\sigma_{\pm}}}{\sqrt{2}}z&\lambda+\dfrac{z}{\sqrt{2}}
\end{bmatrix},
\;{\bf f}_{\pm}(z)=\left[\begin{array}{c}f^{L}_{\pm}(z)\\f^{R}_{\pm}(z)\end{array}\right],\nonumber\\\
{\bf a}_{+}(z)\!\!\!&=&\!\!\!\left[\begin{array}{c}-\lambda\alpha\\ \left(\dfrac{e^{-i\sigma_{+}}}{\sqrt{2}}\alpha-\dfrac{1}{\sqrt{2}}\beta\right)z\end{array}\right],\;
{\bf a}_{-}(z)=\left[\begin{array}{c}\left(\dfrac{1}{\sqrt{2}}\alpha+\dfrac{e^{i\sigma_{+}}}{\sqrt{2}}\beta\right)z^{-1}\\ -\lambda\beta\end{array}\right],\end{eqnarray*}
with $\alpha=\Psi^{L}(0)$ and $\beta=\Psi^{R}(0)$.
Then, we have
\begin{align}
A_{\pm}{\bf f}_{\pm}(z)={\bf a}_{\pm}(z).\label{2phase-hodai1-houteisiki}
\end{align}
\end{lemma}
Taking account of
\begin{align}
\det A_{\pm}=\dfrac{\lambda }{\sqrt{2}z}\left\{z^{2}-\sqrt{2}\left(\dfrac{1}{\lambda}-\lambda\right)z-1\right\},
\label{deta12sou}
\end{align}
we put $\theta_{s}, \> \theta_{l} \in\mathbb{C}$ satisfying
\begin{align}
\det A_{\pm} =\dfrac{\lambda}{ \sqrt{2}z}(z-\theta_{s})(z-\theta_{l}),
\label{deta22sou}
\end{align}
and $|\theta_{s}|\leq1\leq|\theta_{l}|$.
By Eqs. \eqref{deta12sou} and \eqref{deta22sou}, we see $\theta_{s}\theta_{l}=-1$.\\
Hereafter let us derive $f_{\pm}^{L}(z)$ and $f_{\pm}^{R}(z)$ from Lemma \ref{2phase-hodai1}.

\begin{enumerate}
\item Case of $f_{+}^{L}(z)$: Eq.  \eqref{2phase-hodai1-houteisiki} gives
\begin{align*}
f^{L}_{+}(z)
= - \dfrac{ \alpha z}{(z-\theta_{s})(z-\theta_{l})} \left\{ z - \left(\dfrac{1}{\sqrt{2}\lambda}-\dfrac{e^{i\sigma_{+}}\beta}{\sqrt{2}\lambda\alpha}-\sqrt{2}\lambda \right)\right\}.
\end{align*}
Putting $\theta_{s}=\dfrac{1}{\sqrt{2}\lambda}-\dfrac{e^{i\sigma_{+}}\beta}{\sqrt{2}\lambda\alpha}-\sqrt{2}\lambda$, we have
\begin{align*}
f^{L}_{+}(z)
&=-\dfrac{\alpha z\theta_{s}}{1+\theta_{s}z}
\\
&= -\alpha(\theta_{s}z) 
\left\{ 1+(-\theta_{s}z)+(-\theta_{s}z)^{2}+(-\theta_{s}z)^{3}+\cdots \right\}.
\end{align*}
Hence we see 
\begin{align}
f_{+}^{L}(z)=\alpha\sum_{x=1}^{\infty}(-\theta_{s}z)^{x}.
\label{panda12sou}
\end{align}
Equation \eqref{panda12sou} and the definition of $f_{+}^{L}(z)$ give
\begin{align}
\Psi^{L}(x)=\alpha(-\theta_{s})^{x}\;\;\;(x=1,2,\cdots),
\end{align}
where
\begin{align}
\theta_{s}=\dfrac{1}{\sqrt{2}\lambda}-\dfrac{e^{i\sigma_{+}}\beta}{\sqrt{2}\lambda\alpha}-\sqrt{2}\lambda.
\label{onene12sou}
\end{align}
\item Case of $f_{+}^{R}(z)$:
Putting $\theta_{s}= \dfrac{1}{\sqrt{2}\lambda}-\dfrac{e^{-i\sigma_{+}}\alpha}{\sqrt{2}\lambda\beta}$, we have from Eq.  \eqref{2phase-hodai1-houteisiki}
\begin{align}
f^{R}_{+}(z)=\beta\sum_{x=1}^{\infty}(-\theta_{s}z)^{x}.
\label{panda22sou}
\end{align}
Equation \eqref{panda22sou} and the definition of $f_{+}^{R}(z)$ imply
\begin{align}
\Psi^{R}(x)=\beta(-\theta_{s})^{x}\;\;\;(x=1,2,\cdots),
\end{align}
where
\begin{align}
\theta_{s} =\dfrac{1}{\sqrt{2}\lambda}-\dfrac{e^{-i\sigma_{+}}\alpha}{\sqrt{2}\lambda\beta}.
\label{onene22sou}
\end{align}
\item Case of $f_{-}^{L}(z)$:
Putting $\theta_{s}=\dfrac{\alpha+\beta e^{i\sigma_{+}}} {\sqrt{2}\lambda\left\{\alpha+\beta(e^{i\sigma_{+}}-e^{i\sigma_{-}})\right\}}$,
Eq. \eqref{2phase-hodai1-houteisiki} gives
\begin{eqnarray}
f^{L}_{-}(z)=\sum^{-\infty}_{x=-1}\left\{\alpha+\beta(e^{i\sigma_{+}}-e^{i\sigma_{-}})\right\}\theta_{s}^{|x|}z^{x}\label{panda32sou}.
\end{eqnarray}
Equation \eqref{panda32sou} and the definition of $f^{L}_{-}(z)$ yield
\begin{align*}
\Psi^{L}(x)= \left\{\alpha+\beta(e^{i\sigma_{+}}-e^{i\sigma_{-}})\right\}\theta_{s}^{|x|}\;\;\;(x=-1,-2,\cdots),
\end{align*}
where
\begin{align}
\theta_{s}=\dfrac{\alpha+\beta e^{i\sigma_{+}}}{\sqrt{2}\lambda\left\{\alpha+\beta(e^{i\sigma_{+}}-e^{i\sigma_{-}})\right\}}.
\label{onene32sou}
\end{align}

\item Case of $f_{-}^{R}(z)$: Putting
$
\theta_{s}=-\sqrt{2}\lambda+\dfrac{1}{\sqrt{2}e^{i\sigma_{-}}\lambda\beta}(\alpha+e^{i\sigma_{+}}\beta),
$
Eq. \eqref{2phase-hodai1-houteisiki} implies
\begin{align}
f^{R}_{-}(z)=\beta\sum^{-\infty}_{x=-1}\theta_{s}^{|x|}z^{x}.
\label{panda42sou}
\end{align}
Therefore, Eq. \eqref{panda42sou} and the definition of  $f^{R}_{-}(z)$ give
\begin{align*}
\Psi^{R}(x)=\beta\theta_{s}^{|x|}\;\;\;(x=-1,-2,\cdots),
\end{align*}
where
\begin{align}
\theta_{s}=-\sqrt{2}\lambda+\dfrac{1}{\sqrt{2}e^{i\sigma_{-}}\lambda\beta}(\alpha+e^{i\sigma_{+}}\beta).
\label{onene42sou}
\end{align}
\end{enumerate} 
Consequently, we obtain

\begin{align}
\Psi(x)=\left\{ \begin{array}{ll}
(-\theta_{s})^{x}
\begin{bmatrix}
\alpha \\ 
\beta
\end{bmatrix} &(x=0,1,2,\ldots),\\
\\
\theta_{s}^{|x|}
\begin{bmatrix}
\alpha + (e^{i \sigma_{+}} - e^{i \sigma_{-}})  \beta\\ \beta 
\end{bmatrix} &(x=-1,-2,\ldots).
\end{array} \right.
\label{araisan2sou}
\end{align}
From the above discussion, we obtain Proposition \ref{prob-meas}, the solutions of the generalized eigenequation \eqref{eigenvalueproblem} as follows:
\begin{proposition}
\label{prob-meas}
Let $\lambda^{(j)}$ be the eigenvalues of the unitary matrix $U^{(s)}$, and $\Psi^{(j)}(0)$ be the generalized eigenvector, with $j=1,2,3,4$.
Putting
\begin{align*}
p=1-e^{-4i\sigma}-4e^{-2i\sigma},\quad
q=1+e^{-4i\sigma}+6e^{-2i\sigma},\quad
r^{(\pm)}=1\pm e^{2i\sigma}
,\end{align*}
and
$c\in\mathbb{R}_{+}$, we obtain the solutions of the generalized eigenequation \eqref{2phase-eigenvalue.prob.} as follows:
\begin{enumerate}
\item For $\lambda^{(1)}=\sqrt{\dfrac{e^{2i\sigma}\{p+e^{-2i\sigma}r^{(-)}\sqrt{q}\}}{2(-e^{-2i\sigma}r^{(-)}-\sqrt{q})}}$ and $\Psi^{(1)}(0)={}^T\![\alpha,\;\beta]={}^T\!\dfrac{c}{\sqrt{2}}
\left[1,\;\dfrac{e^{-i\sigma_{-}}}{2}(r^{(-)} e^{-2i\sigma}+\sqrt{q})\right]$, we have
\begin{eqnarray*}\Psi^{L}(x)\!\!&=&\!\!\left\{
\begin{array}{ll}
\dfrac{c}{\sqrt{2}}\left(\dfrac{e^{-i\sigma}(r^{(+)}-e^{2i\sigma}\sqrt{q})}{\sqrt{-r^{(-)}-e^{2i\sigma}\sqrt{q}}\sqrt{e^{2i\sigma}p+r^{(-)}\sqrt{q}}}\right)^{x}&(x\geq0),\\
\dfrac{c}{2\sqrt{2}}\left\{2+(1-e^{-2i\sigma})(r^{(-)}+e^{2i\sigma}\sqrt{q})\right\}\left(-\dfrac{e^{-i\sigma}(r^{(+)}-e^{2i\sigma}\sqrt{q})}{\sqrt{-r^{(-)}-e^{2i\sigma}\sqrt{q}}\sqrt{pe^{2i\sigma}+r^{(-)}\sqrt{q}}}\right)^{|x|}&(x\leq-1).
\end{array}
\right.\\
\\
\Psi^{R}(x)\!\!&=&\!\!\left\{
\begin{array}{ll}
\dfrac{c}{2\sqrt{2}}e^{-i\sigma_{+}}(r^{(-)}+e^{2i\sigma}\sqrt{q})\left(\dfrac{e^{-i\sigma}(r^{(+)}-e^{2i\sigma}\sqrt{q})}{\sqrt{-r^{(-)}-e^{2i\sigma}\sqrt{q}}\sqrt{e^{2i\sigma}p+r^{(-)}\sqrt{q}}}\right)^{x}&(x\geq0),\\
\dfrac{c}{2\sqrt{2}}e^{-i\sigma_{+}}(r^{(-)}+e^{2i\sigma}\sqrt{q})\left(-\dfrac{e^{-i\sigma}(r^{(+)}-e^{2i\sigma}\sqrt{q})}{\sqrt{-r^{(-)}-e^{2i\sigma}\sqrt{q}}\sqrt{e^{2i\sigma}p+r^{(-)}\sqrt{q}}}\right)^{|x|}&(x\leq-1).
\end{array}
\right.
\end{eqnarray*}
\item For $\lambda^{(2)}=-\sqrt{\dfrac{e^{2i\sigma}\{p+e^{-2i\sigma}r^{(-)}\sqrt{q}\}}{2(-e^{-2i\sigma}r^{(-)}-\sqrt{q})}}$ and $\Psi^{(2)}(0)=\Psi^{(1)}(0)$, we have
\begin{eqnarray*}\Psi^{L}(x)\!\!&=&\!\!\left\{
\begin{array}{ll}
\dfrac{c}{\sqrt{2}}\left(-\dfrac{e^{-i\sigma}(r^{(+)}-e^{2i\sigma}\sqrt{q})}{\sqrt{-r^{(-)}-e^{2i\sigma}\sqrt{q}}\sqrt{e^{2i\sigma}p+r^{(-)}\sqrt{q}}}\right)^{x}&(x\geq0),\\
\dfrac{c}{2\sqrt{2}}\left\{2+(1-e^{-2i\sigma})(r^{(-)}+e^{2i\sigma}\sqrt{q})\right\}\left(\dfrac{e^{-i\sigma}(r^{(+)}-e^{2i\sigma}\sqrt{q})}{\sqrt{-r^{(-)}-e^{2i\sigma}\sqrt{q}}\sqrt{e^{2i\sigma}p+r^{(-)}\sqrt{q}}}\right)^{|x|}&(x\leq-1),
\end{array}
\right.\\
\\
\Psi^{R}(x)\!\!&=&\!\!\left\{
\begin{array}{ll}
\dfrac{c}{2\sqrt{2}}e^{-i\sigma_{+}}(r^{(-)}+e^{2i\sigma}\sqrt{q})\left(-\dfrac{e^{-i\sigma}(r^{(+)}-e^{2i\sigma}\sqrt{q})}{\sqrt{-r^{(-)}-e^{2i\sigma}\sqrt{q}}\sqrt{e^{2i\sigma}p+r^{(-)}\sqrt{q}}}\right)^{x}&(x\geq0),\\
\dfrac{c}{2\sqrt{2}}e^{-i\sigma_{+}}(r^{(-)}+e^{2i\sigma}\sqrt{q})\left(\dfrac{e^{-i\sigma}(r^{(+)}-e^{2i\sigma}\sqrt{q})}{\sqrt{-r^{(-)}-e^{2i\sigma}\sqrt{q}}\sqrt{e^{2i\sigma}p+r^{(-)}\sqrt{q}}}\right)^{|x|}&(x\leq-1).
\end{array}
\right.
\end{eqnarray*}
\item For $\lambda^{(3)}=\sqrt{\dfrac{e^{2i\sigma}\{p-e^{-2i\sigma}r^{(-)}\sqrt{q}\}}{2(-e^{-2i\sigma}r^{(-)}+\sqrt{q})}}$ and $\Psi^{(3)}(0)={}^T\![\alpha,\;\beta]={}^T\!\dfrac{c}{\sqrt{2}}\left[1,\;\dfrac{e^{-i\sigma_{-}}}{2}(r^{(-)} e^{-2i\sigma}-\sqrt{q})\right]$, we have
\begin{eqnarray*}\Psi^{L}(x)\!\!&=&\!\!\left\{
\begin{array}{ll}
\dfrac{c}{\sqrt{2}}\left(\dfrac{e^{-i\sigma}(r^{(+)}+e^{2i\sigma}\sqrt{q})}{\sqrt{-r^{(-)}+e^{2i\sigma}\sqrt{q}}\sqrt{e^{2i\sigma}p-r^{(-)}\sqrt{q}}}\right)^{x}&(x\geq0),\\
\dfrac{c}{2\sqrt{2}}\left\{2+(1-e^{-2i\sigma})(r^{(-)}-e^{2i\sigma}\sqrt{q})\right\}\left(-\dfrac{e^{-i\sigma}(r^{(+)}+e^{2i\sigma}\sqrt{q})}{\sqrt{-r^{(-)}+e^{2i\sigma}\sqrt{q}}\sqrt{e^{2i\sigma}p-r^{(-)}\sqrt{q}}}\right)^{|x|}&(x\leq-1),
\end{array}
\right.\\
\\
\Psi^{R}(x)\!\!&=&\!\!\left\{
\begin{array}{ll}
\dfrac{c}{2\sqrt{2}}e^{-i\sigma_{+}}(r^{(-)}-e^{2i\sigma}\sqrt{q})\left(\dfrac{e^{-i\sigma}(r^{(+)}+e^{2i\sigma}\sqrt{q})}{\sqrt{-r^{(-)}+e^{2i\sigma}\sqrt{q}}\sqrt{e^{2i\sigma}p-r^{(-)}\sqrt{q}}}\right)^{x}&(x\geq0),\\
\dfrac{c}{2\sqrt{2}}e^{-i\sigma_{+}}(r^{(-)}-e^{2i\sigma}\sqrt{q})\left(-\dfrac{e^{-i\sigma}(r^{(+)}+e^{2i\sigma}\sqrt{q})}{\sqrt{-r^{(-)}+e^{2i\sigma}\sqrt{q}}\sqrt{e^{2i\sigma}p-r^{(-)}\sqrt{q}}}\right)^{|x|}&(x\leq-1).
\end{array}
\right.
\end{eqnarray*}
\item For $\lambda^{(4)}=-\sqrt{\dfrac{e^{2i\sigma}\{p-e^{-2i\sigma}r^{(-)}\sqrt{q}\}}{2(-e^{-2i\sigma}r^{(-)}+\sqrt{q})}}$ and $\Psi^{(4)}(0)=\Psi^{(3)}(0)$, we have
\begin{eqnarray*}\Psi^{L}(x)\!\!&=&\!\!\left\{
\begin{array}{ll}
\dfrac{c}{\sqrt{2}}\left(-\dfrac{e^{-i\sigma}(r^{(+)}+e^{2i\sigma}\sqrt{q})}{\sqrt{-r^{(-)}+e^{2i\sigma}\sqrt{q}}\sqrt{e^{2i\sigma}p-r^{(-)}\sqrt{q}}}\right)^{x}&(x\geq0),\\
\dfrac{c}{2\sqrt{2}}\left\{2+(1-e^{-2i\sigma})(r^{(-)}-e^{2i\sigma}\sqrt{q}\right\}\left(\dfrac{e^{-i\sigma}(r^{(+)}+e^{2i\sigma}\sqrt{q})}{\sqrt{-r^{(-)}+e^{2i\sigma}\sqrt{q}}\sqrt{e^{2i\sigma}p-r^{(-)}\sqrt{q}}}\right)^{|x|}&(x\leq-1),
\end{array}
\right.\\
\\
\Psi^{R}(x)\!\!&=&\!\!\left\{
\begin{array}{ll}
\dfrac{c}{2\sqrt{2}}e^{-i\sigma_{+}}(r^{(-)}-e^{2i\sigma}\sqrt{q})\left(-\dfrac{e^{-i\sigma}(r^{(+)}+e^{2i\sigma}\sqrt{q})}{\sqrt{-r^{(-)}+e^{2i\sigma}\sqrt{q}}\sqrt{e^{2i\sigma}p-r^{(-)}\sqrt{q}}}\right)^{x}&(x\geq0),\\
\dfrac{c}{2\sqrt{2}}e^{-i\sigma_{+}}(r^{(-)}-e^{2i\sigma}\sqrt{q})\left(\dfrac{e^{-i\sigma}(r^{(+)}+e^{2i\sigma}\sqrt{q})}{\sqrt{-r^{(-)}+e^{2i\sigma}\sqrt{q}}\sqrt{e^{2i\sigma}p-r^{(-)}\sqrt{q}}}\right)^{|x|}&(x\leq-1).
\end{array}
\right.
\end{eqnarray*}
\end{enumerate}
\end{proposition}
Here $4$ expressions of $\theta_{s}$, that is, Eqs. \eqref{onene12sou}, \eqref{onene22sou},
\eqref{onene32sou}, and \eqref{onene42sou} provide
\begin{eqnarray*}
\theta_{s}
\!\!\!&=&\!\!\!\dfrac{\alpha-2\lambda^{2}\alpha  - e^{i \sigma_{+}} \beta}{ \sqrt{2} \lambda\alpha} =  \dfrac{ \beta- e^{-i \sigma_{+}}  \alpha}{ \sqrt{2}\lambda\beta}\\
\!\!\!&=&\!\!\!\dfrac{\alpha+\beta e^{i\sigma_{+}}}{\sqrt{2}\lambda\left\{\alpha+\beta(e^{i\sigma_{+}}-e^{i\sigma_{-}})\right\}}=-\sqrt{2}\lambda+\dfrac{1}{\sqrt{2}e^{i\sigma_{-}}\lambda\beta}(\alpha+e^{i\sigma_{+}}\beta), 
\end{eqnarray*}
which leads to the conditions of $\lambda^{(j)}$ and $\Psi^{(j)}(0)\;(j=1,2, 3, 4)$ in Proposition \ref{prob-meas} and Theorem \ref{statmeasure}.
Noting that the stationary measure is defined by $\mu(x)=|\Psi^{R}(x)|^{2}+|\Psi^{L}(x)|^{2}\;(x\in\mathbb{R})$, we arrive at Theorem \ref{statmeasure}.
\\
\flushleft
\noindent {\large{\bf Appendix C}}  \\ \noindent
\label{proof_theo}
In Appendix C, we give the proof of Theorem \ref{weaklimit} in a similar way as Appendix B in Ref. \cite{ekymj}.  Now we consider the characteristic function of QW:
\begin{align}E\left[e^{i\xi \frac{X_{t}}{t}}\right]=\int_{x\in\mathbb{Z}}g_{X_{t}/t}(x)e^{i\xi x}dx,\label{density_function}\end{align}
where $g_{X_{t}/t}(x)$ is the density function of random variable $X_{t}/t$. Hereafter, we rewrite $E\left[e^{iX_{t}/t}\right]\;(t\to\infty)$ to obtain the explicit expression of $w(x)f_{K}(x;1/\sqrt{2})$. By a simple argument, we obtain 
\begin{proposition}
\label{keyrelation}
\begin{align}
E\left[e^{i\xi\frac{X_{t}}{t}}\right]\to\int^{2\pi}_{0}\sum_{\theta\in A}e^{-i\xi \theta^{'}(k)}\left\|Res(\hat{\tilde{\Xi}}(k:z): z=e^{i\theta(k)})\right\|^{2}\frac{dk}{2\pi}\qquad(t\to\infty),\label{limitdensity}
\end{align}
where $A$ is the set of the singular points of $\hat{\tilde{\Xi}}(k:z)\equiv\sum_{x\in\mathbb{Z}}\tilde{\Xi}_{x}(z)e^{ikx}$ with $\tilde{\Xi}_{x}(z)=\sum_{t}\Xi_{t}(x)z^{t}$. Note $\theta^{'}(k)=\partial \theta(k)/\partial k$. 
\end{proposition}
The proof of Proposition \ref{keyrelation} is given in Ref. \cite{maman}.
By mainly using Proposition \ref{keyrelation}, we prove Theorem \ref{weaklimit}.

First of all, we derive the singular points of $\hat{\tilde{\Xi}}(k:z)$ and then, compute the residues of $\hat{\tilde{\Xi}}(k:z)$ at the singular points.
By Lemma \ref{kls-2phase}, we can rewrite $\hat{\tilde{\Xi}}(k:z)$ as
\begin{align}\hat{\tilde{\Xi}}(k:z)=&\left\{\dfrac{e^{ik}}{1-e^{ik}\tilde{\lambda}^{(+)}(z)}\begin{bmatrix}\tilde{\lambda}^{(+)}(z)\tilde{f}_{0}^{(+)}(z)\\ z\end{bmatrix}\left[\dfrac{e^{-i\sigma_{+}}}{\sqrt{2}},-\dfrac{1}{\sqrt{2}}\right] \right.\nonumber\\ &\hspace{30mm}\left.+\dfrac{e^{-ik}}{1-e^{-ik}\tilde{\lambda}^{(-)}(z)}\begin{bmatrix}z\\ \tilde{\lambda}^{(-)}(z)\tilde{f}_{0}^{(-)}(z)\end{bmatrix}\left[\dfrac{1}{\sqrt{2}},\dfrac{e^{i\sigma_{+}}}{\sqrt{2}}\right]
+I\right\} \tilde{\Xi}_{0}(z).\label{french}\end{align}
Note that if $|z|<1$, then $|\tilde{\lambda}^{(\pm)}(z)|<1$ holds, and the infinite series $\sum_{x}(\tilde{\lambda}^{(+)}(z))^{|x|-1}e^{ikx}$ and $\sum_{x}(\tilde{\lambda}^{(-)}(z))^{|x|-1}e^{-ikx}$ converge.
According to Ref. \cite{segawa}, we have by taking $z=(1-\varepsilon)e^i{\theta}
$ with $\varepsilon\downarrow 0$,
\begin{align}
\left\{
\begin{array}{l}
\tilde{\lambda}^{(\pm)}(e^{i\theta})=\mp\{\operatorname{sgn}(\cos\theta)\sqrt{2\cos^{2}\theta-1}+i\sqrt{2}\sin\theta\},\\
\\
\tilde{f}_{0}^{(\pm)}(e^{i\theta})=\operatorname{sgn}(\cos\theta)e^{i(\theta\pm\sigma_{\pm})}\{\sqrt{2}|\cos\theta|-\sqrt{2\cos^{2}\theta-1}\},
\end{array}
\right.\label{nippon}
\end{align}
which can be derived in a similar way as relation (4.25) in Ref. \cite{ekymj}.
The singular points derived from $\tilde{\Xi}_{0}(z)$ are related with localization, while
principal singular points for weak convergence come from 
\begin{eqnarray}1-e^{ik}\tilde{\lambda}^{(+)}(z)=0,\label{eqt.1}\end{eqnarray}and
\begin{eqnarray}1-e^{-ik}\tilde{\lambda}^{(-)}(z)=0.\label{eqt.2}\end{eqnarray}
\noindent
For Eq. \eqref{eqt.1}, we see 
\begin{eqnarray}\cos k=-\operatorname{sgn}(\cos\theta^{(+)}(k))\sqrt{2\cos^{2}\theta^{(+)}(k)-1},\label{cosk+}\end{eqnarray}
\begin{eqnarray}\sin k=\sqrt{2}\sin\theta^{(+)}(k),\label{sink+}\end{eqnarray}
and for Eq. \eqref{eqt.2}, we have
\begin{eqnarray}\cos k=\operatorname{sgn}(\cos\theta^{(-)}(k)(k))\sqrt{2\cos^{2}\theta^{(-)}(k)-1},\label{cosk-}\end{eqnarray}
\begin{eqnarray}\sin k=\sqrt{2}\sin\theta^{(-)}(k).\label{sink-}\end{eqnarray}
\noindent
Put $-\partial\theta^{(\pm)}(k)/\partial k=x_{\pm}$ to compute the RHS of Eq. \eqref{limitdensity}.
Derivating Eqs. \eqref{cosk+} and \eqref{cosk-} with respect to $k$, we obtain $\sin k,\;\cos k,\;\sin\theta^{(\pm)}(k)$, and $\cos\theta^{(\pm)}(k)$ as follows:
Equations \eqref{cosk+} and \eqref{sink+} give
\begin{eqnarray}
\left\{
\begin{array}{l}
\cos k=\operatorname{sgn}(\cos k)\dfrac{x_{+}}{\sqrt{1-x_{+}^{2}}},\;\cos\theta^{(+)}(k)=-\operatorname{sgn}(\cos k)\dfrac{1}{\sqrt{2(1-x_{+}^{2})}},\\
\\
\sin k=\operatorname{sgn}(\sin k)\sqrt{\dfrac{1-2x_{+}^{2}}{1-x_{+}^{2}}},\;\;\sin\theta^{(+)}(k)=\operatorname{sgn}(\sin k)\sqrt{\dfrac{1-2x_{+}^{2}}{2(1-x_{+}^{2})}}.
\end{array}
\right.\label{solutions+}
\end{eqnarray} 
Equations \eqref{cosk-} and \eqref{sink-} provide
\begin{eqnarray}
\left\{
\begin{array}{l}
\cos k=\operatorname{sgn}(\cos k)\dfrac{x_{-}}{\sqrt{1-x_{-}^{2}}},\;\cos\theta^{(-)}(k)=\operatorname{sgn}(\cos k)\dfrac{1}{\sqrt{2(1-x_{-}^{2})}},\\
\sin k=\operatorname{sgn}(\sin k)\sqrt{\dfrac{1-2x_{-}^{2}}{1-x_{-}^{2}}},\;\;\sin\theta^{(-)}(k)=\operatorname{sgn}(\sin k)\sqrt{\dfrac{1-2x_{-}^{2}}{2(1-x_{-}^{2})}}.
\end{array}
\right.\label{solutions-}
\end{eqnarray} 
Thereby, we obtain $A$, the set of the singular points of $\hat{\tilde{\Xi}}(k:z)$:
\[A=\{e^{i\theta^{(+)}(k)},e^{i\theta^{(-)}(k)}\},\]
with
\[e^{i\theta^{(+)}(k)}=-\dfrac{\operatorname{sgn}(\cos k)}{\sqrt{2(1-x^{2}_{+})}}+i\operatorname{sgn}(\sin k)\sqrt{\dfrac{1-2x^{2}_{+}}{2(1-x^{2}_{+})}},\]
and 
\[e^{i\theta^{(-)}(k)}=\dfrac{\operatorname{sgn}(\cos k)}{\sqrt{2(1-x^{2}_{-})}}+i\operatorname{sgn}(\sin k)\sqrt{\dfrac{1-2x^{2}_{-}}{2(1-x^{2}_{-})}}.\]

Next, we compute the residue of $\hat{\tilde{\Xi}}(k;z)$ at $e^{i\theta^{(\pm)}(k)}$.
Substituting the singular points to $\tilde{f}_{0}^{(\pm)}(z)$, we get\\

\begin{enumerate}
\item $\tilde{f}_{0}^{(+)}(e^{i\theta^{(+)}(k)})=-\operatorname{sgn}(\cos k)e^{i(\theta^{+}(k)+\sigma_{+})}\dfrac{\sqrt{1-x_{\pm}^{2}}}{1+|x_{\pm}|},$\quad $\tilde{f}_{0}^{(-)}(e^{i\theta^{(+)}(k)})=-\operatorname{sgn}(\cos k)e^{i(\theta^{(+)}(k)-\sigma_{-})}\dfrac{\sqrt{1-x_{\pm}^{2}}}{1+|x_{\pm}|}$,\quad
\item $\tilde{f}_{0}^{(+)}(e^{i\theta^{(-)}(k)})=\operatorname{sgn}(\cos k)e^{i(\theta^{(-)}(k)+\sigma_{+})}\dfrac{\sqrt{1-x_{\pm}^{2}}}{1+|x_{\pm}|},$\quad $\tilde{f}_{0}^{(-)}(e^{i\theta^{(-)}(k)})=\operatorname{sgn}(\cos k)e^{i(\theta^{(-)}(k)-\sigma_{-})}\dfrac{\sqrt{1-x_{\pm}^{2}}}{1+|x_{\pm}|}$.\\
\end{enumerate}
Taking into account Lemma \ref{kls-2phase}, we have
\begin{eqnarray*}\frac{e^{ik}}{1-e^{ik}\tilde{\lambda}^{(+)}(z)}\!\!\!\!\!\!\!\!&&\!\!\!\!\!\!\!\!\begin{bmatrix}\tilde{f}_{0}^{(+)}(z)\tilde{\lambda}^{(+)}(z)\\ z\end{bmatrix}\left[\dfrac{e^{-i\sigma_{+}}}{\sqrt{2}},\;-\dfrac{1}{\sqrt{2}}\right]\tilde{\Xi}_{0}(z)\\
\!\!\!&=&\!\!\!\frac{1}{\tilde{\Lambda}_{0}(z)}\frac{e^{ik}}{1-e^{ik}\tilde{\lambda}^{(+)}(z)}\begin{bmatrix}\tilde{f}_{0}^{(+)}(z)\tilde{\lambda}^{(+)}(z)\\ z\end{bmatrix}
\dfrac{1}{\sqrt{2}}\left\{\alpha\left(e^{-i\sigma_{+}}-\sqrt{2}\tilde{f}_{0}^{(-)}(z)\right)-\beta\right\},\end{eqnarray*}
and the square norm of residue of the first term of Eq. \eqref{french} is written by
\begin{align*}\left|Res\left(\dfrac{e^{ik}}{1-e^{ik}\tilde{\lambda}^{(+)}(z)}\begin{bmatrix}\tilde{f}_{0}^{(+)}(z)\tilde{\lambda}^{(+)}(z)\\ z\end{bmatrix}\left[\dfrac{e^{-i\sigma_{+}}}{\sqrt{2}},\;-\dfrac{1}{\sqrt{2}}\right]\tilde{\Xi}_{0}(z): z=e^{i\theta^{(+)}(k)}\right)\right|^{2}\hspace{40mm}\\
=\left|Res\left(\dfrac{1}{1-e^{ik}\tilde{\lambda}^{(+)}(z)}: z=e^{i\theta^{(+)}(k)}\right)\right|^{2}\left|\begin{bmatrix}\tilde{f}_{0}^{(+)}(e^{i\theta^{(+)}(k)})\tilde{\lambda}^{(+)}(e^{i\theta^{(+)}(k)})\\ e^{i\theta^{(+)}(k)}\end{bmatrix} \right|^{2}\\\times\dfrac{1}{2\left|\tilde{\Lambda}_{0}(e^{i\theta^{(+)}(k)})
\right|^{2}}
\left|\alpha\left(e^{-i\sigma_{+}}-\sqrt{2}\tilde{f}_{0}^{(-)}(e^{i\theta^{(+)}(k)})\right)-\beta\right|^{2}.
\end{align*}
In a similar fashion, the square norm of residue of the second term of Eq. \eqref{french} becomes
\begin{align*}
\left|Res\left(\dfrac{e^{-ik}}{1-e^{-ik}\tilde{\lambda}^{(-)}(z)}\begin{bmatrix}z\\ \tilde{f}_{0}^{(-)}(z)\tilde{\lambda}^{(-)}(z)\end{bmatrix}\left[\dfrac{1}{\sqrt{2}}, \dfrac{e^{i\sigma_{+}}}{\sqrt{2}}\right]\tilde{\Xi}_{0}(z): z=e^{i\theta^{(-)}(k)}\right)\right|^{2}\hspace{40mm}\\
=\left|Res\left(\dfrac{1}{1-e^{-ik}\tilde{\lambda}^{(-)}(z)}: z=e^{i\theta^{(-)}(k)}\right)\right|^{2}\left|\begin{bmatrix}e^{i\theta^{(-)}(k)}\\ \tilde{f}_{0}^{(-)}(e^{i\theta^{(-)}(k)}) \tilde{\lambda}^{(-)}(e^{i\theta^{(-)}(k)})\end{bmatrix}
\right|^{2}\\\times\dfrac{1}{2\left|\tilde{\Lambda}_{0}(e^{i\theta^{(-)}(k)})\right|^{2}}\left|\alpha+\beta\left(e^{i\sigma_{+}}-\sqrt{2}\tilde{f}^{(+)}(e^{i\theta^{(-)}(k)})\right)\right|^{2}.\end{align*}
Thereby, we obtain
\begin{eqnarray}
\left\|Res(\hat{\tilde{\Xi}}(k:z): z=e^{i\theta^{(\pm)}(k)})\right\|^{2}
&=\left|Res\left(\dfrac{1}{1-e^{ik}\tilde{\lambda}^{(+)}(z)}: z=e^{i\theta^{(+)}(k)}\right)\right|^{2}\left|\begin{bmatrix}\tilde{f}_{0}^{(+)}(e^{i\theta^{(+)}(k)})\tilde{\lambda}^{(+)}(e^{i\theta^{(+)}(k)})\\ e^{i\theta^{(+)}(k)}\end{bmatrix} \right|^{2}\nonumber\\
&\times\dfrac{1}{2\left|\tilde{\Lambda}_{0}(e^{i\theta^{(+)}(k)})
\right|^{2}}
\left|\alpha\left(e^{-i\sigma_{+}}-\sqrt{2}\tilde{f}_{0}^{(-)}(e^{i\theta^{(+)}(k)})\right)-\beta\right|^{2}\nonumber\\
&+\left|Res\left(\dfrac{1}{1-e^{-ik}\tilde{\lambda}^{(-)}(z)}: z=e^{i\theta^{(-)}(k)}\right)\right|^{2}\left|\begin{bmatrix}e^{i\theta^{(-)}(k)}\\ \tilde{f}_{0}^{(-)}(e^{i\theta^{(-)}(k)}) \tilde{\lambda}^{(-)}(e^{i\theta^{(-)}(k)})\end{bmatrix}
\right|^{2}\nonumber\\
&\times\dfrac{1}{2\left|\tilde{\Lambda}_{0}(e^{i\theta^{(-)}(k)})\right|^{2}}\left|\alpha+\beta\left(e^{i\sigma_{+}}-\sqrt{2}\tilde{f}^{(+)}(e^{i\theta^{(-)}(k)})\right)\right|^{2}.
\label{kekka}
\label{mare}\end{eqnarray}
Hereafter, we will write the items below with respect to $x_{+}$ or $x_{-}$, and then substitute those in Eq. \eqref{kekka}.

\begin{itemize}
\item $\left|Res\left(\dfrac{1}{1-e^{ik}\tilde{\lambda}^{(+)}(z)}: z=e^{i\theta^{(+)}(k)}\right)\right|^{2}$ and $\left|Res\left(\dfrac{1}{1-e^{-ik}\tilde{\lambda}^{(-)}(z)}: z=e^{i\theta^{(-)}(k)}\right)\right|^{2}$.\\
\item $\dfrac{1}{\left|\tilde{\Lambda}_{0}(e^{i\theta^{(\pm)}(k)})\right|^{2}}$.\\
\item $\dfrac{1}{2}\left|\alpha\left(e^{-i\sigma_{+}}-\sqrt{2}\tilde{f}_{0}^{(-)}(e^{i\theta^{(+)}(k)})\right)-\beta\right|^{2}$ and $\dfrac{1}{2}\left|\alpha+\beta\left(e^{i\sigma_{+}}-\sqrt{2}\tilde{f}^{(+)}(e^{i\theta^{(-)}(k)})\right)\right|^{2}$.\\
\item $\left\|\begin{bmatrix}\tilde{\lambda}^{(+)}(e^{i\theta^{(+)}(k)})\tilde{f}_{0}^{(+)}(e^{i\theta^{(+)}(k)})\\ e^{i\theta^{(+)}(k)}\end{bmatrix}\right\|^{2}$ and $\left\|\begin{bmatrix}e^{i\theta^{(-)}(k)}\\ \tilde{\lambda}^{(-)}(e^{i\theta^{(-)}(k)})\tilde{f}_{0}^{(-)}(e^{i\theta^{(-)}(k)})\end{bmatrix}\right\|^{2}$.\\
\end{itemize}
\noindent
(I) Derivation of $\left|Res\left(\dfrac{1}{1-e^{ik}\tilde{\lambda}^{(+)}(z)}: z=e^{i\theta^{(+)}(k)}\right)\right|^{2}$ and $\left|Res\left(\dfrac{1}{1-e^{-ik}\tilde{\lambda}^{(-)}(z)}: z=e^{i\theta^{(-)}(k)}\right)\right|^{2}$:\\
Putting $g^{(\pm)}(z)=1-e^{\pm ik}\tilde{\lambda}^{(\pm)}(z)$, and we have
\[Res\left(\frac{1}{1-e^{\pm ik}\tilde{\lambda}^{(\pm)}(z)}: z=e^{i\theta^{(\pm)}(k)}\right)=\left.\frac{1}{\dfrac{\partial g^{(\pm)}(z)}{\partial z}}\right|_{z=e^{i\theta^{(\pm)}(k)}}.\]
Owing to Eq. \eqref{nippon}, we see
\[\left.\dfrac{\partial g^{(\pm)}(z)}{\partial z}\right|_{z=e^{i\theta^{(\pm)}(k)}}=\mp i\dfrac{\operatorname{sgn}(\cos k)}{\sqrt{1-x_{\pm}^{2}}}e^{-i(\theta^{(\pm)}(k)\mp k)}\left\{\operatorname{sgn}(\cos k\sin k)\dfrac{\sqrt{1-2x_{\pm}^{2}}}{x_{\pm}}+i\right\},\]
which lead to
\begin{eqnarray}
\left|Res\left(\dfrac{1}{1-e^{ik}\tilde{\lambda}^{(+)}(z)}: z=e^{i\theta^{(\pm)}(k)}\right)\right|^{2}=x_{+}^{2},\quad
\left|Res\left(\dfrac{1}{1-e^{-ik}\tilde{\lambda}^{(-)}(z)}: z=e^{i\theta^{(\pm)}(k)}\right)\right|^{2}=x_{-}^{2}.
\end{eqnarray}
(I\hspace{-.1em}I) Derivation of $1/\left|\tilde{\Lambda}_{0}(e^{i\theta^{(\pm)}(k)})\right|^{2}$:
Taking into account Lemma \ref{kls-2phase}, we have 
\begin{align}
\left|\tilde{\Lambda}_{0}(e^{i\theta})\right|^{2}&=1+\left|\tilde{f}_{0}^{(+)}(e^{i\theta})\right|^{2}+\left|\tilde{f}_{0}^{(+)}(e^{i\theta})\right|^{2}\left|\tilde{f}_{0}^{(-)}(e^{i\theta})\right|^{2}-\sqrt{2}\Re\left\{e^{-i\sigma_{+}}\tilde{f}_{0}^{(+)}(e^{i\theta})\right\}-\sqrt{2}\Re\left\{e^{i\sigma_{+}}\tilde{f}_{0}^{(-)}(e^{i\theta})\right\}\nonumber\\
&+2\Re\left\{\tilde{f}_{0}^{(+)}(e^{i\theta})\tilde{f}_{0}^{(-)}(e^{i\theta})\right\}+\Re\{e^{-2i\sigma_{+}}\tilde{f}_{0}^{(+)}(e^{i\theta})\overline{\tilde{f}_{0}^{(-)}(e^{i\theta})}\}
-\sqrt{2}\Re\left\{e^{-i\sigma_{+}}\left|\tilde{f}_{0}^{(+)}(e^{i\theta})\right|^{2}\overline{\tilde{f}_{0}^{(-)}(e^{i\theta})}\right\}\nonumber\\
&-\sqrt{2}\Re\left\{e^{i\sigma_{+}}\left|\tilde{f}_{0}^{(-)}(e^{i\theta})\right|^{2}\overline{\tilde{f}_{0}^{(+)}(e^{i\theta})}\right\},\label{tokyo}
\end{align}
for $\theta\in\mathbb{R}$. Thereby, substituting the singular points into Eq. \eqref{tokyo}, we obtain
\begin{eqnarray}
\left|\dfrac{1}{\tilde{\Lambda}_{0}(e^{i\theta^{(\pm)}(k)})}\right|^{2}=\dfrac{(1\pm x_{\pm})^{2}}{2(\sin^{2}\sigma+x_{\pm}^{2}\cos 2\sigma)}.
\end{eqnarray}
(I\hspace{-.1em}I\hspace{-.1em}I) Derivation of $\left|\alpha\left(e^{-i\sigma_{+}}-\sqrt{2}\tilde{f}_{0}^{(-)}(e^{i\theta^{(+)}(k)})\right)-\beta\right|^{2}/2$ and $\left|\alpha+\beta\left(e^{i\sigma_{+}}-\sqrt{2}\tilde{f}_{0}^{(+)}(e^{i\theta^{(-)}(k)})\right)\right|^{2}/2$:\\
Let the initial coin state be $\varphi_{0}={}^T\![\alpha,\beta]$, where $\alpha=ae^{i\phi_{1}},\;\beta=be^{i\phi_{2}}$ with $a,b\in\mathbb{R}$, $a^{2}+b^{2}=1$, and $\phi_{j}\in\mathbb{R}\;(j=1,2)$. 
Taking account of
\begin{eqnarray*}\dfrac{1}{2}\left|\alpha\left(e^{-i\sigma_{+}}-\sqrt{2}\tilde{f}_{0}^{(-)}(e^{i\theta^{(+)}(k)})\right)-\beta\right|^{2}=\dfrac{1}{2}+a^{2}\left|\tilde{f}_{0}^{(-)}(e^{i\theta^{(+)}(k)})\right|^{2}-\sqrt{2}a^{2}\Re\{e^{i\sigma_{+}}\tilde{f}_{0}^{(-)}(e^{i\theta^{(+)}(k)})\}\\-\Re\left\{abe^{i\tilde{\phi_{12}}}\left(e^{-i\sigma_{+}} -\sqrt{2}\tilde{f}_{0}^{(-)}(e^{i\theta^{(+)}(k)})\right)\right\},\end{eqnarray*}
and
\begin{eqnarray*}\dfrac{1}{2}\left|\alpha+\beta\left(e^{i\sigma_{+}}-\sqrt{2}\tilde{f}^{(+)}(e^{i\theta^{(-)}(k)})\right)\right|^{2}=\dfrac{1}{2}+b^{2}\left|\tilde{f}^{(+)}(e^{i\theta^{(-)}(k)})\right|^{2}-\sqrt{2}b^{2}
\Re\{e^{-i\sigma_{+}}\tilde{f}^{(+)}(e^{i\theta^{(-)}(k)})\}\\+\Re\left\{ab e^{i\tilde{\phi_{21}}}\left(e^{i\sigma_{+}}-\sqrt{2}\tilde{f}^{(+)}(e^{i\theta^{(-)}(k)})\right)\right\},\end{eqnarray*}
we have
\begin{eqnarray}
\left\{
\begin{array}{l}
\dfrac{1}{2}\left|\alpha\left(e^{-i\sigma_{+}}-\sqrt{2}\tilde{f}_{0}^{(-)}(e^{i\theta^{(+)}(k)})\right)-\beta\right|^{2}
=\dfrac{1}{2}+a^{2}\dfrac{1-x_{+}}{1+x_{+}}-ab\cos\gamma_{-}\\\\-\dfrac{a}{1+x_{+}}\left\{a\left(\cos2\sigma+\operatorname{sgn}(\sin k\cos k)\sqrt{1-2x_{+}^{2}}\sin2\sigma\right)-b\left(\cos\gamma_{+}+\operatorname{sgn}(\sin k\cos k)\sqrt{1-2x_{+}^{2}}\sin\gamma_{+}\right)\right\},\\
\\
\dfrac{1}{2}\left|\alpha+\beta\left(e^{i\sigma_{+}}-\sqrt{2}\tilde{f}^{(+)}(e^{i\theta^{(-)}(k)})\right)\right|^{2}=\dfrac{1}{2}+b^{2}\dfrac{x_{-}}{1-x_{-}}-\dfrac{ab}{1-x_{-}}\left\{x_{-}\cos\gamma_{-}-\operatorname{sgn}(\sin k\cos k)\sin\gamma_{-}\sqrt{1-2x_{-}^{2}}\right\},
\end{array}
\right.
\end{eqnarray} 
where
$\gamma_{+}=\tilde{\phi}_{12}-\sigma_{-}$ and $\gamma_{-}=\tilde{\phi}_{21}+\sigma_{+}$ with $\tilde{\phi}_{12}=\phi_{1}-\phi_{2}$.\\
(I\hspace{-.1em}V) Derivation of $\left\|\begin{bmatrix}\tilde{\lambda}^{(+)}(e^{i\theta^{(+)}(k)})\tilde{f}_{0}^{(+)}(e^{i\theta^{(+)}(k)})\\ 
e^{i\theta^{(+)}(k)}\end{bmatrix}\right\|^{2}$ and $\left\|\begin{bmatrix}e^{i\theta^{(-)}(k)}\\ \tilde{\lambda}^{(-)}(e^{i\theta^{(-)}(k)})\tilde{f}_{0}^{(-)}(e^{i\theta^{(-)}(k)})\end{bmatrix}\right\|^{2}$:\\
By a simple computation, we have
\begin{eqnarray}
\left\{
\begin{array}{ll}
\left\|\begin{bmatrix}\tilde{\lambda}^{(+)}(e^{i\theta^{(+)}(k)})\tilde{f}_{0}^{(+)}(e^{i\theta^{(+)}(k)})\\ e^{i\theta^{(+)}(k)}\end{bmatrix}\right\|^{2}=\left|\tilde{\lambda}^{(+)}(e^{i\theta^{(+)}(k)})\right|^{2}\left|\tilde{f}_{0}^{(+)}(e^{i\theta^{(+)}(k)})\right|^{2}+1=\dfrac{2}{1+x_{+}}&(x_{+}>0),\\
\\
\left\|\begin{bmatrix}e^{i\theta^{(-)}(k)}\\ \tilde{\lambda}^{(-)}(e^{i\theta^{(-)}(k)})\tilde{f}_{0}^{(-)}(e^{i\theta^{(-)}(k)})\end{bmatrix}\right\|^{2}=1+\left|\tilde{\lambda}^{(-)}(e^{i\theta^{(-)}(k)})\right|^{2}\left|\tilde{f}_{0}^{(-)}(e^{i\theta^{(-)}(k)})\right|^{2}=\dfrac{2}{1-x_{-}}&(x_{-}<0).
\end{array}
\right.
\end{eqnarray}
Remark 
\begin{align}-\frac{\partial\theta^{(\pm)}(k)}{\partial k}=x_{\pm},\label{hensu}\end{align}
which give
\begin{align}x_{+}=\frac{|\cos k|}{\sqrt{1+\cos^{2}k}},\qquad
x_{-}=-\frac{|\cos k|}{\sqrt{1+\cos^{2}k}}.\label{x}\end{align}
Henceforth, we can treat $x_{+}$ and $x_{-}$ as a valuable $x$:
\[
x=\left\{ \begin{array}{ll}
x_{+} & (x>0),\\
x_{-} & (x<0).
\end{array} \right.\] 
Combining Eqs. \eqref{solutions+} and \eqref{solutions-} with Eq. \eqref{x}, and noting Eq. \eqref{hensu}, we see
\[\frac{dx}{dk}=-\operatorname{sgn}(x)\operatorname{sgn}(\sin k\cos k)(1-x^{2})\sqrt{1-2x^{2}},\]
and thereby, we obtain
\begin{align}
dk=\left\{ \begin{array}{ll}
-\operatorname{sgn}(\sin k\cos k)f_{K}(x;1/\sqrt{2})\pi dx&(x>0), \\
\operatorname{sgn}(\sin k\cos k)f_{K}(x;1/\sqrt{2})\pi dx& (x<0).
\end{array} \right.
\end{align} 
\noindent
Substituting the items given in (I) to (I\hspace{-.1em}V) into Eq. \eqref{mare} and combining with Eq. \eqref{limitdensity}, we obtain Theorem \ref{weaklimit}.

\flushleft
\noindent {\large{\bf Appendix D}}  \\ \noindent
In Appendix D, we derive Eqs.\ (\ref{eq:TRS}) - (\ref{eq:chiral}).

It is known that the relevant symmetries for topological phases require Hamiltonian $H$ to satisfy
 the following relations\cite{bernevig13,chiu15,hasan10,qi11}:
\begin{subequations}
\begin{eqnarray}
 T\, H\, T^{-1}\quad\,\,\,\, =&
  +H &\text{  (Time-reversal symmetry)},
\label{eq:TRS H}\\ 
 P\, (H-E_P)\, P^{-1}\, =&
 -(H-E_P) & \text{  (Particle-hole symmetry)},
\label{eq:PHS H}\\ 
\Gamma\, (H-E_\Gamma)\, \Gamma^{-1} =&
  -(H-E_\Gamma) &\text{  (Chiral symmetry) }.
\label{eq:chiral H}
\end{eqnarray}
\label{eq:symmetry H}
\end{subequations}
Here, the operators $T$ and $P$ are anti-unitary operators
(i.e., they should contain a complex conjugate operator $K$), while
$\Gamma$ is a unitary one.
Therefore,
\begin{equation*}
 T^2=\pm 1, \quad P^2=\pm 1, \quad \Gamma^2=+1.
\end{equation*}
In Eqs.\ (\ref{eq:PHS H}) and (\ref{eq:chiral H}), we assume that the
Hamiltonian $H$ satisfies the eigenvalue equation $(H-E_X) {\bm v} =
E{\bm v}$, where $X$ stands for $P$ or $\Gamma$, with an eigenvector ${\bm v}$ and $E, E_X \in \mathbb{R}$. If so, Eqs.\ (\ref{eq:PHS H}) and (\ref{eq:chiral H})
guarantee that $(H-E_X) \cdot X {\bm v} = -E\cdot X {\bm v}$, where $X$
stands for $P$ or $\Gamma$.  Thereby, a pair of eigenvalues  with
opposite signs $\pm E$ around the symmetric point $E_X$ appears.
On the contrary, no special energy is needed to define time-reversal
 symmetry Eq.\ (\ref{eq:TRS H}).

The time-independent Hamiltonian $H$ and the time-evolution operator $U^{(s)}$ for a single
time-step are related by
\begin{equation}
 U^{(s)} = e^{-i H t/\hbar},
\label{eq:U and H}
\end{equation}
where $t$ and $\hbar$ represent a time interval of the single time-step
 operation and a reduced Planck constant, respectively. 
Hereafter, we simply assume $t=\hbar=1$.
Because of Eq.\ (\ref{eq:U and H}), quasi-energy $\varepsilon \in
 {\mathbb R}$, which has $2\pi$ periodicity, is
 introduced from the eigenvalue $\lambda$ of the time-evolution
 operator $U^{(s)}$ in Eq.\ (\ref{2phase-eigenvalue.prob.}):
\begin{equation}
 \lambda = e^{-i\varepsilon}.
\label{eq:quasi-energy}
\end{equation}
When we derive the constraint of the relevant symmetries
for topological phases on the time-evolution operator from Eq.\
 (\ref{eq:symmetry H}),
we replace $E$, $E_P$, and $E_\Gamma$ of the Hamiltonian with $\varepsilon$,
$\varepsilon_P$, and $\varepsilon_\Gamma$,
respectively, in order to emphasize $2\pi$ periodicity of quasi-energy. 

Using Eqs.\ (\ref{eq:symmetry H}) and (\ref{eq:U and H}) and
 considering that only symmetric operators
 $T$ and $P$ contain the complex conjugate operator $K$, we obtain the
 relations in Eqs.\ (\ref{eq:TRS}) - (\ref{eq:chiral}).

\flushleft
\noindent {\large{\bf Appendix E}}  \\ \noindent

In Appendix E, we clarify the presence or absence of 
particle-hole symmetry Eq.\ (\ref{eq:PHS}) and
chiral symmetry Eq.\ (\ref{eq:chiral})
of  the time-evolution  operator of the complete two-phase QW.

To begin with, we rewrite the time-evolution operator so that the
 argument on the symmetries makes easy.
For simplicity, we ignore the position dependence of the phase
 $\sigma_\pm$ and write it as $\sigma_0$ in this appendix.
(The symmetries in case of the position dependent phases $\sigma_\pm$ are
 discussed in the main text.)
We also prefer to introduce an additional parameter $\theta$ into the coin
operator for the sake of general arguments of the topological phase.
Therefore, we focus on the following coin in this appendix:
\begin{equation}
 U_{\sigma_0,\theta} = \begin{bmatrix}
	 \cos(\theta) & e^{i\sigma_0}\sin(\theta) \\
         e^{-i\sigma_0}\sin(\theta)  & -\cos(\theta)	 \end{bmatrix}.
\label{eq:coin}
\end{equation}
Note that $U_+=U_{\sigma_+,\pi/4}$ and $U_-=U_{\sigma_-,\pi/4}$.

We also introduce  the split-shift operator $S_{\pm}$ defined as
\begin{subequations}
\begin{align}
 S_+&=\sum_x 
\left(
\ket{x}\bra{x}\otimes \ket{L}\bra{L} +
\ket{x}\bra{x-1}\otimes \ket{R}\bra{R}
\right),
\label{eq:split shift+}
\\
 S_-&=\sum_x 
\left(
\ket{x}\bra{x+1}\otimes \ket{L}\bra{L} +
\ket{x}\bra{x}\otimes \ket{R}\bra{R}
\right).
\label{eq:split shift-}
\end{align}
\end{subequations}
Note that multiplying $S_+$ and $S_-$ gives the standard shift operator: $S=S_+ S_- = S_- S_+$.

In order to study the relevant symmetries for topological phases, it would be better to introduce
Pauli matrices: 
\begin{eqnarray*}
 \tau_1 =
\begin{bmatrix}
0 & 1 \\
1 & 0
\end{bmatrix},\quad
 \tau_2 =
\begin{bmatrix}
0 & -i \\
i & 0
\end{bmatrix},\quad
 \tau_3 =
\begin{bmatrix}
1 & 0 \\
0 & -1
\end{bmatrix},
\end{eqnarray*}
as well as the identity matrix $\tau_0 ={\mathbb I}_2$, which act on
 the coin space.
The above three Pauli matrices are basic elements of $SU(2)$ matrices.
They satisfy the following algebra:
\begin{subequations}
\begin{align}
\text{i)}\quad\quad (\tau_i)^2&=\tau_0\quad
 (i=0,1,2,3),\label{eq:Pauli square}\\
\text{ii)}\quad\quad  \tau_i \tau_0 &= \tau_0 \tau_i =\tau_i\quad
 (i=1,2,3),\label{eq:Pauli 0i}\\
\text{iii)}\quad\quad  \tau_i \tau_j &= -\tau_j \tau_i \quad (i,j=1,2,3,\  i\ne j).
\end{align}
\end{subequations}
By expanding the exponential function and using Eqs.\
 (\ref{eq:Pauli square}) and (\ref{eq:Pauli 0i}), it is straight forward to derive the following relation:
\begin{equation}
 e^{i a \tau_i} = \cos(a)\tau_0 + i \sin(a) \tau_i\quad (i=0,1,2,3),
\label{eq:Pauli exp}
\end{equation}
where $a \in \mathbb{R}$.

We can express the shift and coin operators by Pauli matrices.
The split-shift operators in Eqs.\ (\ref{eq:split shift+}) 
and (\ref{eq:split shift-}) are written as follows:
\begin{eqnarray*}
 S_{\pm} &=& \frac{1}{2} \big[
\left(
\ket{x}\bra{x} + \ket{x}\bra{x\mp 1}
\right)\tau_0
\pm
\left(
\ket{x}\bra{x}-\ket{x}\bra{x\mp 1}
\right)\tau_3
\big].
\end{eqnarray*}

By using Eq.\ (\ref{eq:Pauli exp}),
the coin operator in Eq.\ (\ref{eq:coin}) is written as
\begin{subequations}
\begin{align}
U_{\sigma_0,\theta} &= R_{\sigma_0,\theta} \cdot \tau_3
=  e^{-i \phi} \cdot R_{\sigma_0,\theta} \cdot e^{-i \chi 
\tau_3},\quad \chi=-\phi=\pi/2,
\label{eq:coin Pauli}
\\
R_{\sigma_0,\theta}&=
\begin{bmatrix}
\cos(\theta) & -e^{i\sigma_0} \sin(\theta) \\
e^{-i\sigma_0} \sin(\theta) & \cos(\theta)
\end{bmatrix}
=e^{-i\theta\left[\sin(\sigma_0)\tau_1 + \cos(\sigma_0)\tau_2\right]}.
\label{eq:coin rotation}
\end{align}
\end{subequations}
Therefore, the time-evolution operator can be written as
 follows:

\begin{equation}
 U^{(s)} = S (\mathbb{I}_\text{p}\otimes U_{\sigma_0,\theta}) = 
e^{-i \phi}  S_-  S_+  (\mathbb{I}_\text{p}\otimes R_{\sigma_0,\theta})
  \cdot e^{-i \chi \tau_3},\quad \chi=-\phi=\pi/2.
\label{eq:U appendix}
\end{equation}

\indent

Hereafter, we examine the relevant symmetries for topological phases of
the time-evolution operator in Eq.\ (\ref{eq:U appendix}).
First, we focus on identifying chiral symmetry Eq.\ (\ref{eq:chiral}) by applying
 the method developed in Refs.\ \cite{asboth13,obuse15}.
We understand that Eq.\ (\ref{eq:chiral}) is satisfied when the time-evolution operator $U^{(s)}$ is decomposed as
 follows:
\begin{equation}
 U^{(s)} = e^{-i \varepsilon_\Gamma} F \cdot \Gamma^{-1} F^{-1} \Gamma,
\label{eq:F}
\end{equation}
with the help of the relation $\Gamma=\Gamma^{-1}$.
In order to confirm Eq.\ (\ref{eq:F}) for the time-evolution
 operator of the complete  two-phase QW,
we need to shift the origin of time by a half of the coin operator to
 fit into  a {\it symmetry time frame} introduced in Ref.\
 \cite{asboth13}.
We also use the commutation relation between  $S_\pm$ and $e^{-i \chi \tau_3}$, since both are described only by $\tau_0$ and
$\tau_3$ components. Thereby, we obtain the single-step time-evolution operator in
 the symmetry time frame:

\begin{eqnarray}
 U^{(s)\prime} &=& 
e^{-i\phi}
(\mathbb{I}_\text{p} \otimes
R_{\sigma_0,\theta/2})   \cdot 
e^{-i (\chi/2)  \tau_3} \cdot S_- \cdot S_+ \cdot e^{-i (\chi/2)
\tau_3}\cdot (\mathbb{I}_\text{p} \otimes R_{\sigma_0,\theta/2}).
\label{eq:U'}
\end{eqnarray}
Note  we use relations $R_{\sigma_0,\theta} =
R_{\sigma_0,\theta/2} \cdot R_{\sigma_0,\theta/2}$ and
$e^{-i \chi \tau_3} =e^{-i (\chi/2) \tau_3}\cdot
e^{-i (\chi/2) \tau_3}$.

Comparing the global phase factors in Eq.\ (\ref{eq:F}) with Eq.\ (\ref{eq:U'}), 
we identify $\varepsilon_\Gamma=\phi=-\pi/2$.
Then, comparing the rest parts, we identify $F=(\mathbb{I}_\text{p} \otimes R_{\sigma_0,\theta/2})   \cdot 
e^{-i (\chi/2)  \tau_3} \cdot S_-$ and  chiral symmetry requires the condition:

\begin{align*}
\Gamma 
(
S_+ \cdot e^{-i (\chi/2)
\tau_3}\cdot (\mathbb{I}_\text{p} \otimes R_{\sigma_0,\theta/2})
)
\Gamma^{-1}
=
((\mathbb{I}_\text{p} \otimes R_{\sigma_0,\theta/2})   \cdot 
e^{-i (\chi/2)  \tau_3} \cdot S_-)^{-1}.
\end{align*}

Taking account of $(R_{\sigma_0,\theta})^{-1}=R_{\sigma_0,-\theta}$, 
chiral symmetry is established if the following conditions are satisfied:
\begin{subequations}

\begin{align}
\Gamma (\mathbb{I}_\text{p} \otimes R_{\sigma_0,\theta}) \Gamma^{-1} &= \mathbb{I}_\text{p} \otimes R_{\sigma_0,-\theta},
\label{eq:condition coin}\\
\Gamma e^{-i (\chi/2)  \tau_3} \cdot S_\pm \Gamma^{-1} &=
 e^{+i (\chi/2)  \tau_3} \cdot (S_\mp)^{-1}.
\label{eq:condition shift}
\end{align}

\end{subequations}
When $\sigma_0=0$, the chiral symmetry operator 
$\Gamma=\sum_x |x\rangle\langle x| \otimes \tau_1$ satisfies the above conditions\cite{obuse15}. However, for the arbitrary value of $\sigma_0$,
the above $\Gamma$  does not satisfy Eq.\ (\ref{eq:condition coin}), while
it does Eq.\ (\ref{eq:condition shift}). The point is
$R_{\sigma_0,\theta}$ in Eq.\ (\ref{eq:coin rotation}) contains the $
\tau_1$ component which commutes with the chiral symmetry operator
 $\Gamma$. 
This problem is solved by removing the $\tau_1$ component by a unitary
 transformation before $\tau_1$ acts, and then
recovering it by another unitary transformation. 
This problem can be solved by including additional unitary
 transformations into the chiral symmetry operator.
In summary, $U^{(s)\prime}$ has chiral symmetry with the  following chiral symmetry
operator $\Gamma$;
\begin{subequations}
\begin{align}
& \Gamma\, e^{i \varepsilon_\Gamma}U^{(s)\prime}\, \Gamma^{-1} = 
\left(e^{i \varepsilon_\Gamma}U^{(s)\prime}\right)^{-1},\quad \text{with
 }\varepsilon_\Gamma=-\frac{\pi}{2},
\label{eq:chiral appendix}\\ 
&\Gamma=\sum_x |x\rangle \langle x| \otimes V_{\sigma^\prime}\, \tau_1\,
 V_{\sigma^\prime}^{-1},
\quad V_{\sigma^\prime}:=e^{i(\sigma^\prime/2)\tau_3},\quad \sigma^\prime=\sigma_0.
\label{eq:def Gamma}
\end{align}
\end{subequations}
\indent

Next, we focus on particle-hole symmetry Eq.\ (\ref{eq:PHS}).
While the symmetry time frame is unnecessary to define particle-hole
 symmetry, we keep using the time-shifted time-evolution operator $U^{(s)\prime}$.
Taking into account the fact that  $V_{\sigma_0}^{-1} R_{\sigma_0,\theta} V_{\sigma_0} =
 R_{0,\theta}$ and $e^{-i\phi} e^{-i \chi \tau_3}=\sigma_z$ with $\chi=-\phi=\pi/2$ are expressed
 by real numbers,
particle-hole symmetry of $U^{(s)\prime}$   is identified as follows:
\begin{subequations}
\begin{align*}
& P\, e^{i \varepsilon_P}U^{(s)\prime}\, P^{-1} = e^{i
 \varepsilon_P}U^{(s)\prime},\quad \text{with } \varepsilon_P=0,
\label{eq:PHS appendix}\\
&P=\sum_x |x\rangle \langle x| \otimes V_{\sigma^\prime}\, \tau_0 K\,
 V_{\sigma^\prime}^{-1},\quad \sigma^\prime=\sigma_0.
\end{align*} 
\end{subequations}

\flushleft
\noindent {\large{\bf Appendix F}}  \\ \noindent
\label{app:topological invariant}

Once chiral symmetry is identified, the topological number of the
time-evolution operators $U^{(s)\prime}$ can be calculated from the
Berry phase, or the winding number.
In case of the presence of chiral symmetry, an important difference of topological phases of quantum walks
from those of time-independent topological insulators is the existence of two topological
 numbers, $\nu_{\varepsilon_\Gamma}$ and $\nu_{\varepsilon_\Gamma+\pi}$, because of $2 \pi$
 periodicity of quasi-energy.
According to the method  to calculate the two topological numbers developed in Ref.\ \cite{asboth13},
we prepare two time-evolution operators which have different symmetry time frames. Considering Eq.\ (\ref{eq:F}), 
we find the other time-evolution operator
 $U^{(s)}=e^{-i\varepsilon_\Gamma} \Gamma^{-1} F^{-1} \Gamma \cdot F$, in
 which the order of operators is inverted, also satisfies chiral symmetry
 Eq.\ (\ref{eq:chiral}). For explicitly, the other chiral symmetric
 time-evolution operator of the complete two-phase QW is expressed as

\begin{equation*}
U^{(s)\prime\prime} = 
e^{-i\phi}
S_+ \cdot e^{-i (\chi/2)
\tau_3}\cdot 
(\mathbb{I}_\text{p} \otimes R_{\sigma_0,\theta/2}) \cdot
(\mathbb{I}_\text{p} \otimes R_{\sigma_0,\theta/2}) \cdot 
e^{-i (\chi/2)  \tau_3} \cdot S_-,\quad \chi=-\phi=\pi/2.
\end{equation*}
We also apply a unitary transformation into the two time-evolution operators
 so as to make calculations of the Berry phase simple.
 Hereafter, we treat the following two time-evolution operators under the
 unitary transformation:
\begin{align*}
\tilde{U}^{(s)\prime} &= V^{-1}\, U^{(s)\prime}\, V,\\
\tilde{U}^{(s)\prime\prime} &= V^{-1}\, U^{(s)\prime\prime}\, V,\\
V&=V_{\sigma^\prime} e^{-i(\pi/4)\tau_2},\quad \sigma^\prime=\sigma_0.
\end{align*}

Since there are no position dependent parameters in $\tilde{U}^{(s)\prime}$ and $\tilde{U}^{(s)\prime\prime}$,
the system has translation invariance, and then 
we consider the time-evolution operators 
$\tilde{U}^{(s)\prime}(k)$ and $\tilde{U}^{(s)\prime\prime}(k)$
in the momentum (wave-number) space representations
by applying the Fourier transformation. 
The winding number $\nu$ and the Berry phase $\varphi_B$ is calculated 
from the eigen function $\psi(k)$ of the time-evolution operators in the
momentum space representation as
\begin{equation*}
\nu=\frac{\varphi_B}{\pi},\quad \varphi_B = \frac{1}{i}\int_{-\pi}^{\pi} dk\,  \psi^*(k) \frac{d}{dk} \psi(k).
\label{eq:Berry phase}
\end{equation*}
\indent

The time-evolution operator $\tilde{U}^{(s)\prime}$ in the momentum
 representation becomes 
\begin{align*}
\tilde{U}^{(s)\prime}(k) &=
\begin{bmatrix}
c_k c_\theta & -r_k e^{i \phi_k^{\prime}} \\
r_k e^{-i \phi_k^{\prime}} & c_k c_\theta
\end{bmatrix},\\
r_k &= \sqrt{1-c_k^2 c_\theta^2}\ge 0,\quad
e^{i \phi_k^{\prime}} =
 r_k^{-1}(c_k s_\theta + i s_k),
\end{align*}
with the shorthands
\begin{equation*}
 c_x := \cos(x), \quad s_x := \sin(x).
\end{equation*}
Note that we redefine $k-\chi$ as $k$ since the shift of momentum by
$\chi$ does not change the Berry phase.
We also put $\phi=0$ since the phase $\phi$ only shifts the origin of quasi-energy,
 which again does not change the Berry phase.
The eigenvalues of $\tilde{U}^{(s)\prime}$ is 
\begin{equation*}
 \lambda_\pm^\prime = e^{\pm i \varepsilon} =c_k c_\theta \pm i r_k.
\end{equation*}
The corresponding eigenvectors are
\begin{equation*}
 \psi_\pm^{\prime}(k) = \frac{1}{\sqrt{2}}  
\begin{pmatrix}
 \pm i e^{i \phi_k^{\prime}} \\
1
\end{pmatrix} .
\end{equation*}
Then, the winding number $\nu^\prime$ calculated from $\psi_\pm^{\prime}(k)$ is given by
\begin{eqnarray}
\nu^\prime &=& \frac{1}{2\pi} \oint d \phi_k^{\prime}
\nonumber \\
&=&
\left\{
\begin{array}{ll}
+1 &\quad \text{for}\  0<\theta<\pi, \\
-1 &\quad \text{for}\  -\pi<\theta<0.
\end{array}
\right.
\label{eq:ap nu'}
\end{eqnarray}
Note that the direction of the integration path depends on the sign of $s_\theta$ since $r_k e^{i \phi_k^\prime}=x(k)+iy(k)$ with $x(k)=s_\theta c_k$ and $y(k)=s_k$ goes around the origin along the elliptic circle $x^2/s_\theta^2 + y^2 = 1$ in the complex plane.

In the same way, the time-evolution operator
 $\tilde{U}^{(s)\prime\prime}$ becomes
\begin{align*}
\tilde{U}^{(s)\prime\prime} &=
\begin{bmatrix}
c_k c_\theta & -r_k e^{i \phi_k^{\prime\prime}} \\
r_k e^{-i \phi_k^{\prime\prime}} & c_k c_\theta
\end{bmatrix},\\
e^{i \phi_k^{\prime\prime}} &= r_k^{-1}(s_\theta + i s_k c_\theta).
\end{align*}
Again, we redefine $k-\chi$ as $k$ and put $\phi=0$.
The eigenvalue of $\tilde{U}^{(s)\prime\prime}$ is the same with $\lambda_\pm^\prime$ of
$\tilde{U}^{(s)\prime}$ and 
the corresponding eigenvectors are
\begin{equation*}
 \psi_\pm^{\prime\prime}(k) = \frac{1}{\sqrt{2}}  
\begin{pmatrix}
 \pm i e^{i \phi_k^{\prime\prime}} \\
1
\end{pmatrix}. 
\end{equation*}
The winding number $\nu^{\prime\prime}$ calculated from $\psi_\pm^{\prime\prime}(k)$ is given by
\begin{eqnarray}
 \nu^{\prime\prime} &=& \frac{1}{2\pi} \oint d \phi_k^{\prime\prime}
\nonumber \\
&=&
0 \quad \text{for}\  -\pi<\theta<\pi.
\label{eq:ap nu''}
\end{eqnarray}

Ref.\ \cite{asboth13} derives that  the topological numbers $\nu_0$ and
$\nu_\pi$ for quasi-energy
$0$ and $\pi$, respectively, are given by
\begin{eqnarray}
 \nu_0 = \frac{\nu^{\prime\prime}+\nu^{\prime}}{2},\quad\quad
 \nu_\pi = \frac{\nu^{\prime\prime}-\nu^{\prime}}{2},
\label{eq:nu0 nupi}
\end{eqnarray}
if the symmetric point of quasi-energy is $0$.
Hence, by using Eqs.\ (\ref{eq:ap nu'})-(\ref{eq:nu0 nupi}), we obtain the topological number $\nu_0$ and
$\nu_\pi$ for quasi-energy $\varepsilon=0$ and $\pi$, respectively, as follows:
\begin{eqnarray*}
 (\nu_0, \nu_\pi) = \left\{
\begin{array}{lr}
(+1/2,-1/2) & \quad 0 < \theta < \pi, \\
(-1/2,+1/2) & \quad -\pi < \theta < 0.
\end{array}
\right.
\end{eqnarray*}
According to the bulk-edge correspondence, the number of edge states
comes from the absolute value of the difference of topological numbers
of the two adjacent spatial regions. This allows us to add a constant to
all topological numbers. 
Also, we need recover $\phi(=\varepsilon_\Gamma)$ to $-\pi/2$.
Thereby, by adding $1/2$ into the all topological numbers and shifting
the origin of quasi-energy by $-\pi/2$, 
we reach to the following result:
\begin{eqnarray}
 (\nu_{-\pi/2}, \nu_{+\pi/2}) = \left\{
\begin{array}{lr}
(1,0) & \quad 0 < \theta < \pi, \\
(0,1) & \quad -\pi < \theta < 0.
\end{array}
\right.
\label{eq:topological num}
\end{eqnarray}

Finally, we apply the above result to the complete two-phase QW. 
First, we consider the  topological numbers for the complete two-phase QW with the phase
 $\sigma_+$ in the region $x\ge 0$.
By putting $\sigma_0=\sigma^\prime=\sigma_+$ of $R_{\sigma_0,\theta}$ and
 $V_{\sigma^\prime}$, and fixing $\theta=\pi/4$, we obtain
Eq. (\ref{eq:topological invariant +}). 
Then, we focus on those for the complete two-phase QW with the phase $\sigma_-$ in the
 region $x \le -1$. In order to retain chiral symmetry of the whole of
complete two-phase QW, we have to use the same chiral symmetry operator
 $\Gamma$. 
If we choose the phase $\sigma^\prime=\sigma_+$ in $V_{\sigma^\prime}$ in Eq.\ (\ref{eq:def
 Gamma}), 
$\sigma_-$ should be 
\begin{equation*}
 \sigma_-=\sigma_+ + n\pi\quad (n \in {\mathbb Z}). 
\end{equation*}
Taking account of the relation

\begin{equation*}
 V_{\sigma_+}^{-1}\,((\mathbb{I}_\text{p} \otimes R_{\sigma_-, \theta})\,  V_{\sigma_+} =
 V_{\sigma_+}^{-1}\,(\mathbb{I}_\text{p} \otimes R_{\sigma_+ + n \pi, \theta})\,  V_{\sigma_+} = \mathbb{I}_\text{p} \otimes R_{\sigma_+, (-1)^n\theta},
\end{equation*}
and Eq.\ (\ref{eq:topological num}), we obtain Eq.\ 
(\ref{eq:topological invariant -}).

\flushleft
\noindent {\large{\bf Appendix G}}  \\ \noindent
\label{sec:path}

The QW on the path has the finite number of bases
\[
\{|-N,R\rangle,
|-N+1,L\rangle,|-N+1,R\rangle,\cdots,|N-2,L\rangle,|N-2,R\rangle,|N-1,L\rangle\},
\]
and the coin and shift operators are modified as follows:
\begin{eqnarray*}
U&=&
\sum_{x=-N+1}^{N-2} |x\rangle \langle x| \otimes U_x
\nonumber \\
&&
+
\frac{1}{\sqrt{2}}e^{-i\sigma_-} |-N\rangle \langle -N| \otimes
|R\rangle \langle R| 
+
\frac{1}{\sqrt{2}}e^{i\sigma_+} |N-1\rangle \langle N-1| \otimes
|L\rangle \langle L| ,\\
 S &=&
\sum_{x=-N+1}^{N-2} \Big(
|x\rangle \langle x+1| \otimes |L\rangle \langle L| 
+|x\rangle \langle x-1| \otimes |R\rangle \langle R| 
\Big)
\nonumber \\
&&+
|-N\rangle \langle -N+1| \otimes |R\rangle \langle L| 
+
|N-1\rangle \langle N-2| \otimes |L\rangle \langle R| .
\end{eqnarray*}

\end{document}